\documentclass[aps,prd,showpacs,twocolumn,superscriptaddress,nofootinbib,preprintnumbers]{revtex4-1} 

\makeatletter
\def\p@subsection{}
\makeatother

\usepackage{color,graphicx,amsmath,amssymb,lipsum}
\usepackage[colorlinks=true,citecolor=blue,linkcolor=blue,urlcolor=blue, backref=false,pdfborder={0 0 0}]{hyperref}

\usepackage[normalem]{ulem}
\usepackage[utf8]{inputenc} 
\usepackage{mathtools}
\usepackage{wrapfig}

\usepackage{float}

\usepackage{ulem}
\usepackage{diagbox}

\usepackage[dvipsnames]{xcolor}

\newcommand{\be}{\begin{equation}}
\newcommand{\ee}{\end{equation}}
\newcommand{\beqa}{\begin{eqnarray}}
\newcommand{\eeqa}{\end{eqnarray}}

\renewcommand\L{\Lambda}

\newcommand{\kmax}{k_{\rm max}}
\newcommand{\hMpc}{h \text{Mpc}^{-1}}

\newcommand{\bseq}{\begin{subequations}}
\newcommand{\eseq}{\end{subequations}}

\renewcommand{\ln}{\mathop{\rm ln}\nolimits}

\def\gsim{\raise0.3ex\hbox{$\;>$\kern-0.75em\raise-1.1ex\hbox{$\sim\;$}}}
\def\lsim{\raise0.3ex\hbox{$\;<$\kern-0.75em\raise-1.1ex\hbox{$\sim\;$}}}

\def\beqn#1{\begin{equation}\label{#1}}
\def\eeqn{\end{equation}}

\def\beqa#1{\begin{eqnarray}\label{#1}}
\def\eeqa{\end{eqnarray}}

\def\Z2{$\mathcal{Z_2}$}

\newcommand {\ignore}[1]{}

\begin{document}


\title{Cosmological constraints from the power spectrum of eBOSS emission 
line galaxies}

\author{Mikhail M. Ivanov}\email{mi1271@nyu.edu}\affiliation{Center for Cosmology and Particle Physics, Department of Physics, New York University, New York, NY 10003, USA}
\affiliation{Institute for Nuclear Research of the
Russian Academy of Sciences, \\ 
60th October Anniversary Prospect, 7a, 117312
Moscow, Russia
}

\begin{abstract} 
We present cosmological parameter measurements 
from the effective field theory-based full-shape analysis 
of the 
power spectrum of emission line galaxies (ELGs).
First, we perform extensive tests 
on simulations and
determine appropriate scale cuts 
for the perturbative 
description of the ELG power spectrum.
We study in detail non-linear redshift-space distortions (``fingers-of-God'')
for this sample and show that they are somewhat weaker than those of luminous red galaxies.
This difference is not significant for current data, but may become important 
for future surveys like Euclid/DESI.
Then we analyze recent measurements 
of the ELG power spectrum from the extended Baryon acoustic Oscillation Spectroscopic Survey (eBOSS) within
the $\nu\Lambda$CDM model.
Combined with the BBN baryon density prior, the ELG pre- 
and post-reconstructed power 
spectra alone constrain
the matter density 
$\Omega_m=0.257_{-0.045}^{+0.031}$,
the current mass fluctuation amplitude 
$\sigma_8=0.571_{-0.076}^{+0.052}$, and the Hubble constant 
$H_0=84.5_{-7}^{+5.8}$ km/s/Mpc (all at 68\% CL).
Combining with other full-shape and BAO data 
we measure 
$\Omega_m=0.327_{-0.016}^{+0.014}$,
$\sigma_8=0.69_{-0.045}^{+0.038}$, 
and 
\mbox{$H_0=68.6_{-1.1}^{+1}$ km/s/Mpc}.
The total neutrino mass is constrained to be $M_{\rm tot}<0.63$ eV
(95\% CL)
from the BBN, full-shape and BAO data only.
Finally, we discuss the apparent $\sim 3\sigma$ discrepancy 
in the inferred clustering amplitude between 
our full shape analysis and the cosmic microwave background data.
\end{abstract}

\maketitle

\section{Introduction}

Historically, the distribution of matter and its luminous 
tracers on the largest observable scales was one of the earliest sources
of cosmological parameter measurements~\cite{1980lssu.book.....P}.
In particular, some of the first ever limits on 
matter and baryon densities were extracted from the 
shape of the 2dFGRS power spectrum~\cite{Percival:2001hw}. 
Around this time, the cosmic microwave background (CMB)
measurements started to become progressively more and more precise,
which soon made them the leading source of cosmological 
information. The program of 
measuring cosmological parameters from the CMB 
anisotropies 
has culminated 
with the \textit{Planck} satellite data,
which determined parameters of the minimal $\L$CDM model 
with percent precision~\cite{Aghanim:2018eyx}.
Similar precisions were recently reached by other CMB
experiments, e.g. ACT~\cite{Aiola:2020azj} and SPT~\cite{Aylor:2017haa,Bianchini:2019vxp}.
During this program, the large scale structure data was crucial 
in combined analyses with the CMB, allowing one to  
break parameter degeneracies in the 
standard model
extensions. 
However, the 
quality of 
large-scale structure (LSS) data 
has also improved in the last years
to the extent that now several different datasets 
are
able to constrain 
our Universe
without any input
from the CMB
anisotropies, and with competitive precision, see e.g. results of photometric
surveys 
KIDS~\cite{Asgari:2020wuj}, DES~\cite{Abbott:2017wau,Abbott:2021bzy}, and the Baryon acoustic Oscillation Spectroscopic Survey (BOSS)~\cite{Alam:2016hwk,Alam:2020sor}.

Over the last decade, 
the volume of spectroscopic 
galaxy clustering surveys has been 
continuously growing, 
which calls for a constant 
reassessment of 
analysis techniques.
Early cosmological parameter constraints from 
redshift space clustering data were based on consistent analyses of the galaxy power 
spectrum shape, which were carried out in 
the same fashion as the 
CMB power spectrum shape 
analysis~\cite{Tegmark:2001xb,Tegmark:2003uf,Cole:2005sx,Percival:2006gt}.
These works were based on 
approximate phenomenological models
for nonlinear effects
and fits from simulations (e.g. the ``Halofit'' prescription~\cite{Smith:2002dz}),
which were computationally efficient, but 
raised concerns about systematic 
uncertainties of these methods. 
After the first detection of the baryon acoustic oscillations (BAO)
in the clustering of SDSS and 2dFGRS
galaxies~\cite{Percival:2007yw}, the focus  
shifted from the power 
spectrum shape to this feature, 
which is widely believed to be more robust  than the shape
w.r.t. nonlinearities
and observational systematics. 
At some point it became standard 
to analyze the entire power spectrum 
just like 
the 
BAO feature: by using the fixed 
linear power spectrum shape 
and varying cosmology only by means of 
scaling
parameters capturing the 
Alcock-Paczynski (AP) effect~\cite{Alcock:1979mp}
and redshift-space distortions (RSD), see e.g.~\cite{Cabre:2008sz,Blake:2011ep,Samushia:2011cs,Chuang:2011fy}.
We dub this approach 
the ``RSD/AP fixed template method''.

It is important to emphasize that it has a number of limitations.
First, a number of works have shown that 
the measurements of scaling parameters have a non-negligible dependence
on the choice of the underlying linear power spectrum shape template, 
see e.g. Fig.~3 of Ref.~\cite{deMattia:2020fkb} and Fig.~14 of ~\cite{Smith:2020stf}. 
The typical approach is to compute 
the linear power spectrum 
for a set of fiducial cosmological 
parameters that agree with the \textit{Planck} CMB 
measurements. This is justified as long as
we are to combine the RSD/AP measurements 
with the \textit{Planck} CMB data, which
determines these parameters much better than the galaxy power spectrum
shape itself. 
This was roughly true for surveys like BOSS and eBOSS, but 
will not be the case for future surveys like DESI~\cite{Aghamousa:2016zmz} and Euclid~\cite{Laureijs:2011gra}.
The second related problem 
is an  
interpretation of the 
AP/RSD parameter constraints 
in cosmological models different from 
the one used to fix the template.
Indeed, the linear power spectrum 
shape depends both on the details of early and 
late time evolution. 
The early-type processes (e.g. details of the matter-radiation
equality)
modulate the spectrum at high redshifts, 
while e.g.
massive neutrinos lead 
to late-time shape distortions~\cite{Lesgourgues:2006nd}.
Therefore, it is not clear to what extent the standard 
AP/RSD constraints are valid 
for cosmologies different from the fiducial one.
In other words, using the fixed template is equivalent to imposing 
strong \textit{Planck}/$\L$CDM-motivated priors in the analysis,
which may bias the result of cosmological inference for 
extended cosmological models. 
Recent examples of such situations can be found in 
analyses of the early dark energy model~\cite{Poulin:2018cxd},
which alternates the matter power spectrum shape at high redshifts
as compared to $\L$CDM~\cite{Hill:2020osr,Ivanov:2020ril,DAmico:2020ods}, and hence
raises concerns about the legitimacy
of the standard RSD/AP constraints.

The third drawback of the standard RSD/AP analysis 
is that it ignores the shape signal.
Historically this was partly motivated by the absence of a 
robust model for nonlinear effects. However, 
recent theoretical efforts in the effective field theory (EFT) of 
large-scale structure~\cite{Baumann:2010tm,Carrasco:2012cv}
and its extensions~\cite{Porto:2013qua,Vlah:2015sea,Blas:2015qsi,Chen:2020zjt} 
have demonstrated that an accurate 
and efficient
analytic description of galaxy clustering
is possible. 
In the quasi-linear regime all effects of
structure formation simplify and can be taken 
into account by means of a systematic
expansion similar to the derivative expansion in field theory.
In this regime one can accurately capture quasi-linear clustering 
of underlying matter~\cite{Carrasco:2013sva,Carrasco:2013mua,Foreman:2015lca,Baldauf:2015aha,Baldauf:2015zga,Baldauf:2015tla,Konstandin:2019bay}, galaxy bias~\cite{Senatore:2014eva,Angulo:2015eqa,Mirbabayi:2014zca,Assassi:2014fva,Desjacques:2016bnm}, redshift-space distortions~\cite{Senatore:2014vja,Lewandowski:2015ziq,Perko:2016puo,Ivanov:2018gjr,Vlah:2018ygt}, 
baryonic effects~\cite{Lewandowski:2014rca}, 
massive neutrinos~\cite{Senatore:2017hyk}, 
non-linear smearing of the BAO~\cite{Senatore:2014via,Vlah:2015zda,Blas:2016sfa,Senatore:2017pbn,Baldauf:2015xfa}
and other effects needed to understand galaxy survey data. 
Thus, EFT provides a consistent and accurate model
for the whole power spectrum shape
without necessity to decompose the signal into 
partial information channels such as
RSD, BAO, or AP. Besides, it offers a 
tool to estimate theory systematic errors~\cite{Baldauf:2016sjb,Chudaykin:2020hbf}.

Within the EFT approach, the power spectrum shape itself becomes a rich 
source of cosmological information~\cite{Ivanov:2019pdj,DAmico:2019fhj,Philcox:2020vvt,Philcox:2020xbv}. 
Extracting this information
is clearly 
not possible with 
the standard RSD/AP fixed template methodology.
The recent EFT analyses of the BOSS 
data have shown that the galaxy power spectrum 
can be used for precise measurements of 
fundamental cosmological 
parameters, such as the Hubble constant $H_0$
or the matter density
fraction $\Omega_m$, both in the $\Lambda$CDM model
and in its extensions. In particular, 
the EFT-based full-shape BOSS likelihood can constrain
$w$CDM~\cite{DAmico:2020kxu,DAmico:2020tty}, 
$w_0 w_a$CDM, $o\Lambda$CDM~\cite{Chudaykin:2020ghx}, 
early dark energy models~\cite{Ivanov:2020ril,DAmico:2020ods},
ultralight axion dark matter models~\cite{Lague:2021frh}, as well as models with
massive neutrinos and extra relativistic degrees of freedom~\cite{Ivanov:2019hqk}.

Recent forecasts suggests that full-shape cosmological constraints can dramatically improve in the era
of future spectroscopic surveys such as DESI and Euclid~\cite{Chudaykin:2019ock,Philcox:2020xbv,Sailer:2021yzm}.
An important technical aspect of these surveys is that
they will mainly target emission line galaxies (ELGs), whereas the 
recent surveys like BOSS and eBOSS used luminous red galaxies (LRGs)
as their main samples~\cite{Alam:2016hwk,Alam:2020sor}. 
Current data suggests that emission line galaxies 
reside in relatively low-mass halos and hence are less biased than the LRGs, see e.g.~\cite{Avila:2020rmp}. 
This is important in the context of the EFT where the accuracy of  
the perturbative expansion is controlled by a wavenumber cutoff, $k_M$,
that can be estimated as the inverse
Lagrangian size of typical halos hosting 
ELGs~\cite{Senatore:2014eva,Desjacques:2016bnm}. 
Lighter halos imply a larger cutoff and hence the EFT modeling,
at face value, may work better for that sample.
The situation becomes more complicated 
when we take into account 
redshift space distortions, which is the dominant 
source of nonlinearity in the observed data.
The effective cutoff of the redshift space mapping 
is dominated by the galaxy velocity dispersion, 
$k^{r}_{\rm NL}\sim \sigma_v^{-1}$,
which is quite large for LRGs~\cite{Ivanov:2019pdj,Chudaykin:2020hbf}. For ELGs the current data suggests
the velocity dispersion may be lower than
that of the LRG sample~\cite{Orsi:2009mj,Okumura:2015lvp,Orsi:2017ggf,deMattia:2020fkb},
 and hence one 
might hope that the EFT modeling can be pushed to shorter scales
as compared to previous analyses based on LRGs.

\begin{figure*}[htb!]
\begin{minipage}{0.5\textwidth}
\centering
\includegraphics[width=0.95\textwidth]{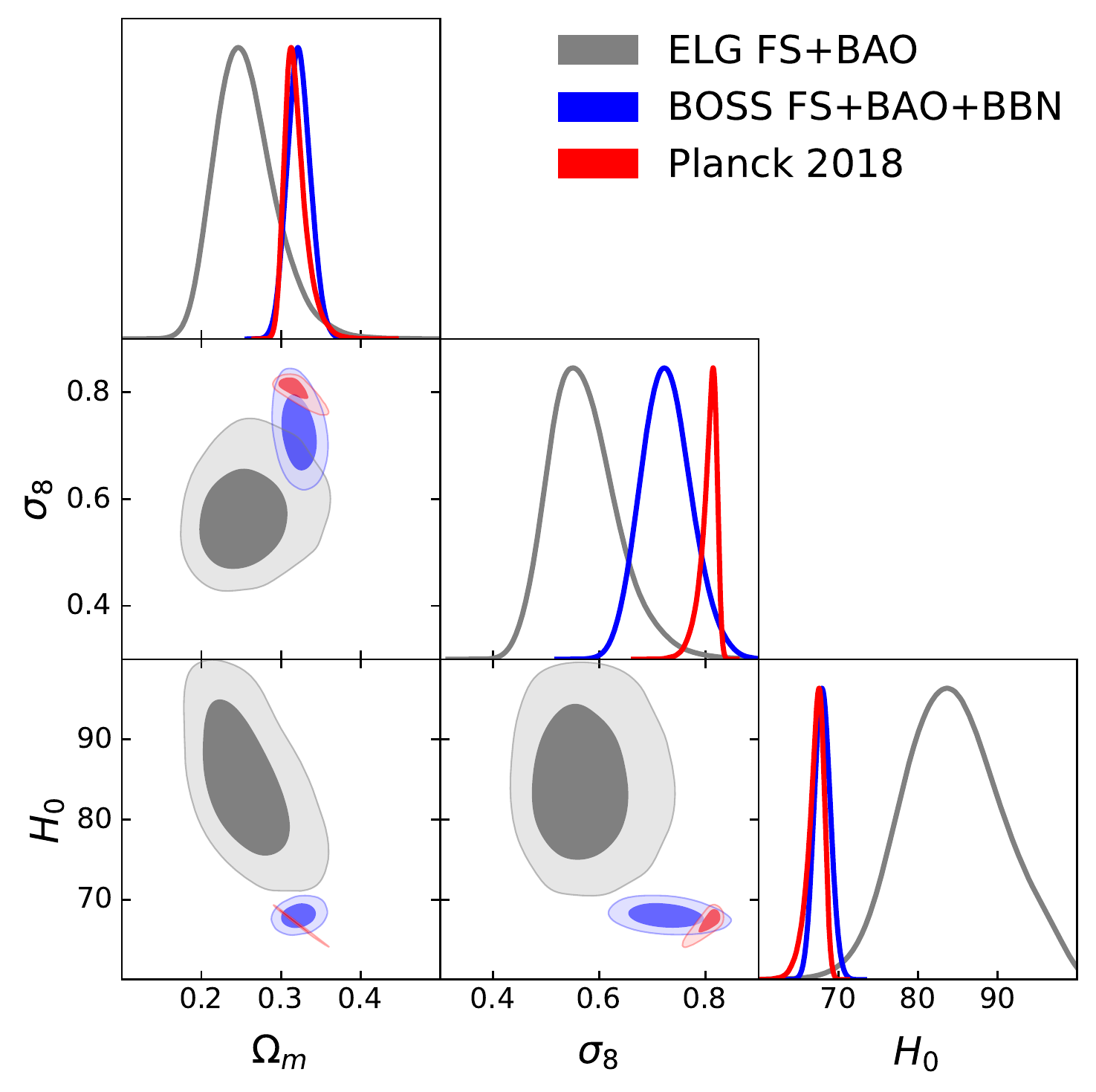}
\end{minipage}%
\centering
 \begin{minipage}{0.5\textwidth}
\includegraphics[width=0.95\textwidth]{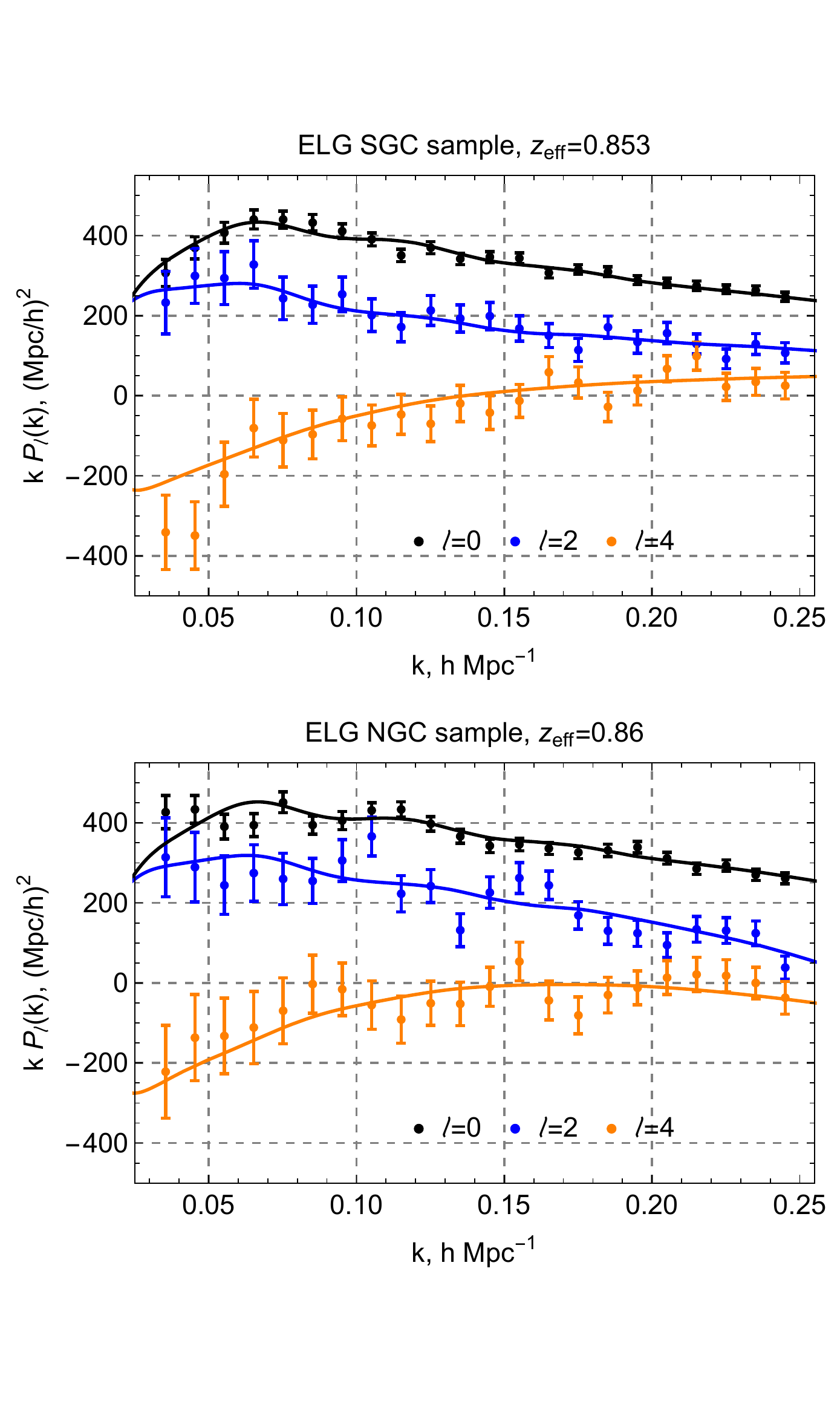}
\end{minipage}
\vspace{-1cm}
\caption{
\textit{Left panel:} marginalized constraints on 
mass fluctuation amplitude $\sigma_8$ 
and matter fraction $\Omega_m$ in $\L$CDM 
from the power spectrum of eBOSS emission line galaxies (ELGs).
For comparison we also show the \textit{Planck} CMB 2018~\cite{Aghanim:2018eyx}  
and BOSS DR12 combined full-shape (FS) + BAO + BBN contours~\cite{Chudaykin:2020ghx}.
\textit{Right panel:} ELG power spectra 
monopole ($\ell=0$), quadrupole ($\ell=2$)
and hexadecapole ($\ell=4$)
moments measured from South Galactic Cap and North Galactic Cap 
eBOSS data chunks. 
The best-fit $\Lambda$CDM
model from the combined analysis of both chunks
is plotted in solid lines. The shown error bars correspond to 
the square root of
diagonal elements of the covariance matrix.
The large-scale structure behavior 
of multipoles is very different from the linear theory
Kaiser prediction because of 
the pixelation scheme used to mitigate angular
systematics. 
\label{fig:money} } 
\end{figure*}
\begin{table*}[t!]
  \begin{tabular}{|c||c|c|c|c|} \hline
   \diagbox{ {\small Param.}}{\small Dataset}  
   &  ELG FS+BAO
   & BOSS FS+BAO+BBN
   & BOSS+ELG FS+BAO+BBN 
   &  \textit{Planck} 2018
      \\ [0.2cm]
\hline
$\omega_{cdm}$   
& $0.1538_{-0.028}^{+0.021}$
& $0.1237_{-0.0095}^{+0.0081}$
& $0.1286_{-0.0097}^{+0.0081}$ 
& $0.1201_{-0.0014}^{+0.0013}$
\\ 
\hline
$10^2\omega_{b}$   
& $-$
& $-$
& $-$
& $2.238_{-0.015}^{+0.016}$
\\ 
\hline 
  $h$   & $0.845_{-0.07}^{+0.058}$
  & $0.6801_{-0.011}^{+0.01}$
  & $0.6858_{-0.011}^{+0.01}$
  & $0.6714_{-0.0072}^{+0.013}$
  \\ \hline
$\ln(10^{10}A_s)$   & $2.155_{-0.31}^{+0.29}$
& $2.877_{-0.21}^{+0.17}$
& $2.733_{-0.19}^{+0.15}$
& $3.045_{-0.016}^{+0.014}$
\\ 
\hline
$n_s$  
& $-$
& $0.9774_{-0.075}^{+0.068}$
& $0.9339_{-0.068}^{+0.063}$
& $0.9646_{-0.0045}^{+0.0045}$
\\ 
\hline
$M_{\rm tot}$, eV
& $-$
& $<0.71~$(95\%CL)
& $<0.63~$(95\%CL)
& $<0.26~$(95\%CL)
\\
\hline
\hline
$\Omega_m$   & $0.2564_{-0.045}^{+0.031}$
& $0.3225_{-0.015}^{+0.014}$
& $0.3267_{-0.016}^{+0.014}$
& $0.3191_{-0.016}^{+0.0085}$
\\ \hline
$\sigma_8$   & $0.5713_{-0.076}^{+0.052}$
& $0.7263_{-0.052}^{+0.044}$ 
& $0.6896_{-0.045}^{+0.038}$
& $0.8053_{-0.0091}^{+0.0089}$ 
\\ 
\hline
\end{tabular}
\caption{Mean values and 68\% CL minimum credible
intervals for the parameters of the $\nu\L$CDM model.
For the total neutrino mass $M_{\rm tot}$ we present $95\%$ 
upper bounds.
 The BBN prior on $\omega_b$ is assumed in the three LSS analyses, and the corresponding posterior is not displayed.
 Dashes indicate parameters that cannot be significantly 
 constrained.
The top group are the parameters directly sampled in the MCMC chains, the bottom ones
are derived parameters.}
\label{table1}
\end{table*}

\begin{figure*}[!htb]
\centering
\includegraphics[width=0.95\textwidth]{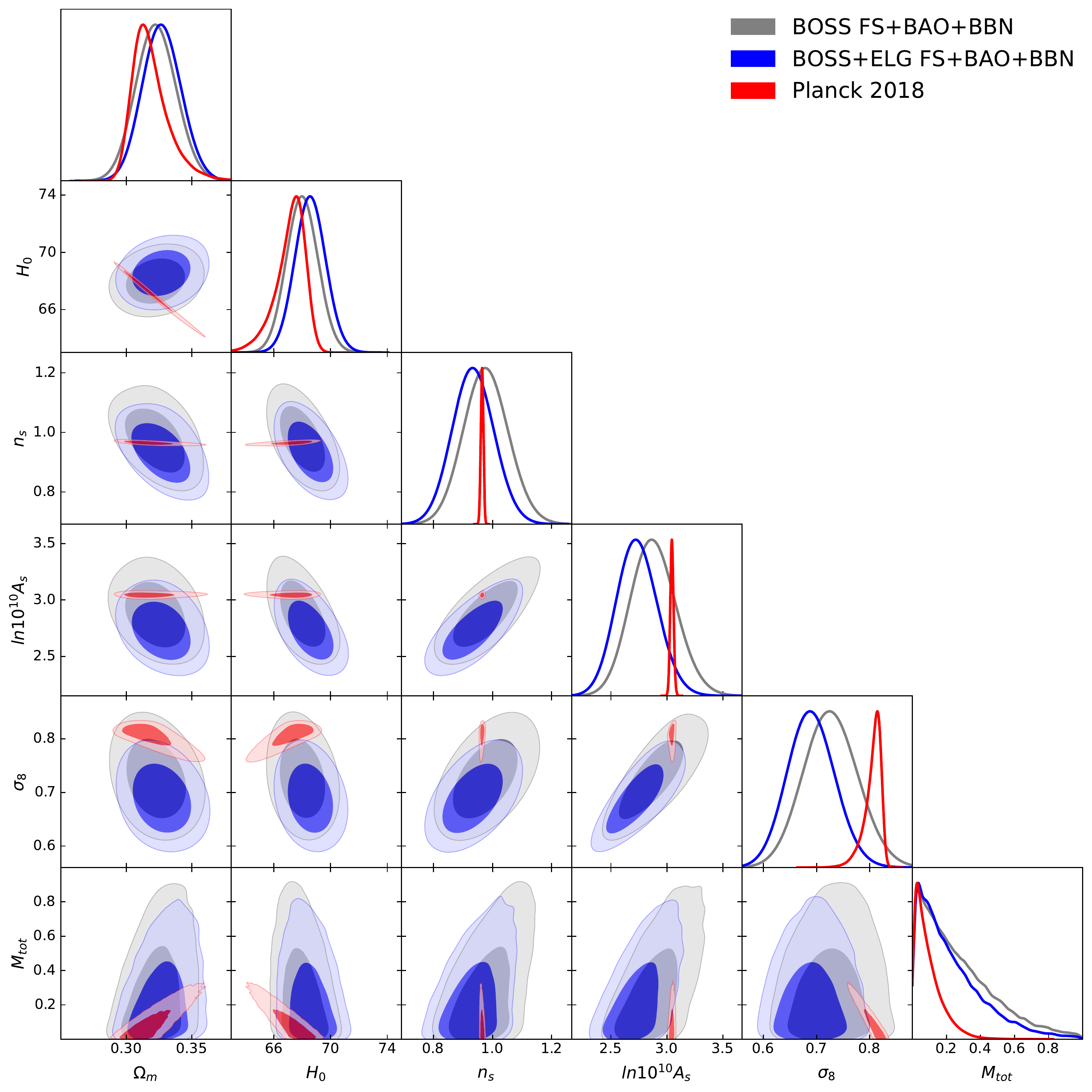}
\caption{Triangle plots for 
the posterior distributions of the 
combined BOSS full-shape  + BAO + BBN (BOSS FS+BAO+BBN) dataset 
and BOSS+ELG FS+BAO+BBN 
data. The \textit{Planck} 2018 CMB contours for the 
same $\nu\Lambda$CDM model 
are shown for comparison.
\label{fig:money2} } 
\end{figure*}

In this paper we explicitly check this expectation with 
data from the recent
eBOSS survey~\cite{Alam:2020sor},
which has collected the largest to date spectroscopic 
sample of emission 
line galaxies~\cite{Ross:2020lqz}.
To that end we re-analyze the power spectrum of emission line galaxies 
from eBOSS using the EFT-based full-shape methodology. 
The motivation for this analysis is two-fold. 
On the one hand,  it is important to determine the regime of validity 
of the EFT for this sample, which then can be used to estimate the 
performance of the full-shape 
analysis for the future DESI-like surveys. 
On the other hand, it is interesting to see what constraints 
one can get from this sample in a consistent analysis 
beyond the RSD/AP fixed template method.

Our paper is structured as follows. We start with the 
main results in Sec.~\ref{sec:main}, then discuss 
technical aspects of the analysis in Sec.~\ref{sec:data}.
Sec.~\ref{sec:sims} describes tests of our pipeline 
on Outer Rim ELG mock
catalogs. We detail our baseline analyses in Sec.~\ref{sec:res}  
and conclude in Sec.~\ref{sec:disc}.
Appendix \ref{app} contains
some additional information.

\section{Summary of Main Results}
\label{sec:main}

It is convenient to start with
a short summary of 
our main results. We have analyzed 
the publicly available
redshift-space power spectrum monopole ($\ell=0$), 
quadrupole ($\ell=2$)
and hexadecapole ($\ell=4$)
moments of 
eBOSS emission line galaxies~\cite{deMattia:2020fkb}. 
This analysis is based 
on the pre-reconstructed power spectrum measurements 
from two samples stretched across the redshift rage 
$0.7<z<1.1$ (effective redshift $z_{\rm eff}\approx 0.86$)
collected from two different patches of sky;
North Galactic Cap (NGC) and South Galactic Cap (SGC).
We combine them with the post-reconstructed BAO 
data from the redshift range $0.6<z<1.1$
and take into account its covariance with the 
pre-reconstructed full-shape measurements.
Our full-shape + BAO analysis 
pipeline is essentially the same as the 
one used in Refs.~\cite{Ivanov:2019pdj,Philcox:2020vvt}.
We fit the eBOSS ELG full shape data 
with the one-loop EFT model assuming the 
minimal flat $\Lambda$CDM cosmology.
In our baseline ELG FS+BAO analysis, we constrain 
the current physical baryon density $\omega_b$ to the 
BBN prior $\omega_b=0.02268\pm 0.0038$, 
and vary all other cosmological parameters 
in our MCMC chains:
the reduced Hubble constant\footnote{Throughout our paper,
we will use both the ``usual'' Hubble constant $H_0$ and its dimensionless version $h=H_0/(100~\text{km/s/Mpc})$.} $h$, 
total neutrino mass $M_{\rm tot}$,
physical dark matter density $\omega_{cdm}$,
the primordial scalar power spectrum 
amplitude $A_s$ and its tilt $n_s$. 
The neutrino sector is approximated with
three degenerate mass states. Along with the cosmological 
parameters, we also vary a set of 26 nuisance 
coefficients (13 per each cap) capturing 
fiber collisions, 
non-linear effects of matter clustering, galaxy
bias and redshift-space distortions.

Using large Outer Rim simulations
we have checked that our pipeline
based on the one-loop EFT model 
applied at the scale cut \mbox{$\kmax=0.25~\hMpc$}
can recover correct values 
of $H_0$, 
$\Omega_m$ to within $0.1\sigma$
and $\sigma_8$
to within $0.3\sigma$ of the full eBOSS errors.
This is similar to the scale cut used 
in previous analyses of the BOSS LRG power spectra data chunks
with similar effective volumes, see e.g.~\cite{Ivanov:2019pdj}.
We have found that the ELGs are indeed less affected by non-linearities
than LRGs, but this difference is 
not very significant for the eBOSS data.

The posterior distribution of 
the ELG FS+BAO data in
$\sigma_8-\Omega_m-H_0$ planes is 
shown in the left panel of Fig.~\ref{fig:money},
along with contours from the combined  
BOSS FS+BAO+BBN analysis~\cite{Philcox:2020vvt},
and the \textit{Planck} 2018 data~\cite{Aghanim:2018eyx}
analyses, run for the same baseline 
$\nu \Lambda$CDM model.
The right panel of Fig.~\ref{fig:money}
shows the eBOSS ELG $P_\ell(k)$ data
along with the best-fitting theoretical models 
extracted from our MCMC chains. The 1d marginalized 
constraints are listed in table~\ref{table1}.

The analysis of the combined BOSS+ELG FS+BAO+BBN data
are shown in Fig.~\ref{fig:money2}
along with the \textit{Planck} CMB posteriors~\cite{Aghanim:2018eyx} for the same 
cosmological
model. 
We see that the ELG data marginally tightens
parameter constraints, in particular,
on the total neutrino mass.

\section{Data and Methodology}
\label{sec:data}

\subsection{Data}

\textbf{ELG FS + BAO.}
Our main analysis is based 
on the power spectrum  
from the ELG SDSS DR16 clustering catalog~\cite{Raichoor:2020vio},
see Ref.~\cite{Raichoor:2017nuz} for selection details.
The ELG sample was selected from the DECam Legacy Survey
(DECaLS) \textit{grz} photometry~\cite{Karim:2020qao}. 
We use the publicly available
baseline power spectrum measurements
of Ref.~\cite{deMattia:2020fkb}. 
Note that these measurements 
made use of the so-called ``shuffled'' scheme 
where randoms are assigned data redshifts,
which suppresses the radial modes on large scales~\cite{deMattia:2019vdg}. 
The eBOSS ELG sample may  
also be affected by 
unknown angular systematics,
which is mitigated by canceling the angular modes 
with the so-called pixelated scheme. 
We refer the reader to 
Refs.~\cite{Raichoor:2020vio,Tamone:2020qrl,deMattia:2020fkb}
for details on the eBOSS ELG catalog 
and its systematics.

The NGC and SGC patches used in the full-shape
analysis have different 
effective redshifts $z_{\rm eff}$, number densities $\bar n$,
and effective geometric volumes $V_{\rm eff}$,
\be 
\begin{split}
 \text{NGC}:~&z_{\rm eff}=0.86\,,~ \frac{1}{\bar n}=5.4\cdot 10^3~[\text{Mpc}/h]^3\,,\\
&  V_{\rm eff}=0.42~h^{-3}\text{Gpc}^3\,,
\end{split}
\ee
\be 
\begin{split}
\text{SGC}:~&z_{\rm eff}=0.853\,,~ \frac{1}{\bar n}=4.8\cdot 10^3~[\text{Mpc}/h]^3\,,\\
&  V_{\rm eff}=0.41~h^{-3}\text{Gpc}^3\,.
\end{split}
\ee

The ELG BAO signal is extracted from the post-reconstructed
power spectra measurements 
in the range $0.6<z<1.1$. 
Reconstruction~\cite{Eisenstein:2006nk} was performed  
assuming the logarithmic growth parameter 
$f = 0.82$, the linear bias $b_1 = 1.4$,
and using a Gaussian smoothing kernel of width $15$~Mpc/$h$.

In our analysis we will use the publicly available 
pre- and post-reconstructed
eBOSS ELG power spectrum multipoles 
computed 
from the FKP-weighted density field~\cite{Feldman:1993ky} 
with the Yamamoto estimator~\cite{Yamamoto:2005dz}
implemented in the \texttt{nbodykit} code~\cite{Hand:2017pqn}.
The power spectrum is binned in spherical shells 
of width $\Delta k=0.01~\hMpc$.
The redshifts and angles are converted 
into comoving distances assuming a  
flat $\Lambda$CDM cosmology with $\Omega_m=0.31$.

\textbf{BOSS FS+BAO+BBN.}
Our additional analyses will be based 
on the  
BOSS DR12 LRG full-shape power spectrum 
data from~\cite{Alam:2016hwk} combined with the 
complete SDSS BAO measurements and BBN. 
We will use the latest version of the 
EFT-based full-shape LRG power spectrum 
likelihood described in detail 
in Refs.~\cite{Ivanov:2019pdj,Philcox:2020vvt,Chudaykin:2020aoj,Wadekar:2020hax}. 
In this work, we will employ the same likelihood 
as in Ref.~\cite{Chudaykin:2020aoj},
but use the new window-free power spectrum
measurements of Ref.~\cite{Philcox:2020vbm}.\footnote{Publicly available at \url{https://github.com/oliverphilcox/BOSS-Without-Windows}.
} 
Our likelihood is based on the 
monopole, quadrupole, and hexadecapole moments with $k_{\rm min}=0.01~\hMpc$ and $\kmax=0.2~\hMpc$.
We include the hexadecapole  for 
completeness, even though it
contains very little
cosmological information for the BOSS LRG sample~\cite{Chudaykin:2020aoj}.

The tail of the 
BOSS LRG full-shape data has a small overlap with the ELG sample 
over the redshift rage $0.7<z<0.75$.
This overlap is quite insignificant and we neglect 
a correlation between the eBOSS ELG and 
BOSS LRG samples in this work. 
This assumption was validated 
in Refs.~\cite{Alam:2020sor,Zhao:2020tis}.

As far as the BAO data is concerned, 
we use measurements from BOSS DR12~\cite{Beutler:2016ixs}, 
and include their cross-covariance with the full-shape 
data as described in Ref.~\cite{Philcox:2020vvt}.
We compute the cross-covariance between 
the BAO and full-shape measurements from the
Patchy mocks~\cite{Kitaura:2015uqa}.
Additionally, we use the BAO from 6DF ($z_{\rm eff}=0.106$)~\cite{Ross:2014qpa}, 
SDSS DR7 MGS ($z_{\rm eff}=0.15$)~\cite{Beutler:2011hx}, 
and Lyman-$\alpha$ forest, which is
based on auto and cross-correlation with quasars 
at $z_{\rm eff}=2.33$~\cite{duMasdesBourboux:2020pck,Alam:2020sor}.
Finally, 
we also use the BAO measurements from the 
eBOSS quasar sample ($z_{\rm eff}=1.48$)~\cite{Neveux:2020voa}.

We will combine all our large-scale structure likelihoods with
the BBN physical baryon
density prior
$\omega_b = 0.02268\pm 0.00038$, extracted 
from a conservative standard analysis of data from~\cite{Aver:2015iza,Cooke:2017cwo}.

Finally, let us briefly comment on the data nomenclature. 
For brevity 
we do not mention ``BBN'' in the ELG FS+BAO
likelihood, but it should be recalled that it always contains the BBN prior.

\subsection{Theory model}

We fit the ELG power spectrum  multipoles' data using 
the one-loop perturbation theory model described in detail
in Refs.~\cite{Ivanov:2019pdj,Chudaykin:2020aoj,Chudaykin:2020hbf,Chudaykin:2020ghx}. The theoretical templates 
are computed using the \texttt{CLASS-PT} 
code~\cite{Chudaykin:2020aoj}.
In this work we 
vary all the one-loop power spectrum 
nuisance parameters in our MCMC chains, including 
a full set of stochastic counterterms, 
\be
\{b_1,b_2,b_{\mathcal{G}_2},b_{\Gamma_3},c_0,c_2,c_4,\tilde{c},a_0,a_2,P_{\rm shot}\}.
\ee
Even though some of 
these contributions cannot be determined by the data 
itself, and they could be set to constant 
values (typically zero, see e.g.~\cite{Ivanov:2019pdj}), we believe that it is more appropriate
to marginalize over all of them within physically motivated
priors. These priors will be discussed shortly.

\subsection{Window function}

\begin{figure}[ht]
\begin{center}
\includegraphics[width=0.49\textwidth]{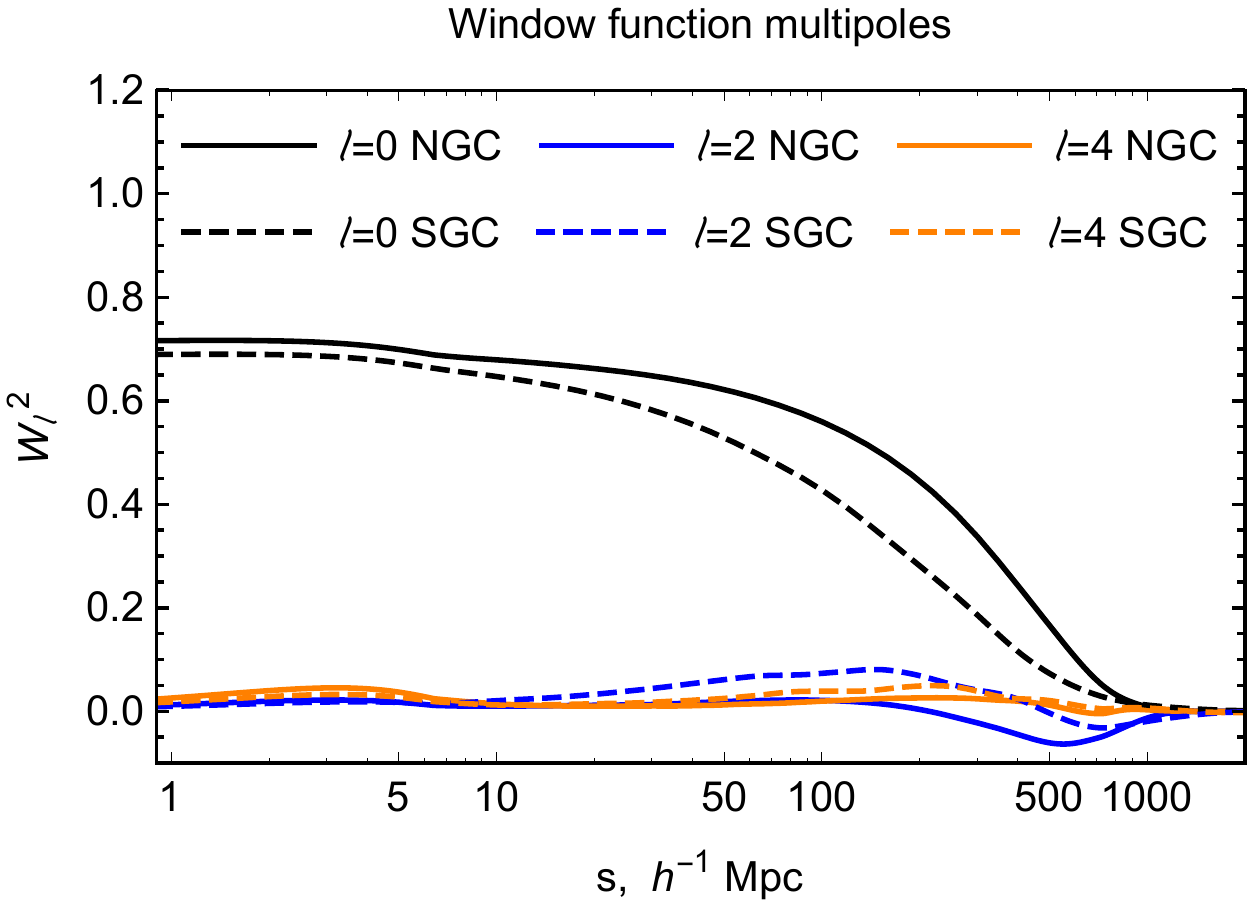}
\end{center}
\vspace{-0.5cm}
\caption{Window function multipoles of 
the eBOSS baseline full shape ELG sample.
\label{fig:win0} } 
\end{figure}

\textbf{Survey geometry.}
The publicly available eBOSS ELG power spectrum 
has been obtained using the FKP estimator, which
measures the true spectrum convolved with 
the survey window function. 
We use the publicly available measurements
of the survey window function~\cite{deMattia:2020fkb}.
It is important to mention that the eBOSS ELG
window function
includes fine-grained veto masks, which make the 
window function kernels scale-dependent even on small
scales (see Fig.~\ref{fig:win0} or Fig. 1 of Ref.~\cite{deMattia:2020fkb}).

\begin{figure}[ht]
\begin{center}
\includegraphics[width=0.49\textwidth]{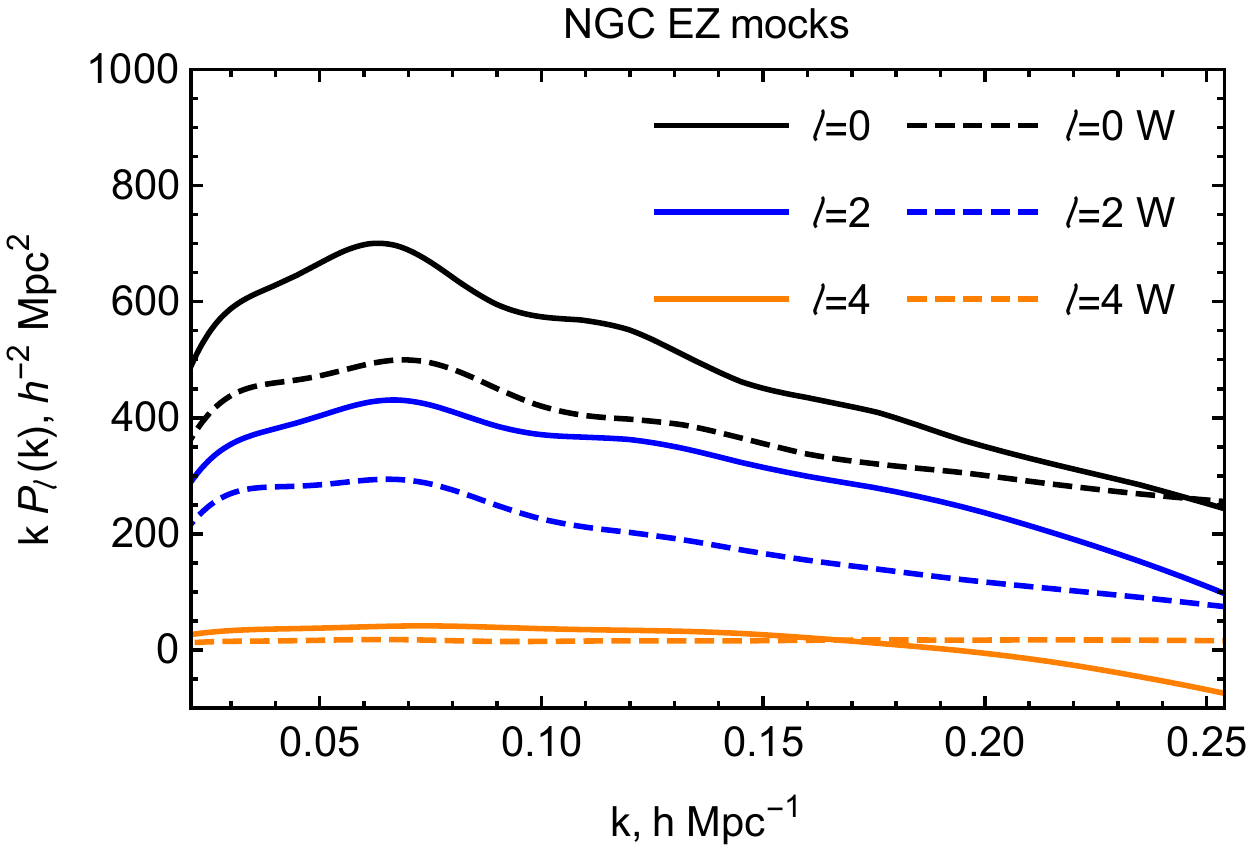}
\includegraphics[width=0.49\textwidth]{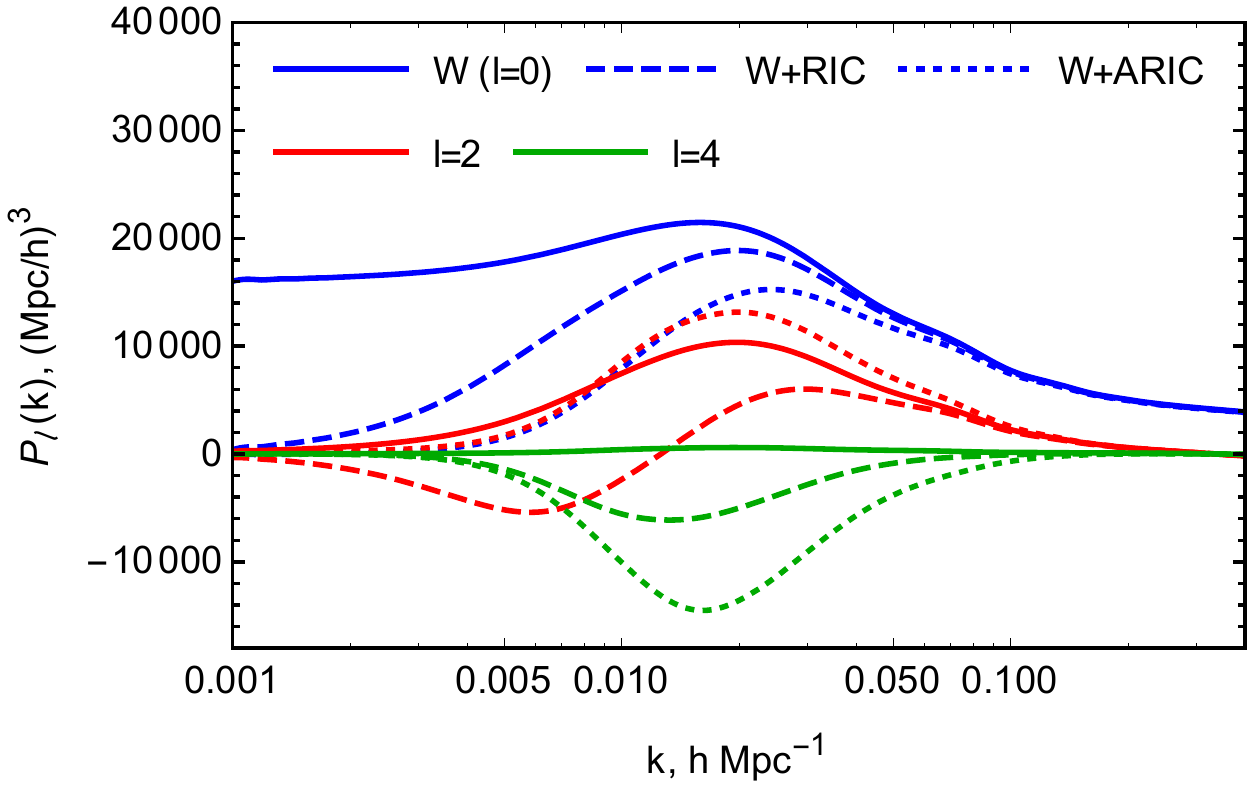}
\end{center}
\vspace{-0.5cm}
\caption{\textit{Upper panel:} typical power spectrum multipoles 
of the reference EZ mocks before (solid lines) and after 
(dashed lines) convolution with the survey window function. \\
\textit{Lower panel:} the impact of the integral constraints
on the $\ell=0$ (blue), $\ell=2$ (red), $\ell=4$ (green) 
moments. We show the windowed spectra without integral 
constraint corrections 
(``W'', in solid),
the effect of the global and radial constrains (``W+RIC'', dashed)
and the cumulative effect of 
the global, angular and radial constrains (``W+ARIC'', dotted),
which have to be implemented if the pixelated scheme is used 
to mitigate the angular systematics.
\label{fig:winez0} } 
\end{figure}

We take the window function 
effects into account by employing the method 
of Refs.~\cite{Wilson:2015lup,Beutler:2016arn} 
based on the global plane parallel approximation.
This method amounts to a simple multiplication
of the true position space 
correlation function multipoles $\xi^{\rm true}_\ell$
with window function multipoles $W_\ell^2$, i.e. 
for our analysis involving moments up to the hexadecapole we have
\be
\begin{split}
& \xi^{\rm win}_0=\xi^{\rm true}_0 W_0^2 
+ \frac{1}{5}\xi^{\rm true}_2 W_2^2+\frac{1}{9}\xi^{\rm true}_4 W_4^2\,,\\
& \xi^{\rm win}_2=\xi^{\rm true}_0 W_2^2 + 
\xi^{\rm true}_2\left[
W_0^2 + \frac{2}{7}W_2^2 + \frac{2}{7}W_4^2 
\right] \\
&\quad \quad \quad+ \xi^{\rm true}_4\left[
\frac{2}{7} W_2^2 + \frac{100}{693}W_4^2 
\right] \,,\\
& \xi^{\rm win}_4=\xi^{\rm true}_0 W_4^2 + 
\xi^{\rm true}_2\left[
\frac{18}{35}W_2^2 + \frac{20}{77}W_4^2 
\right]\\
& \quad \quad \quad + 
\xi^{\rm true}_4\left[W_0^2+
\frac{20}{77}W_2^2 + \frac{162}{1001}W_4^2 
\right]\,.
\end{split} 
\ee
We have checked that higher order window
function miltipoles do not affect our results,
in agreement with conclusions of Ref.~\cite{deMattia:2020fkb}.
The effect of the window function is illustrated 
in the upper panel of Fig.~\ref{fig:winez0}.

\textbf{Integral constraints.}
We also implement the global, radial and angular 
integral constraints 
following Refs.~\cite{deMattia:2019vdg,deMattia:2020fkb}.
Schematically, the integral constraint corrected (``icc'')
correlation functions have the following form,
\be
\label{eq:ic}
\begin{split}
\xi_\ell^{\rm icc}(s)&=\xi_\ell^{\rm win}(s)\\
&-\sum_{\ell'}
\frac{4\pi}{2\ell'+1}\int s'^2ds'~\mathcal{W}_{\ell \ell'}(s,s')
\xi_{\ell'}^{\rm true}(s')\,,
\end{split} 
\ee
where $\mathcal{W}_{\ell \ell'}$
are Legendre multipole moments of the 3-point correlator
of the window function~\cite{deMattia:2019vdg}.
Note that it is crucial to use in Eq.~\eqref{eq:ic}
the masked multipoles
$\xi_\ell^{\rm true}$ that contain the full pair count 
shot noise correction, i.e. Fourier transforms
of the full power spectrum \textit{with} the $1/\bar n$
shot noise term. 
The effect of integral constraint corrections is illustrated 
in the lower panel of Fig.~\ref{fig:winez0}.
One can notice that the 
radial constraint corrections 
suppress the quadrupole, while the angular constraint
term enhances it. The effect on the monopole 
is a suppression, which becomes stronger
with the angular integral constraint.

Importantly, we see a relatively large effect 
of the integral constraint on the hexadecapole.
The situation here is qualitatively 
different from the case of the BOSS LRG sample,
which is not strongly affected by the integral 
constraints~\cite{deMattia:2019vdg}. Because of this the hexadecapole 
of the BOSS LRG sample has a relatively weak amplitude
and does not improve cosmological constraints. 
In contrast to that, the eBOSS ELG hexadecapole 
is dominated by the leakage from the monopole
and quadrupole moments through the integral constraints.
This results in a relatively 
strong amplitude of the hexadecapole signal. 
Because of this effect we may expect 
that the ELG $\ell=4$ moment should be important for parameter
constraints and therefore we take this moment into account.

\textbf{Fiber collisions.}
Finally, we implement the effective window method in order 
to account for fiber collisions~\cite{Hahn:2016kiy}. This method consists of two steps. 
First, one corrects the position space correlation
function 
with a steplike 
kernel proportional to the 
fraction of collided pairs $f_s$, 
\be
\label{eq:fc}
\begin{split}
&\xi^{\text{fc}}_\ell = \xi^{\rm icc}_\ell -f_s\Delta \xi^{\rm fc}_\ell\,, \\
&\Delta \xi^{\rm fc}_\ell= \frac{2\ell+1}{2}\int_{-1}^1 
d\mu~W_{\rm fc}(s_\perp)
(1+\xi^{\rm icc}(s, \mu))L_\ell(\mu)\,,
\end{split}
\ee
where $\textbf{s}$ 
is the pair separation vector, $s=|\textbf{s}|$,
$\mu=(\hat{\textbf{z}},\hat{\textbf{s}})$,
$\textbf{z}$ is the unit line-of-sight vector,
$s_\perp=s\sqrt{1-\mu^2}$,
$L_\ell$ is the Legendre polynomial of order $\ell$,
and $W_{\rm fc}(x)$ is the position space 
top-hat function with the step scale $D_{\rm fc}$.
For eBOSS ELGs we have~\cite{deMattia:2020fkb}
\be
f_s(\text{NGC})=0.46\,,\quad f_s(\text{SGC}) =0.38\,.
\ee
The length scale of 
the fiber collision window
is $D_{\rm fc} =0.62~\text{Mpc}/h$. This is the
comoving size that corresponds to the 
fiber collision angular scale $62''$
at $z_{\rm eff}=0.86$. 
After the convolution with the effective window 
kernel and the consequent Fourier space 
transform one must marginalize over 
a set of nuisance parameters
that renormalize the Fourier integral.
Specifically, one needs to add the following 
higher derivative stochastic
corrections $\Delta P_\ell^{\rm fc, corr}$
for each multipole~\cite{Hahn:2016kiy}:
\be 
\label{eq:Cs}
\begin{split}
& \Delta P_0^{\rm fc, corr}=C_{0,0} + C_{0,2} k^2 +...\,, \\
&\Delta P_2^{\rm fc, corr} = C_{2,2} k^2 +...\,,\\
&\Delta P_4^{\rm fc, corr}=C_{4,4} k^4 +...\,,\\
\end{split}
\ee
and marginalize over new nuisance parameters $C_{\ell,n}$.
If the survey window function were scale-independent 
on large scales, 
for the monopole and quadrupole 
these parameters would exactly coincide with
the EFT stochastic counterterms $P_{\rm shot}$, 
$a_0$ and $a_2$ that are already 
marginalized over in our analysis.
For the hexadecapole there is a new higher
order stochastic counterterm $\sim k^4$. 
The parameters \eqref{eq:Cs} that we use 
in our analysis will be discussed in detail in the next section.

\begin{figure}[ht]
\begin{center}
\includegraphics[width=0.49\textwidth]{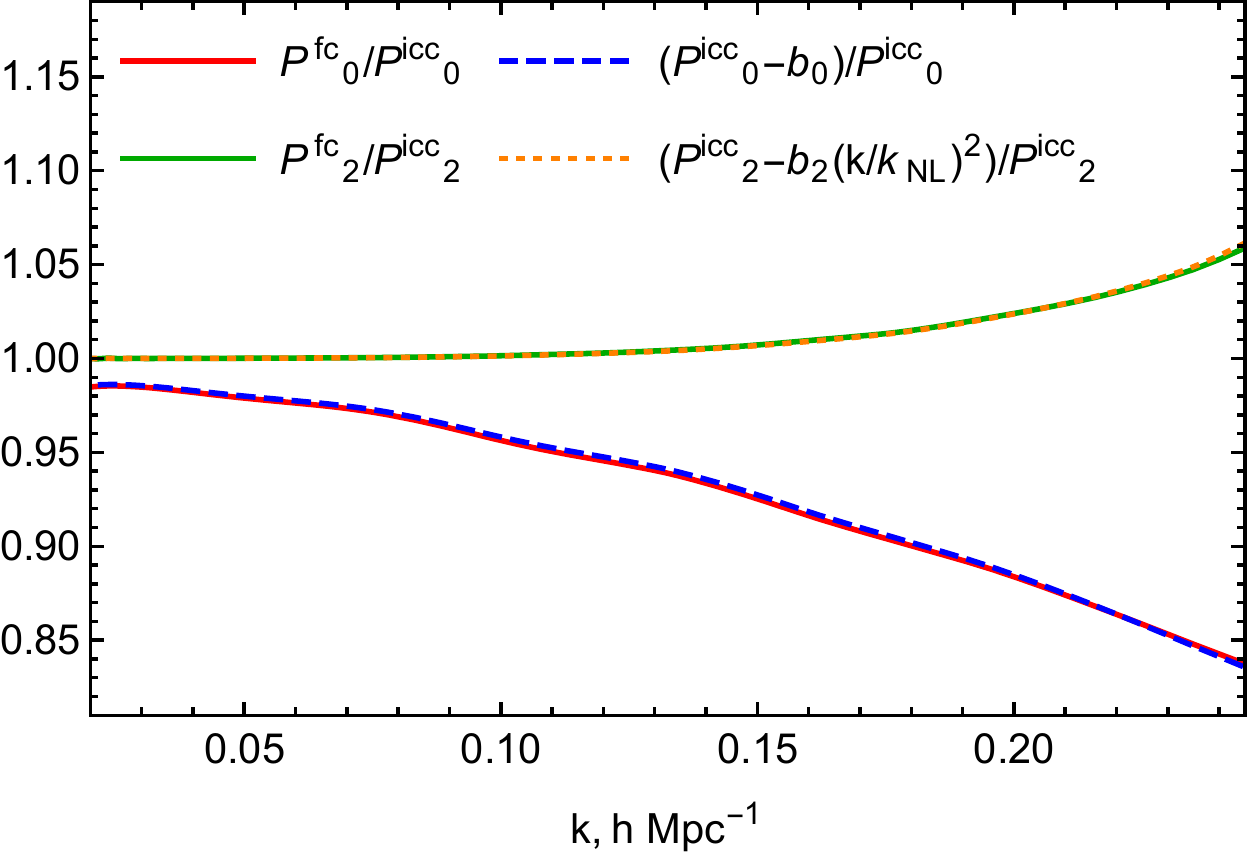}
\includegraphics[width=0.49\textwidth]{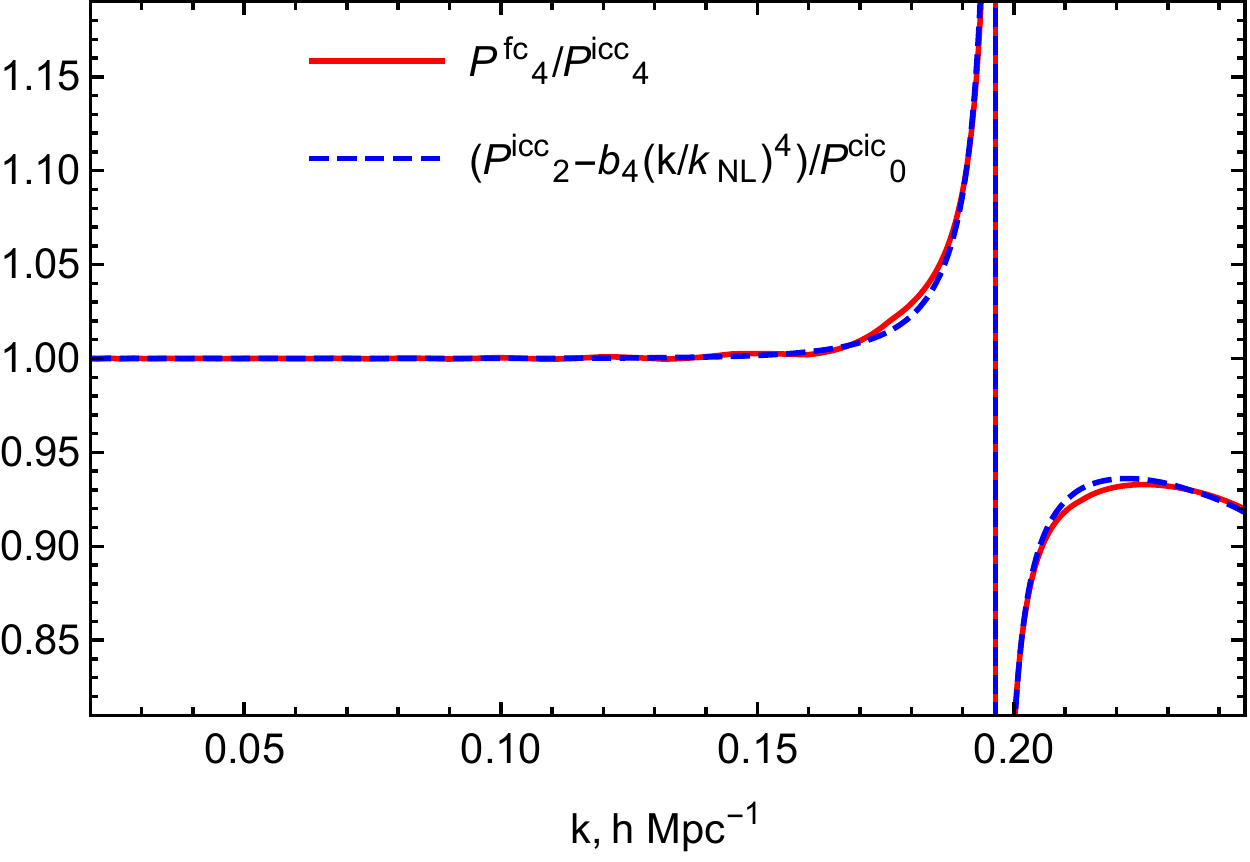}
\end{center}
\vspace{-0.5cm}
\caption{Relative impact of the fiber-collision effective window function 
on the $\ell=0,\ell=2$ moments (upper panel) and the $\ell=4$
moment (lower panel) of the eBOSS NGC ELG data.
The 
hexadecapole plot exhibits a spike 
around $k= 0.2~\hMpc$ where this moment crosses zero.
\label{fig:fc} } 
\end{figure}

In Fig.~\ref{fig:fc} we illustrate the impact of the 
effective window approach to account for fiber
collisions. We show pixelated power 
spectrum multipoles after correcting 
them as prescribed by Eq.~\eqref{eq:fc}. 
All nuisance parameters are kept fixed 
during this procedure. 
The difference between the ``usual''
and the fiber-collision corrected spectra 
are well approximated by a constant 
for the monopole, by a $k^2$ term for the 
quadrupole, and by a $k^4$ term for the 
hexadecapole. 

The general approximate 
formula for the effective window corrections can be written as
\be
\label{eq:effW}
\begin{split}
& P_\ell^{\rm fc}=P_\ell^{\rm icc}-\Delta P_\ell^{\rm fc}\,,\quad \Delta P_\ell^{\rm fc} = b_\ell\left(\frac{k}{k_{\rm NL}}\right)^{\ell}\frac{1}{\bar n}\,,
\end{split} 
\ee
where $k_{\rm NL}=0.45~\hMpc$. The best-fitting values 
for the NGC footprint are given 
\be
b_0=0.032\,,\quad b_2=-0.014 \,,\quad b_4=0.0028\,.
\ee
We see that for the monopole and quadrupole 
the effective window 
corrections just renormalize 
the one-loop 
EFT nuisance parameters. 
The hexadecapole fiber collision correction 
renormalizes the two-loop stochastic 
counterterm $\sim k^4 \mu^4$, 
which is higher order in our nomenclature 
but we still include it in our model in
order to accurately capture fiber collisions.

The degeneracy between the effective window scheme
and EFT nuisance parameters was pointed out in Ref.~\cite{Ivanov:2019pdj},
which argued that marginalization 
over the stochastic EFT parameters 
completely
captures the effect of fiber collisions
even without an explicit implementation
of the effective window function 
corrections.
However, including these corrections 
has no impact on computational time,
and we prefer to keep them in
our analysis for completeness.

\subsection{Parameters and priors}

Let us discuss priors on cosmological and nuisance parameters
that we impose in our ELG full-shape analysis.
These priors are similar to those used in the DES year 3 
cosmological analysis~\cite{Abbott:2021bzy},
\be
\label{eq:priorcosm}
\begin{split}
& h\in \text{flat}[0.4,1]\,,\quad 
\omega_{cdm}\in \text{flat}[0.03,0.7]\,,\\
&\ln(10^{10}A_s)\in \text{flat}[0.1,10]\,,\quad n_s\in \text{flat}[0.87,1.07]\,,\\
& M_{\rm tot}\in \text{flat}[0,1]~\text{eV}\,.
\end{split} 
\ee

As far as the nuisance parameters are concerned, 
we use the following priors for biased parameters~\cite{Chudaykin:2020aoj,Wadekar:2020hax},
\be
\begin{split}
& b_1\in \text{flat}[0,4]\,, \quad b_2\sim \mathcal{N}(0,2^2)\,, 
\quad b_{\mathcal{G}_2}\sim \mathcal{N}(0,2^2) \,, \\
& b_{{\Gamma}_3}\sim \mathcal{N}\left(\frac{23}{42}(b_1-1),1^2\right) \,,
\end{split}
\ee
higher derivative terms
(counterterms)
\be
\begin{split}
& \frac{c_0}{[\text{Mpc}/h]^2} \sim \mathcal{N}(0,30^2)\,,\quad 
\frac{c_2}{[\text{Mpc}/h]^2} \sim \mathcal{N}(10,30^2)\,,\\
& \frac{c_4}{[\text{Mpc}/h]^2} \sim \mathcal{N}(0,30^2)\,,\quad 
\frac{\tilde{c}}{[\text{Mpc}/h]^4} \sim \mathcal{N}(200,500^2)\,,
\end{split}
\ee
and stochastic contributions 
\be
\label{eq:shotpr}
\begin{split}
P_{\rm shot} \sim \mathcal{N}(0,4^2)\,,\quad a_{0}
\sim \mathcal{N}(0,4^2)\,,\quad a_2\sim \mathcal{N}(0,2^2)\,,
\end{split}
\ee
where we have introduced the notation 
$\mathcal{N}(\bar a,\sigma^2_a)$ for a Gaussian distribution
of a variable $a$
with mean $\bar a$ and variance $\sigma^2_a$.
Note that our convention for the stochastic contributions is given by,
\be
\label{eq:Pstoch}
P_{\rm stoch}=\frac{1}{\bar n}\left(1+
P_{\rm shot} + a_0 \left(
\frac{k}{k_{\rm NL}}\right)^2 
+ a_2\mu^2 \left(
\frac{k}{k_{\rm NL}}\right)^2 
\right)\,,
\ee
with $k_{\rm NL}=0.45~\hMpc$. 
The publicly available eBOSS measurements 
have the windowed Poissonian shot noise contribution 
subtracted from the data~\cite{deMattia:2020fkb}.
To account to this we subtract the 
$W_0^2(0)/\bar n$ 
term from the  
window-convolved 
monopole templates in our likelihood.

As part of the effective window method
to capture fiber collisions, we need to marginalize 
the windowed theory model 
over stochastic contributions given in Eq.~\eqref{eq:Cs}.
Because of the scale-dependence of the window function, 
the fiber-collision induced stochastic terms are not exactly 
degenerate with the 
``fundamental'' stochastic contribution~\eqref{eq:Pstoch}.
Hence, we 
additionally
marginalize over the stochastic 
fiber collision corrections~\eqref{eq:Cs}.
Based on tests with mocks 
with injected fiber collisions,
we found that it is sufficient to use the three new parameters 
$C_{0,0},C_{2,2}$ and $C_{44}$ with the following priors:
\be
\begin{split}
& \frac{C_{0,0}}{\bar n^{-1}}\sim \mathcal{N}(0,1^2)\,,
\quad 
 \frac{C_{2,2}}{\bar n^{-1}k_{\rm NL}^{-2}}\sim \mathcal{N}(-1,0.2^2)\,,
\\ 
& \frac{C_{4,4}}{\bar n^{-1}k_{\rm NL}^{-4}}\sim \mathcal{N}(0,0.1^2)\,.
\end{split} 
\ee

Note a few differences
in the choice of priors w.r.t. former LRG analyses~\cite{Chudaykin:2020aoj,Chudaykin:2020ghx}. 
We double the standard deviations in 
priors for $b_2$ and $b_{\mathcal{G}_2}$,
which is motivated by recent simulation-based evidence that these 
biases deviate from those of host halos for
weakly biased galaxies such as ELGs~\cite{Eggemeier:2021cam,Barreira:2021ukk}. 
The second point is that we use smaller mean
values in priors for fingers-of-God counterterms 
$c_2$ and $\tilde{c}$, 
which reflects the fact that the velocity 
dispersion of ELGs is expected to be 
somewhat smaller than that 
of LRGs. The third point is that we use a more 
conservative model for the stochastic contributions. 
We explicitly marginalize over the higher derivative 
stochastic counterterm $a_0$ now and also significantly 
increase the standard deviation in the prior for 
the constant shot noise contribution, which is chosen 
in order to account for a possible super-Poissonian sampling,
which can take place for weakly biased tracers such as ELGs~\cite{Schmittfull:2018yuk,Schmittfull:2018yuk}.

Since the SGC and NGC samples have different sky calibrations,
we use a separate set of nuisance parameters 
for each footprint.

\subsection{Likelihood}

We sample the Gaussian power spectrum likelihood 
\be\begin{split}
& -2\ln \mathcal L_P = 
\sum_{i,j}(C^{\ell \ell'}_{ij})^{-1}
\Delta P_{\ell}(k_j)
\Delta P_{\ell'}(k_j)\,,\quad \text{where} \\
& \Delta P_{\ell}(k_i)\equiv 
P^{\rm theory}_{\ell}(k_j)- P^{\rm data}_{\ell}(k_j)
\,,
\end{split}
\ee
and $\ell,\ell'=0,2,4$.
We use the publicly available covariance matrix estimated from
1000 realizations of EZ mocks~\cite{Zhao:2020bib}.
Our combined FS+BAO analysis will be based 
on the datavector $[P_0,P_2,P_4,\alpha]$, where $\alpha$
is the isotopic BAO scaling parameter. The covariance 
of this datavector will be estimated from 
the same EZ mocks using the method of Ref.~\cite{Philcox:2020vvt}.

\section{Tests on simulations}
\label{sec:sims}

\subsection{Outer Rim mocks}

We determine the maximal scale cut $\kmax$
of our theory model by analyzing 
Outer Rim ELG mocks. These mocks were designed 
to match observed properties of the 
eBOSS ELG sample~\cite{Avila:2020rmp} and were used 
to validate the baseline eBOSS analysis pipeline~\cite{Alam:2020jvh,Rossi:2020wxx}.
The mocks are generated 
from the Outer Rim dark matter simulation~\cite{Heitmann:2019ytn},
which has 27~$h^{-3}$Gpc$^3$ volume
and has been run for the following fiducial 
$\L$CDM
cosmology:
\be
\begin{split}
& h=0.71\,,\quad \omega_{cdm}=0.1109\,,\quad  \omega_b = 0.02258\,,\\
& n_s= 0.963\,,\quad \sigma_8 = 0.8\,, \quad M_{\rm tot}=0~\text{eV}\,.
\end{split} 
\ee
The simulation snapshot at $z=0.865$
was populated with mock ELGs 
using several different halo occupation distribution
(HOD) models. In our analysis we focus on the HOD-3 model,
which was a default choice of 
the eBOSS collaboration~\cite{Avila:2020rmp}. 
Within this model, the large-scale power spectrum
primarily depends on the satellite galaxy fraction  
$f_{\rm sat}$. This is an important parameter 
in the EFT context as well, as it controls the strength 
of the fingers-of-God effect, 
and hence is linked with the cutoff of the redshift-space
mapping expansion~\cite{Hand:2017ilm,Chudaykin:2019ock,Ivanov:2019pdj,Chudaykin:2020aoj}. Given this reason, we analyze three 
sets of HOD-3 mocks that differ by the satellite fraction, 
$f_{\rm sat}=0,0.3,0.45$. These mocks have large 
number densities of $\bar n^{-1}\simeq 500$ [Mpc$/h]^3$,
which is approximately 10 times larger than the number density
of the actual ELG sample.

\begin{figure}[ht!]
\begin{center}
\includegraphics[width=0.49\textwidth]{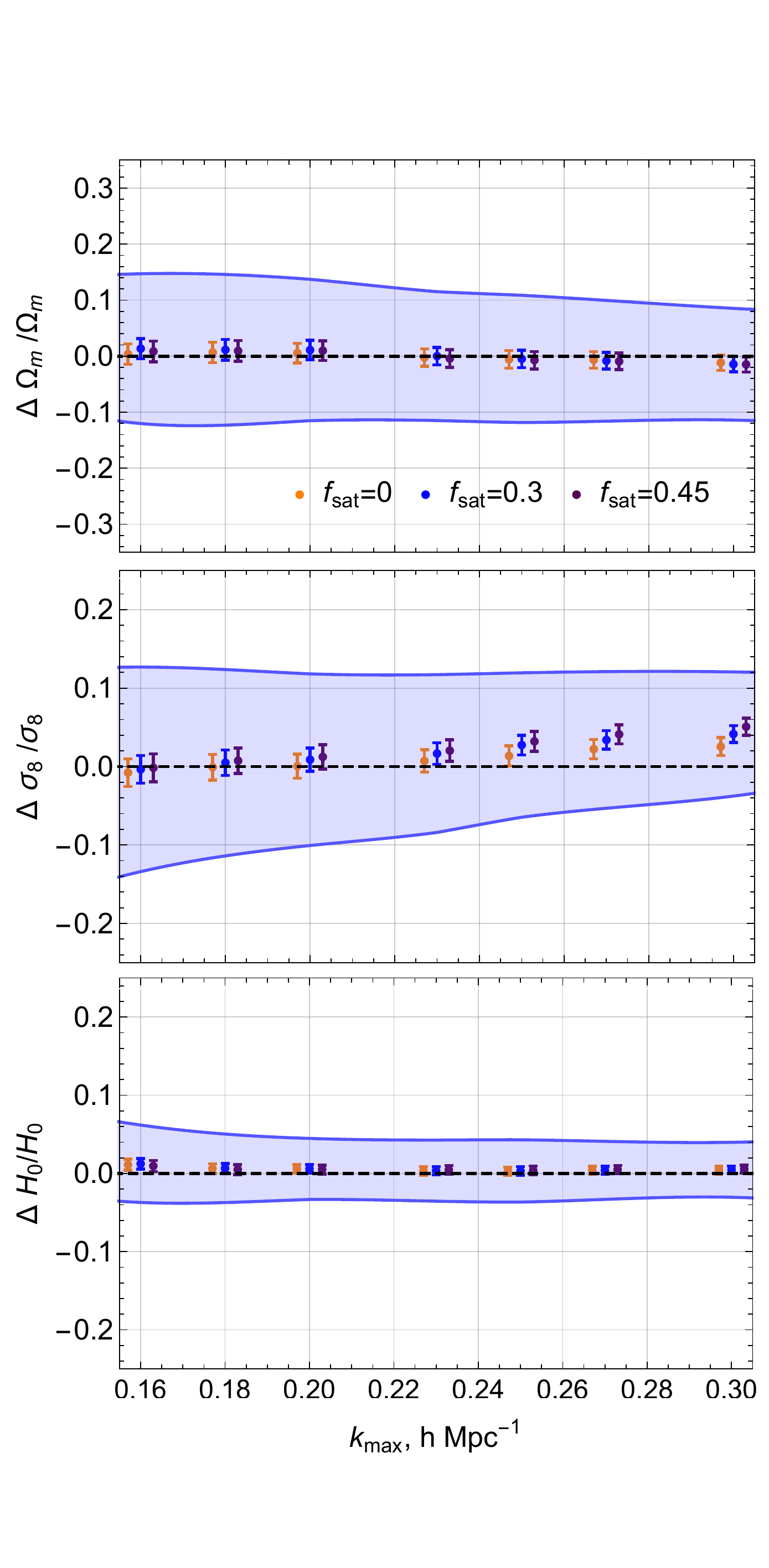}
\end{center}
\vspace{-2cm}
\caption{
One dimensional marginalized $68\%$ confidence limits on the parameters 
$\Omega_m,H_0$ and $\sigma_8$  
extracted from full-shape analyses of  
Outer Rim mocks
as a function of scale cut $\kmax$. 
We use three mock catalogs generated for the HOD-3 
model of Ref.~\cite{Avila:2020rmp}
for different choices of the satellite fraction $f_{\rm sat}$.
The shaded regions mark the errobars 
of the $f_{\rm sat}=0.3$ mocks,
which are scaled to match the effective volume of the eBOSS ELG sample.
The measurements are offset along the $x$-axis for better visibility. 
\label{fig:OR} } 
\end{figure}

\begin{figure}[!ht!]
\begin{center}
\includegraphics[width=0.49\textwidth]{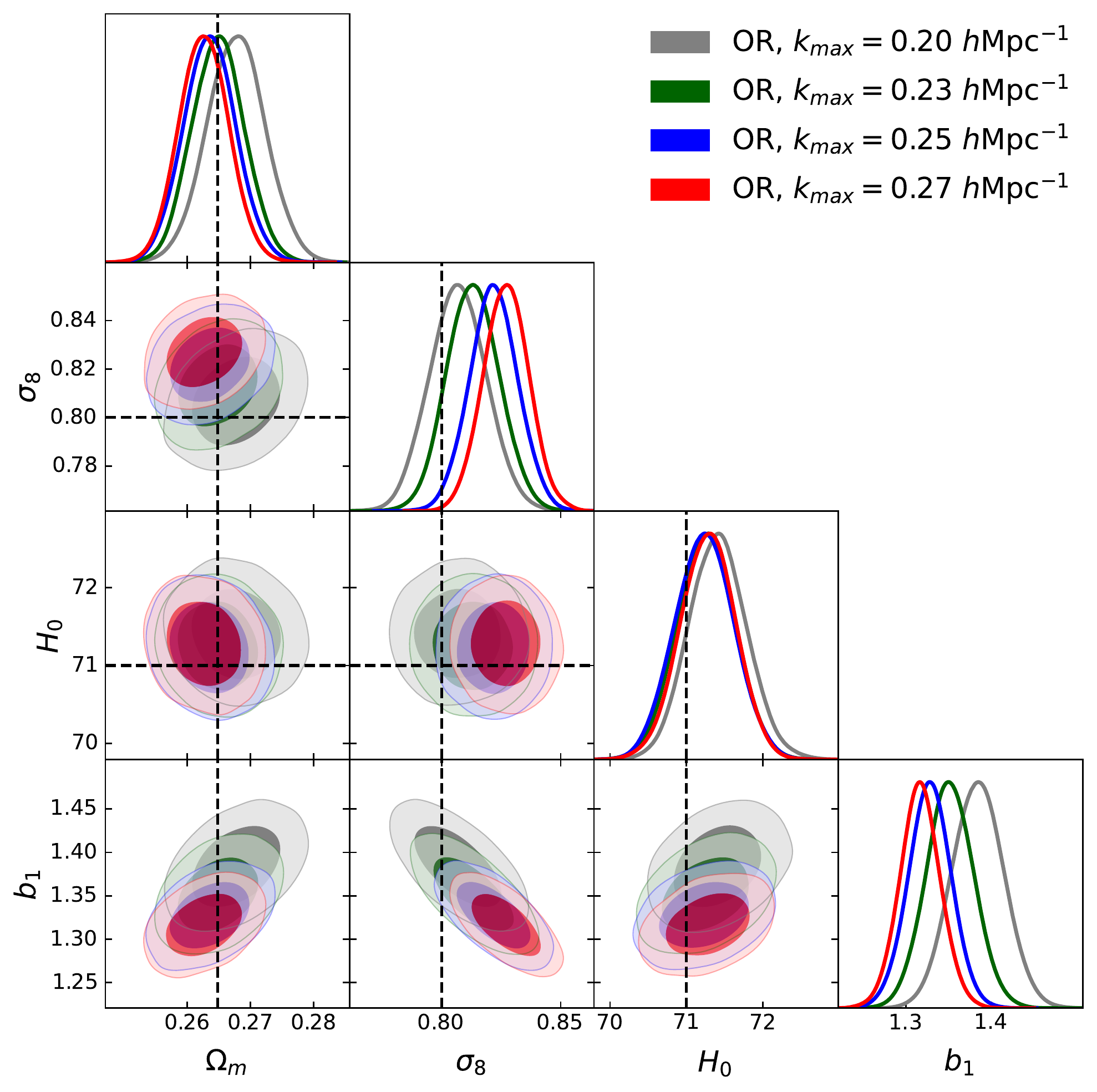}
\end{center}
\vspace{-0.5cm}
\caption{Triangle plot for parameters $\Omega_m,H_0,\sigma_8$
and linear bias $b_1$ extracted 
from the Outer Rim HOD-3 $f_{\rm sat}=0.3$ mocks
for several choices of $\kmax$.
Dashed lines mark the true values used in simulations.
\label{fig:OR2} } 
\end{figure}

\begin{table}[!ht!]
  \begin{tabular}{|c||c|c|c|} \hline
   \diagbox{ { $\kmax $}}{\small Param}  
   &  $\Omega_m$
   & $H_0$
   & $\sigma_8$
      \\ [0.2cm]
\hline
$0.16 \hMpc$   
& $0.2683_{-0.005}^{+0.0049}$
& $71.87_{-0.5}^{+0.49}$
& $0.7973_{-0.015}^{+0.015}$
\\ 
\hline
$0.18 \hMpc$  
&  $0.2678_{-0.0051}^{+0.0047}$
 & $71.47_{-0.44}^{+0.44}$
  & $0.8041_{-0.013}^{+0.013}$
\\ 
\hline 
 $0.20 \hMpc$    & 
$0.2677_{-0.0048}^{+0.0046}$
  & $71.41_{-0.4}^{+0.37}$
  & $0.8071_{-0.012}^{+0.012}$
  \\ \hline
$0.23 \hMpc$   & 
$0.2649_{-0.0044}^{+0.0041}$
& $71.26_{-0.4}^{+0.38}$
& $0.8133_{-0.012}^{+0.011}$
\\ 
\hline
$0.25 \hMpc$ 
& $0.2636_{-0.0042}^{+0.0042}$
& $71.23_{-0.4}^{+0.39}$
& $0.8219_{-0.01}^{+0.01}$
\\ \hline
$0.27 \hMpc$   & 
$0.2626_{-0.004}^{+0.004}$
& $71.28_{-0.37}^{+0.37}$ 
& $0.827_{-0.0097}^{+0.0098}$
\\ \hline
$0.30 \hMpc$   & 
$0.261_{-0.0037}^{+0.0036}$
& $71.33_{-0.35}^{+0.35}$
& $0.8331_{-0.0088}^{+0.0088}$
\\ 
\hline
\end{tabular}
\caption{1d marginalized constrains for 
parameters $\Omega_m,H_0,\sigma_8$ extracted 
from the Outer Rim HOD-3 $f_{\rm sat}=0.3$ mocks
for several choices of $\kmax$.
The fiducial values are $\Omega_m=0.2648$, $H_0=71$ km/s/Mpc,
$\sigma_8=0.8$.
}
\label{table2}
\end{table}

The
corresponding mock catalogs are publicly available
in the form of 
27 non-overlapping subboxes of size 1 $h^{-3}$Gpc$^{3}$,
which were
cut out from the full Outer Rim box.
We combine these public subboxes into the original large box 
with periodic boundary conditions, and compute 
the mock power spectrum multipoles
with the \texttt{nbodykit} code~\cite{Hand:2017pqn} 
using a $512^3$-grid and Triangular-Shape-Cloud mass assignment.
We bin 
the power spectrum 
in uniform $k$-shells of width $\Delta k=0.01~\hMpc$,
and assume the true fiducial $\L$CDM cosmology with 
$\Omega_m=0.2648$ in the redshift/angles-to-distance
conversion.


For each HOD we fit 
the mock power spectrum from the total Outer Rim box,
which has the 
volume of $27~h^{-3}$Gpc$^3$.
Additionally, in order to reduce statistical 
scatter in the velocity field we average power spectrum multipoles
over three independent line-of-sight directions following 
Ref.~\cite{Smith:2020dpo}.
At this level of precision, we also account for binning 
and discreteness
effects along the lines of Ref.~\cite{Nishimichi:2020tvu}.
We employ the Gaussian periodic box boundary condition
approximation
for the covariance matrix with the effective volume 
$27~h^{-3}$Gpc$^3$.
We have computed the covariance both 
using the analytic linear theory disconnected part 
expression~\cite{Wadekar:2019rdu} 
and the empirical 
estimator for the diagonal elements from $27$ sub-boxes.
In the latter case we divided 
the covariance elements by 27 to match the total simulation
volume.\footnote{The mean from individual subboxes is expected 
to deviate  
from the power spectrum computed from the full box at the level 
of $10^{-5}$ due to super-sample effects~\cite{dePutter:2011ah}, which is totally negligible for our purposes.
This was explicitly checked in Ref.~\cite{Rossi:2020wxx}. 
}
The two options give statistically 
consistent posteriors, 
and here we present 
results obtained with the analytic Gaussian 
covariance. 
As an additional test, we have repeated our analysis 
for 
the covariance that includes the connected non-Gaussian 
contribution 
computed in perturbation theory
along the lines of Ref.~\cite{Wadekar:2019rdu}.
In agreement with previous studies~\cite{Wadekar:2020hax,Philcox:2020zyp}, we have found the impact of the non-Gaussian covariance  elements 
to be negligibly small.\footnote{We thank Jay Wadekar for his help
with this analysis.}

We have fitted the Outer Rim ELG mock power
spectrum multipoles $\ell=0,2,4$ 
using the full one-loop perturbation theory model of Refs.~\cite{Ivanov:2019pdj,Chudaykin:2020aoj,Chudaykin:2020ghx} with the priors on nuisance 
parameters specified above. We have fixed 
$n_s$, $M_{tot}$ and $\omega_b$ to their fiducial values
and explicitly varied $\omega_{cdm}$, $h$ and $\ln(10^{10}A_s)$
in our MCMC chains. 
Fixing $n_s$, $M_{tot}$ and $\omega_b$ is optional. We do 
so in order to access the accuracy of the other 
parameter recovery in the situation when they are most
tightly constrained.
Varying $n_s$, $M_{tot}$ and $\omega_b$ in the fit 
would inflate the posteriors and might obscure
the model performance evaluation.
We choose to present our main results for
$H_0$ and derived parameters
$\sigma_8$ and $\Omega_m$, which we expect to 
constrain from the actual data 
and which are more common
in the large-scale structure literature.

\textbf{Strength of fingers-of-God.}
As a first step, we have verified
that the ELG power spectrum requires the next-to-leading order
fingers-of-God~\cite{Jackson:2008yv}
operator $\tilde{c}k^4 \mu^4 P_{\rm lin}(k)$ just like the LRG sample, but its amplitude is 
smaller than that 
of the LRGs (taken e.g. from the analysis of Ref.~\cite{Nishimichi:2020tvu}\footnote{In contrast 
to the original East Coast Team analysis of Ref.~\cite{Nishimichi:2020tvu}, 
in order to extract $\tilde{c}$ here
we fitted the full set of nuisance parameters, including the stochastic 
counterterms. This is crucial for unbiased estimation of nuisance 
parameters, which are highly correlated among each other. }):
\be
\begin{split}
 & \tilde{c}^{\rm ELG}(z=0.85)=(300\pm 140)~[h^{-1}\text{Mpc}]^4\,,\\
 & \tilde{c}^{\rm LRG}(z=0.61)=(734\pm 130) ~[h^{-1}\text{Mpc}]^4\,.
 \end{split}
\ee
Here we quote $\tilde{c}^{\rm ELG}$ obtained
from the analysis of the $f_{\rm sat}=0.3$ mocks 
at $\kmax=0.16~\hMpc$. 
Recalling the relationship between $\tilde{c}$
and the cutoff of the redshift-space mapping in the EFT 
$\tilde{c}\propto (k^{r}_{\rm NL})^{-4}$ we obtain estimates
\be
\label{eq:est}
\begin{split}
& k^{r}_{\rm NL}(\text{LRG}, z=0.61)\sim 0.2~\hMpc\,,\\
&k^{r}_{\rm NL}(\text{ELG}, z=0.85)\sim 0.25~\hMpc\,.
 \end{split}
\ee
Thus, we see that the ELG RSD cutoff $k^{r}_{\rm NL}$ 
is of the same order as the cutoff 
of the LRG sample. A similar conclusion holds true 
even for mocks with $f_{\rm sat}=0$.

It is also instructive to compare the cutoffs extracted
from the quadrupole counterterms ($\sim k^2\mu^2$) of ELG and LRG samples.
We have 
\be
\begin{split}
 & (c_2)^{\rm ELG}(z=0.85)=(25\pm 10)~[h^{-1}\text{Mpc}]^2\,,\\
 & (c_2)^{\rm LRG}(z=0.61)=(35\pm 3) ~[h^{-1}\text{Mpc}]^2\,,
 \end{split}
\ee
which gives the same estimates for the 
nonlinear scale as in Eq.~\eqref{eq:est}. Note 
that these counterterms 
can be interpreted as the short-scale 
velocity dispersion $\sigma_v$, implying
\be 
\begin{split}
& \sigma_v^{\rm ELG}(z=0.85)\simeq 5~\text{Mpc}/h\,,\\
& \sigma_v^{\rm LRG}(z=0.61)\simeq 6~\text{Mpc}/h\,.
\end{split}
\ee

All in all, we conclude that the fingers-of-God effect is indeed 
somewhat weaker 
for the ELG sample, but it is not sufficiently small 
to allow us to set $\tilde{c}=0$ even for the vanishing 
satellite fraction $f_{\rm sat}=0$.

\textbf{Cosmological constraints.}
As a second step, we fit the mock spectra
for seven data cut choices 
\[
\kmax=\{0.16,0.18,0.2,0.23,0.25,0.27,0.30\}~\hMpc\,.
\]
Our results for the 1d marginalized 
posteriors of 
$\Omega_m,\sigma_8$ and $H_0$ are displayed in Fig.~\ref{fig:OR}. 
We show the fractional difference between the recovered and 
the true values, e.g. $\sigma_8^{\rm fit}/\sigma_8^{\rm true}-1$.
The blue band represents the errorbar estimates 
for the $54$ times smaller survey
volume, which is roughly equal to the ratio 
of volumes of the mocks and the actual eBOSS ELG sample. 
The triangle plot for the parameters measured in the
 $f_{\rm sat}=0.3$ case are shown in Fig.~\ref{fig:OR2},
 the 1d marginalized posteriors are displayed in
 Table~\ref{table2}.


Looking at the top of Fig.~\ref{fig:OR}, 
we see that the true
values of $\Omega_m$ and $H_0$ are 
recovered within (68-95)\%~CL for all $\kmax$ and HOD variants.\footnote{The difference w.r.t. the previous version of this paper that had some bias on $\Omega_m$ is due to the subboxes' treatment. 
Previously, we 
averaged spectra from  
subboxes. However, when measuring these spectra
we incorrectly assumed periodicity of the individual subboxes.
In contrast, now we combine the subboxes into the original simulation box and then measure the spectrum from the total catalog. 
This 
procedure allows us to reconstruct the correct spectrum 
of the simulation satisfying periodic boundary conditions,
thereby
removing the bias on $\Omega_m$. }
The $\sigma_8$ recovery is unbiased 
until $\kmax=(0.23-0.25)~\hMpc$, after which the 
measured $\sigma_8$ exhibits a bias growing with $\kmax$. 
The situation here is similar to the behavior
seen in the analyses of large-volume LRG mock data~\cite{Ivanov:2019pdj,DAmico:2019fhj,Nishimichi:2020tvu,Chudaykin:2020hbf}.
In the LRG case the $\sigma_8$ recovery is unbiased up to $\kmax\simeq 0.16~\hMpc$,
after which the bias grows and 
reaches the level of $\simeq 3\%$ at $\kmax=0.2~\hMpc$.
It is important to mention
that this LRG trend hardly depends on redshift (in the range $z=0.3-0.6$), 
which suggests that it would also hold true at the redshift 
of our mock ELG sample $z=0.85$.
We see that the ELG sample is somewhat better because 
the analysis is unbiased up to $\kmax=0.23~\hMpc$, 
and the $3\%$ bias appears at $\kmax\simeq 0.27~\hMpc$. Even though this difference 
is quite insignificant for eBOSS, it may be important for future surveys 
like DESI/Euclid, for which our results suggest that the EFT analysis indeed may be pushed to 
smaller scales.

To put this claim in context, let us notice that several 
forecasts, e.g.~\cite{Aghamousa:2016zmz,Ivanov:2019hqk,Sailer:2021yzm} suggest 
that
DESI will probe 
the velocity fluctuation amplitude $f\sigma_8$
with $\sim 0.5\%$ precision if the analysis 
is carried out at scales $\kmax\simeq 0.20~\hMpc$.
Achieving this goal is feasible with the 
ELG sample within our analysis, 
which produces unbiased results
at this $\kmax$. 
Notice also that at this $\kmax$ 
DESI is not affected by observational or instrumental systematics, and 
not dominated by shot noise~\cite{Aghamousa:2016zmz}.
However, if the strength of non-linear RSD  
were comparable to that of the LRG sample, we would be forced to stop 
at $\kmax\simeq 0.14\hMpc$~\cite{Nishimichi:2020tvu} by theoretical 
systematics,  
so that the $f\sigma_8$ precision 
would degrade by a factor of two, i.e. down to $\sim 1\%$~\cite{Aghamousa:2016zmz,Nishimichi:2020tvu}.

The bias on $\sigma_8$ is  
somewhat
stronger for mocks with larger $f_{\rm sat}$,
in agreement with arguments from the previous literature, see e.g.~\cite{Chudaykin:2019ock,Chen:2020fxs}.
However, we see a similar bias even for mocks
without satellites, which suggests that 
they are not the main culprit 
of strong fingers-of-God.
In the EFT context, the apparent bias on $\sigma_8$ is a signature of two-loop corrections
omitted in our theory model. 
In principle, the two-loop corrections can be partly
taken into account 
by the theoretical error covariance~\cite{Chudaykin:2020hbf}.
However, in terms of the true eBOSS ELG 
error bars, the bias due to two-loop corrections 
is quite small. 
This bias reaches $0.3\sigma$ of the true statistical
error on $\sigma_8$ from the eBOSS data
at $\kmax=0.25~\hMpc$.
Therefore, we believe that the choice 
$\kmax=0.25~\hMpc$ accomplishes a good 
balance between the statistical and systematic errors.


All in all, we conclude that our theory model 
should describe the actual ELG data accurately enough
up to $\kmax=0.25~\hMpc$.

\subsection{Reference EZ mocks}

\begin{figure}[ht!]
\begin{center}
\includegraphics[width=0.49\textwidth]{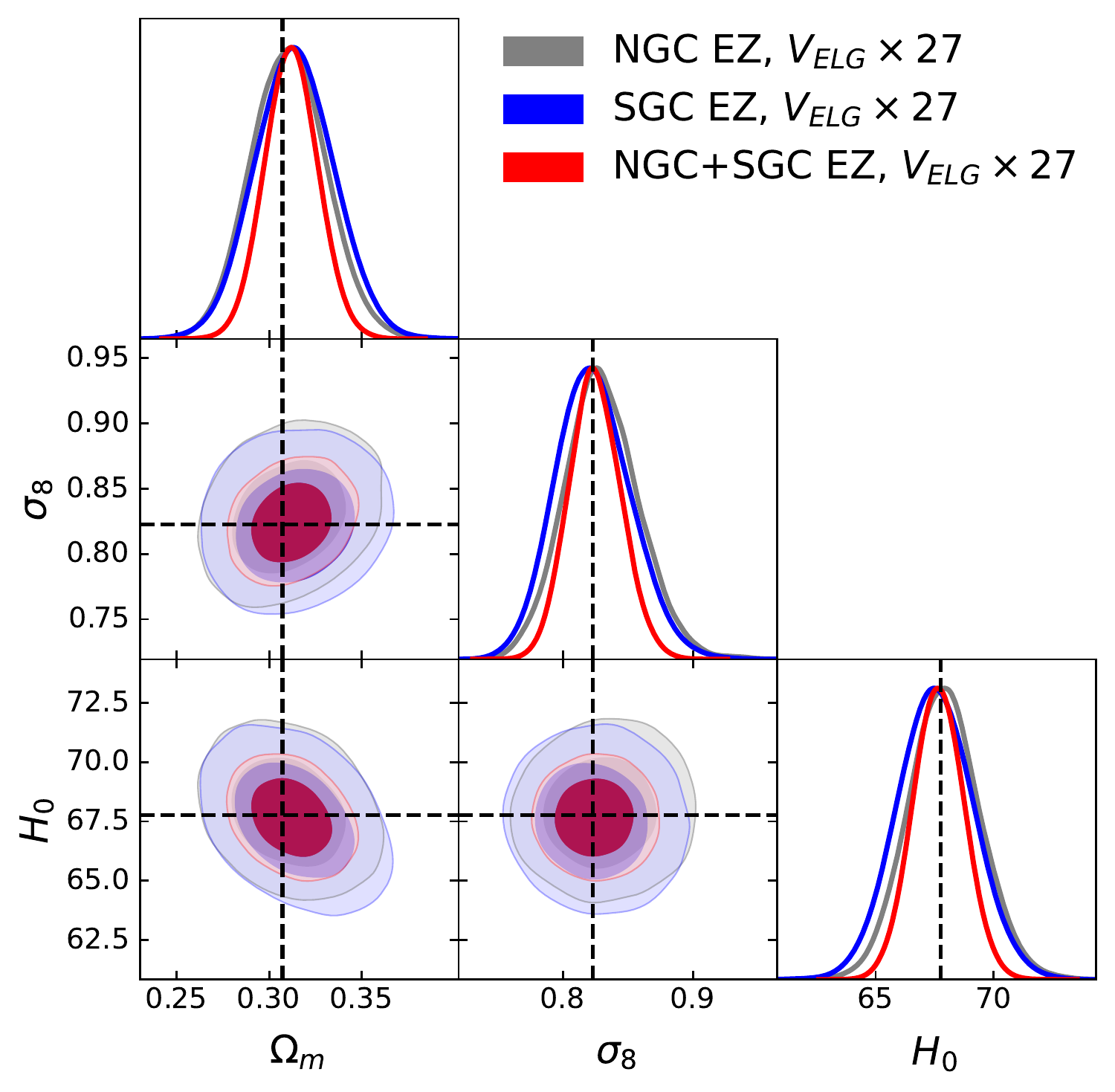}
\end{center}
\caption{
2d and 1d marginalized posteriors
from the analysis of the reference EZ mocks.
\label{fig:ezref} } 
\end{figure}
\begin{table}[!ht!]
  \begin{tabular}{|c||c|c|c|} \hline
   \diagbox{ { Sample}}{\small Param}  
   &  $\Omega_m$
   & $H_0$
   & $\sigma_8$
      \\ [0.2cm]
\hline
NGC  
& $0.3106_{-0.021}^{+0.02}$
& $67.91_{-1.6}^{+1.5}$
& $0.83_{-0.031}^{+0.027}$
\\ 
\hline
SGC    &  
$0.3137_{-0.022}^{+0.021}$
& $67.54_{-1.7}^{+1.6}$
  & $0.8236_{-0.032}^{+0.027}$
\\ 
\hline 
 SGC+NGC   & $0.3118_{-0.015}^{+0.014}$
  & $67.67_{-1.1}^{+1.1}$
  & $0.8244_{-0.021}^{+0.02}$
  \\ \hline
\end{tabular}
\caption{1d marginalized intervals of parameters $\Omega_m,H_0$ (in units km/s/Mpc), and $\sigma_8$ extracted 
from the reference EZ mocks.
In all analyses the covariance was divided by 27
to match the OR simulation volume.
The fiducial values are $\Omega_m=0.307115$, $H_0=67.77$ km/s/Mpc,
$\sigma_8=0.8225$.
}
\label{table3}
\end{table}

In this section we test our pipeline on 
approximate 
EZ lightcone mocks. These mocks 
are based on the Zel'dovich approximation for the gravity solver,
but they capture the 
lightcone effects, 
survey geometry, veto masks, 
and other observational features present in the 
actual data~\cite{Zhao:2020bib}.
We start by analyzing the reference set of mocks,
which include the survey geometry 
and veto masks, but not the observational 
systematics. The reference set 
consists of 1000 catalog realizations 
mimicking ELG clustering 
in 
northern and southern galactic caps
in the redshift range $0.6<z<1.1$.
Following~\cite{deMattia:2020fkb}, 
we model corresponding spectra 
as if they were taken at a single snapshot 
with $z_{\rm eff}=0.845$. The mock fiducial cosmology 
is given by 
\be
\begin{split}
&h=0.6777\,,\quad \Omega_m=0.307115\,,\quad \omega_b = 0.02214\,,\\
&\omega_{cdm}=0.1189\,,\quad \sigma_8=0.8225\,,\quad n_s=0.9611\,.
\end{split} 
\ee

We fit the mean of 1000 mocks 
with the corresponding empirical sample covariance 
rescaled by $27$ in
order to match the volume of the Outer Rim mocks.
Since the reference mocks do not contain
observational systematics, we do not 
implement the integral constrains
and fiber collision corrections in our theory templates.
We restrict our analysis to $k_{\rm min}=0.03~\hMpc$
(following Ref.~\cite{deMattia:2020fkb})
and $k_{\rm max}=0.25~\hMpc$. As in the previous section, we vary the cosmological parameters
$\omega_{cdm},A_s$ and $h$ in our chains and fix all other 
cosmological parameters to their fiducial values.
In Fig.~\ref{fig:ezref}
and table~\ref{table3} we show results of our analysis for SGC and
NGC chunks separately and their combination. For clarity 
we show only the parameters $\Omega_m,H_0$ and $\sigma_8$.
The first important observation is that our pipeline
recovers true values of cosmological parameters
without any noticeable bias.
The second observation is that the precision of 
cosmological parameter measurements from the EZ mocks 
is noticeably worse than those from the OR mocks without
the survey window function, especially for $\Omega_m$
and $H_0$.

\begin{figure}[ht]
\begin{center}
\includegraphics[width=0.49\textwidth]{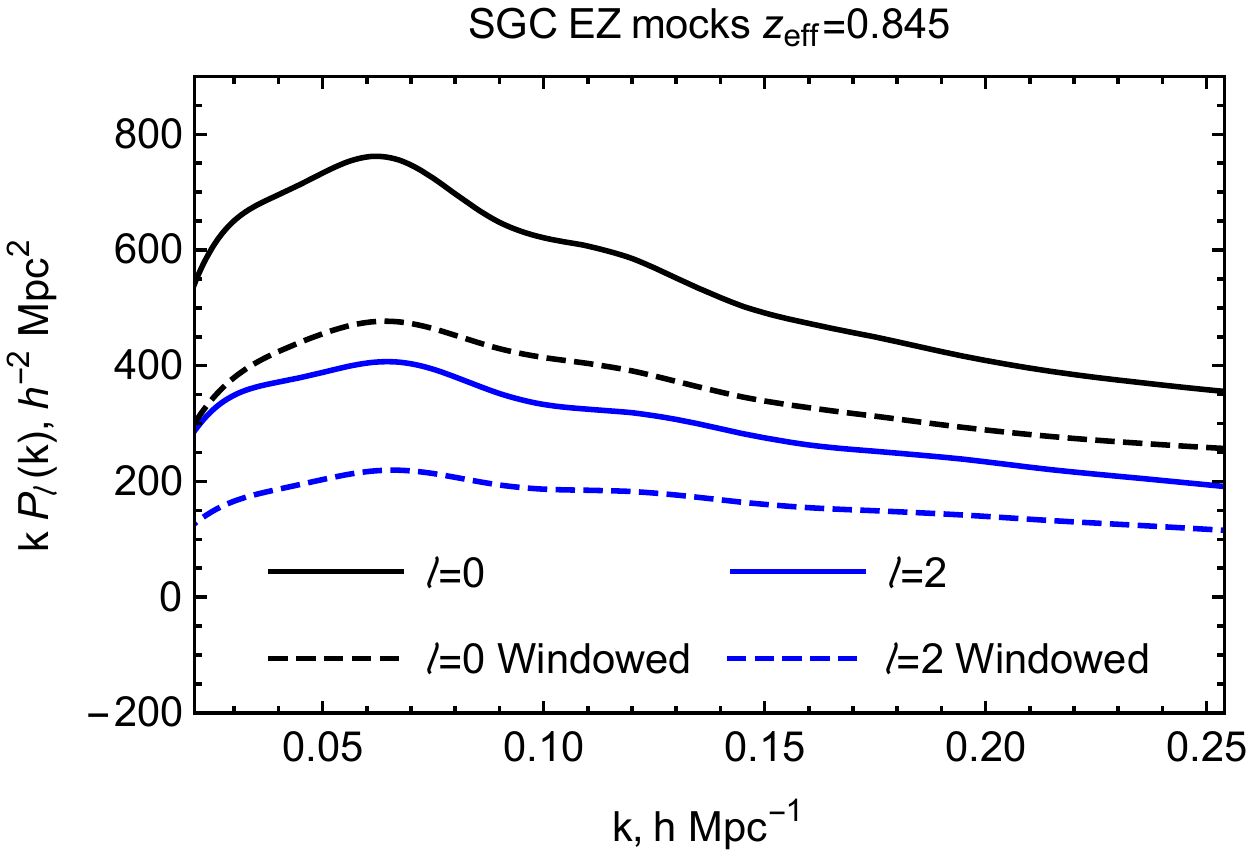}
\includegraphics[width=0.49\textwidth]{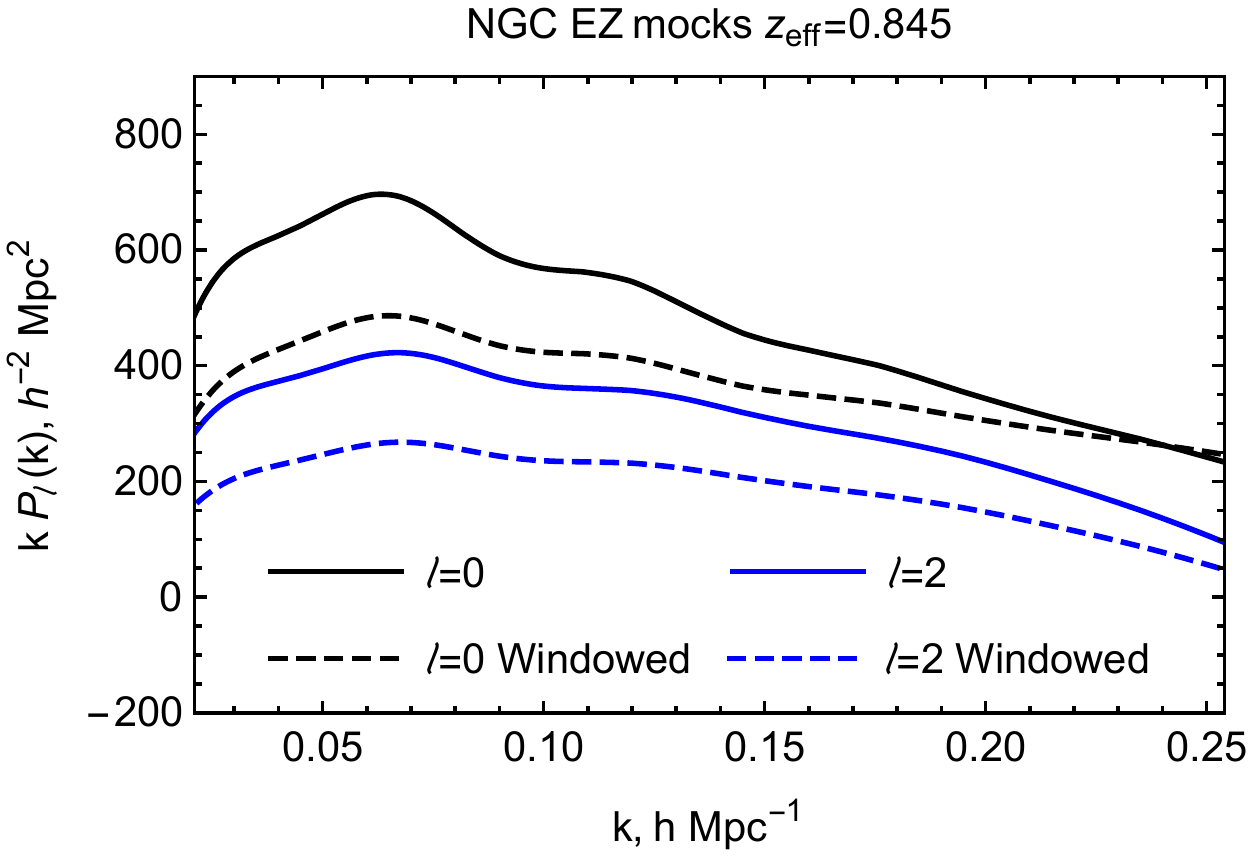}
\end{center}
\vspace{-0.5cm}
\caption{Typical power spectrum multipoles 
of the reference EZ mocks before (solid lines) and after 
(dashed lines) convolution with the survey window function. 
\label{fig:winez} } 
\end{figure}

In the next section we will see
the actual ELG full-shape data alone 
cannot strongly constrain the Hubble parameter.
This happens because of several reasons.
First, the effective volume of this sample, 
is relatively small.
Second, this sample has a specific shape of the window function,
which is noticeably 
suppressed for pair separations $s>100~h^{-1}$Mpc.
This suppression translates into a suppression
of the BAO feature, which is crucial 
for the $H_0$ measurement 
from the full shape~\cite{Ivanov:2019pdj}.
Because of these effects, 
previous analyses 
were not able to 
confidently detect the BAO 
feature in ELG sample even after 
applying the reconstruction technique~\cite{deMattia:2020fkb}.
The suppression effect can be observed in Fig.~\ref{fig:winez},
which displays a typical ELG power spectra from our MCMC chains before 
and after the convolution with the window function.
We see that the unwindowed power spectrum monopole exhibits 
a clear BAO signature, which becomes suppressed after
the window function application. 
As far as the quadrupole moment is concerned, we
do not expect to have a strong BAO feature without reconstruction.
The BAO in the quadrupole is suppressed 
due to IR resummation
to begin with~\cite{Ivanov:2018gjr}, and the 
relatively high shot noise level 
of the eBOSS ELG sample 
makes it unlikely that the 
residual BAO signature could be detected 
even if the effects of the window function
were completely absent.

\begin{figure*}[ht]
\begin{center}
\includegraphics[width=0.49\textwidth]{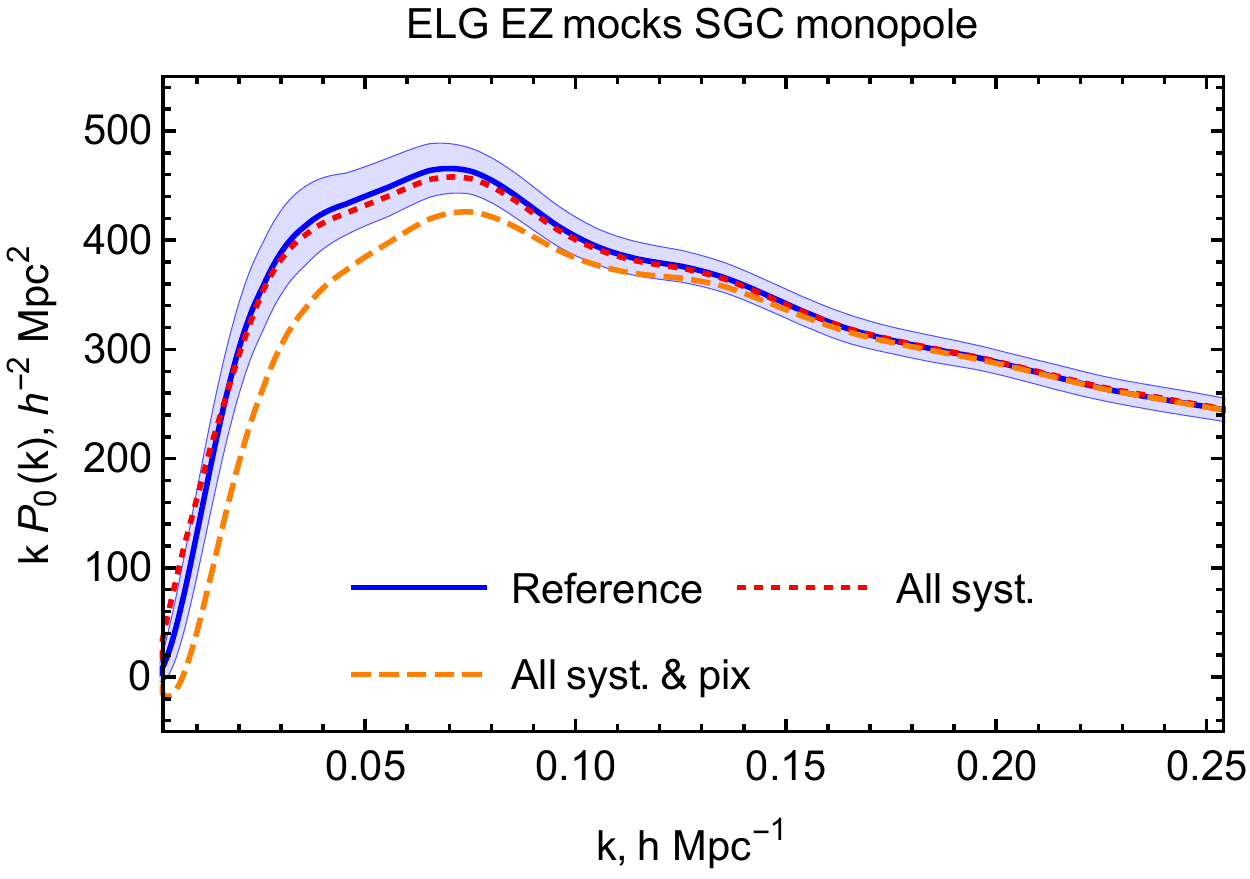}
\includegraphics[width=0.49\textwidth]{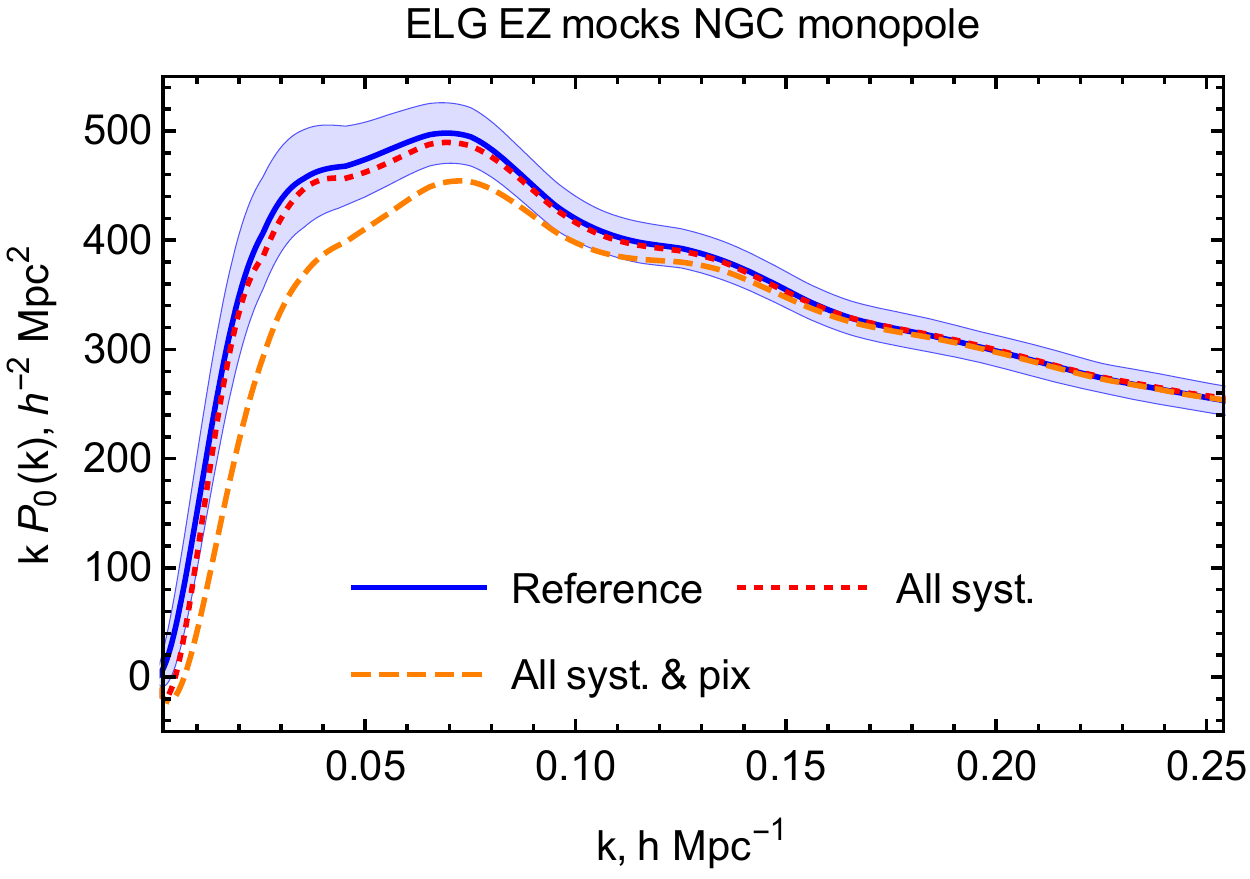}
\includegraphics[width=0.49\textwidth]{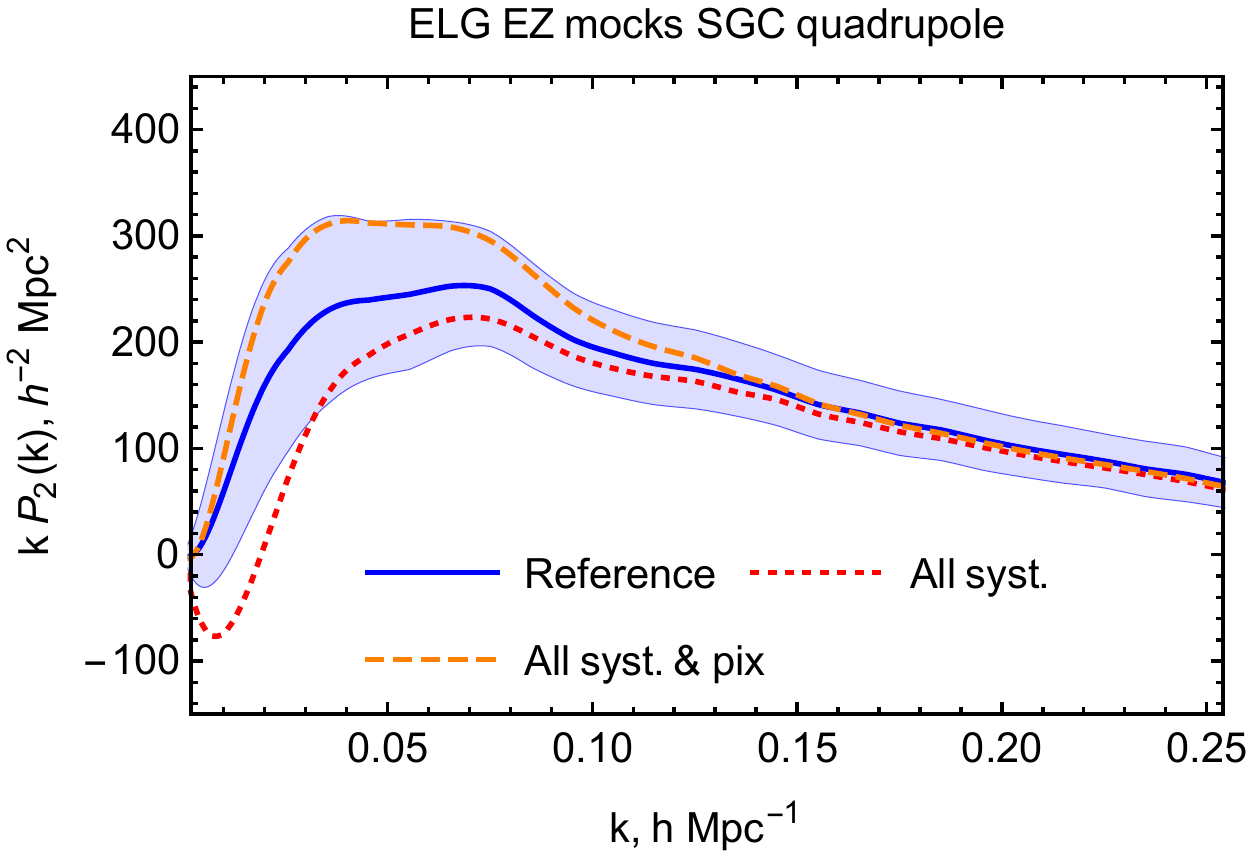}
\includegraphics[width=0.49\textwidth]{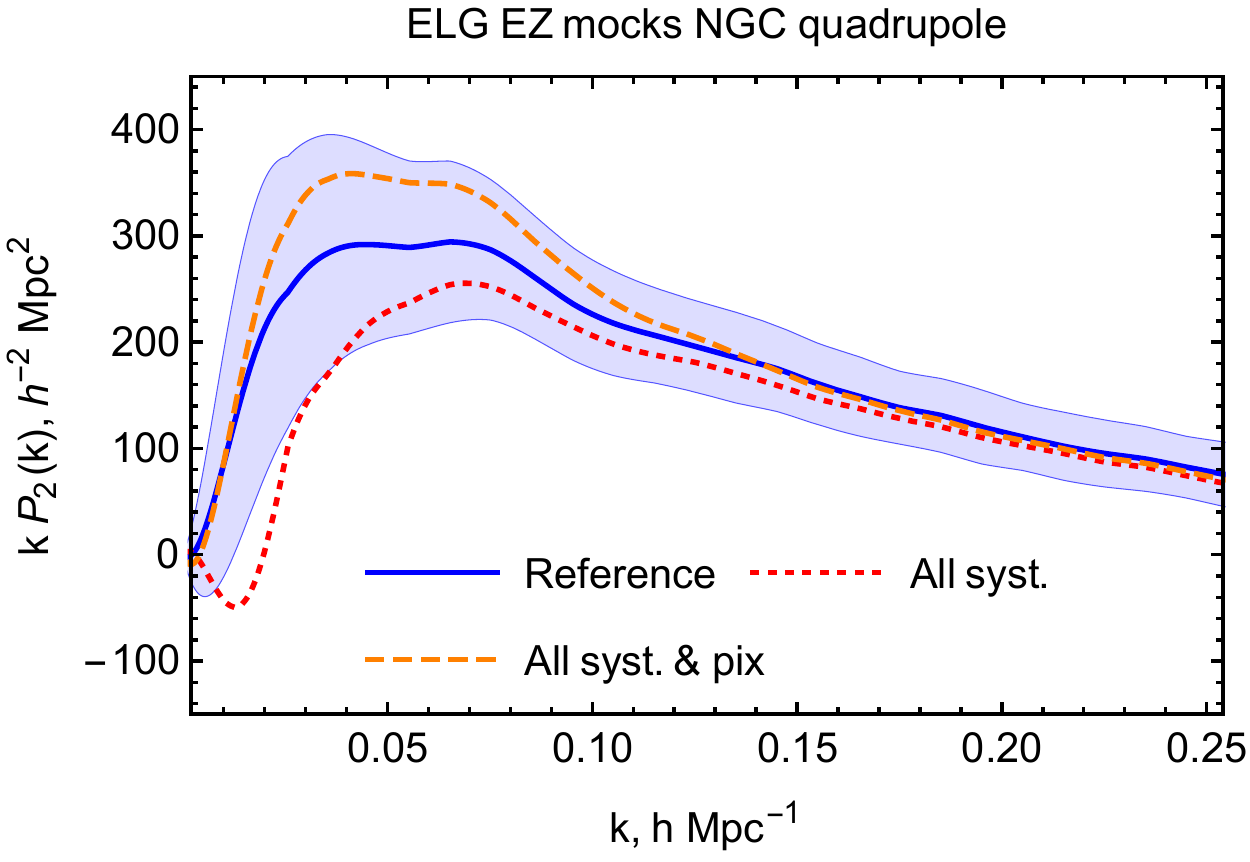}
\includegraphics[width=0.49\textwidth]{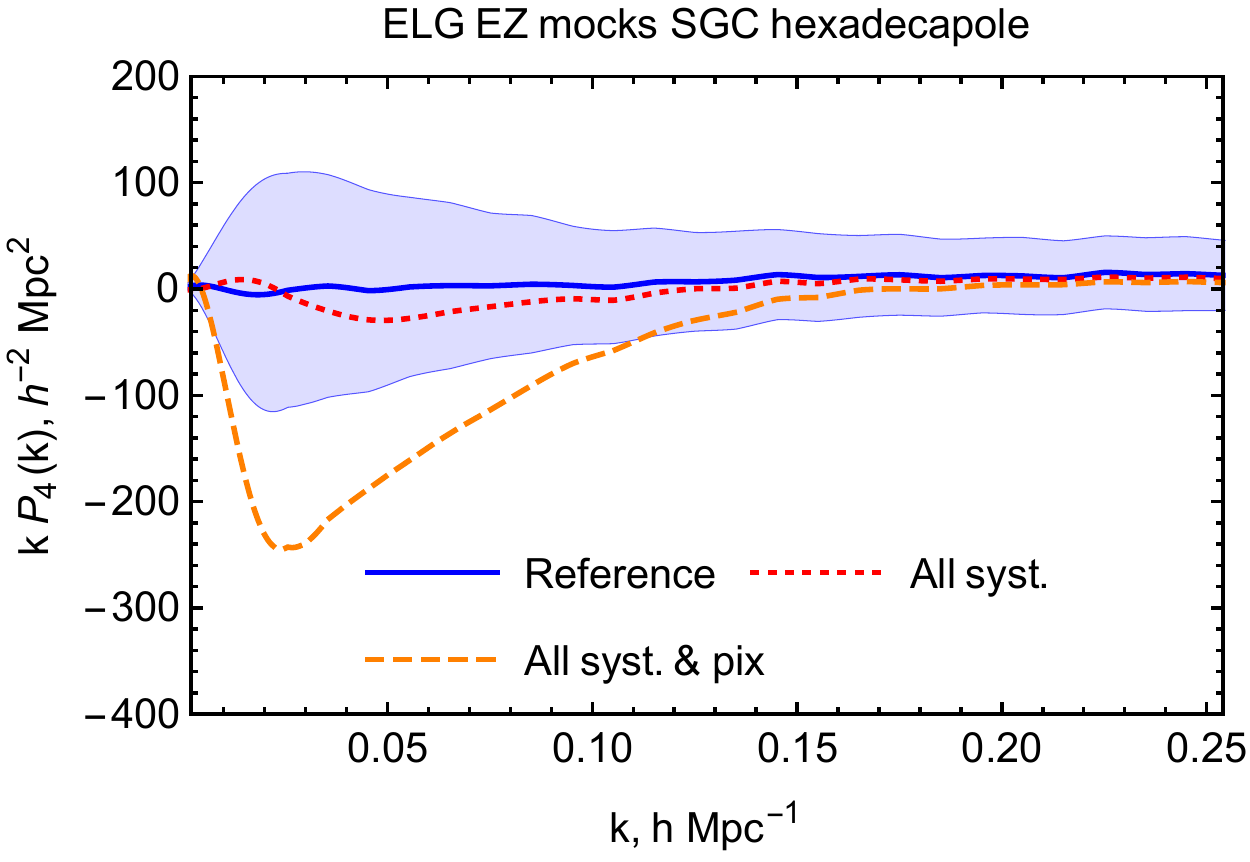}
\includegraphics[width=0.49\textwidth]{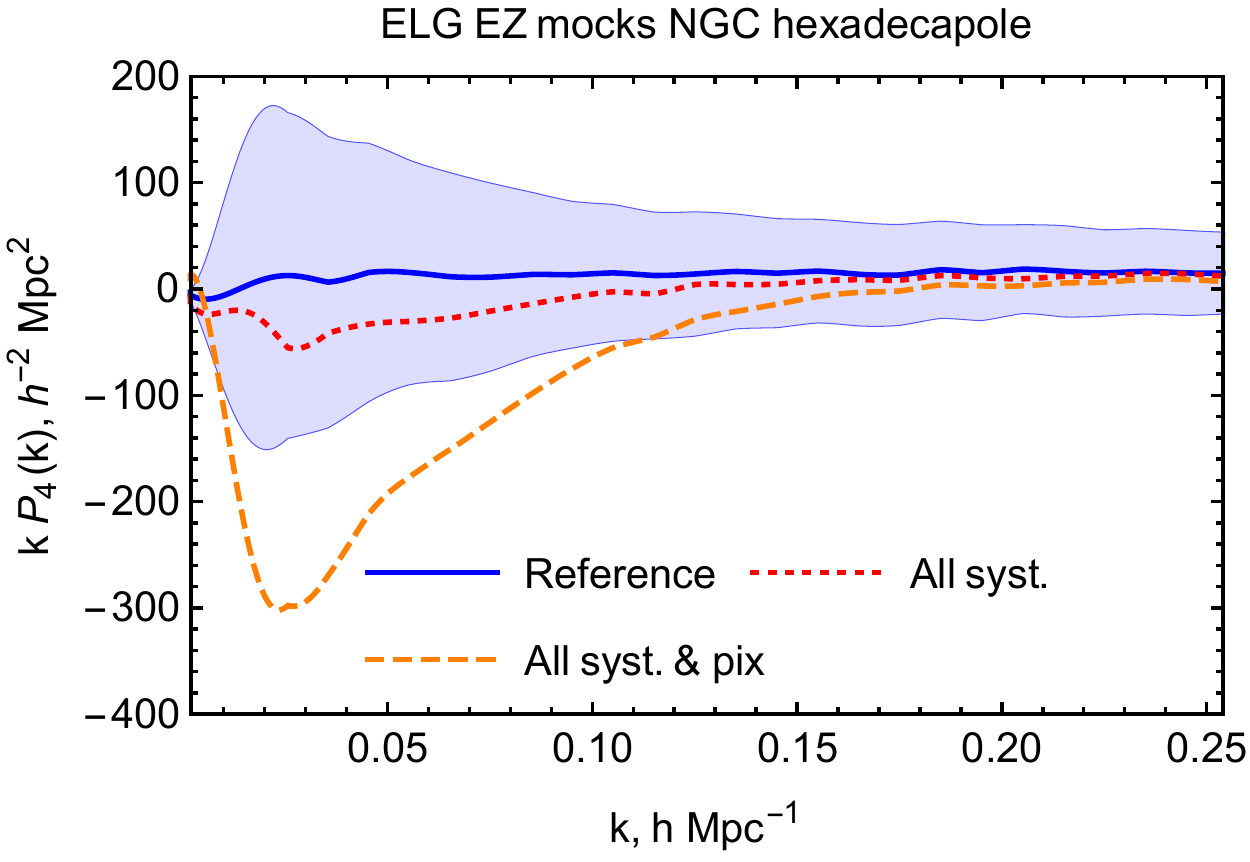}
\end{center}
\vspace{-0.5cm}
\caption{Mean of EZ mocks without observational systematics (solid blue), with all systematics (dotted red)
and with all systematics after pixelation (dashed orange).
Left and right panels show the SGC and NGC footprints, respectively.
The shaded blue band represents the variance over reference mocks without systematics.
\label{fig:syst} } 
\end{figure*}

\subsection{Contaminated EZ mocks}

In order to test the perfomance of our pipeline
on actual observations, we analyze now the ``contaminated''
EZ mocks
with injected systematic effects. The dominant 
effects are angular photometric uncertainties,
shuffling that is used to generate random
catalogs, 
and fiber collisions. 
In order to mitigate the angular systematics, 
the correction (``pixelation'') scheme is used, which 
introduces new alternations to the power spectrum shape.  
The cumulative systematic effect with and without pixelation, 
as modelled in
the mocks, can be seen in Fig.~\ref{fig:syst}.
There we display the power spectrum multipoles
of the contaminated mock catalogs, 
along with their pixelated
version, and the reference mocks discussed before. 
Note that both contaminated and pixelated 
mocks correspond to slightly different redshift 
interval $0.7<z<1.1$, which is chosen 
in order to exclude data with large variations of the redshift
density with photometric depth present in the range 
$0.6<z<0.7$. The effective redshift of the contaminated 
mocks is $z_{\rm eff}=0.86$.

\begin{figure}[ht!]
\begin{center}
\includegraphics[width=0.49\textwidth]{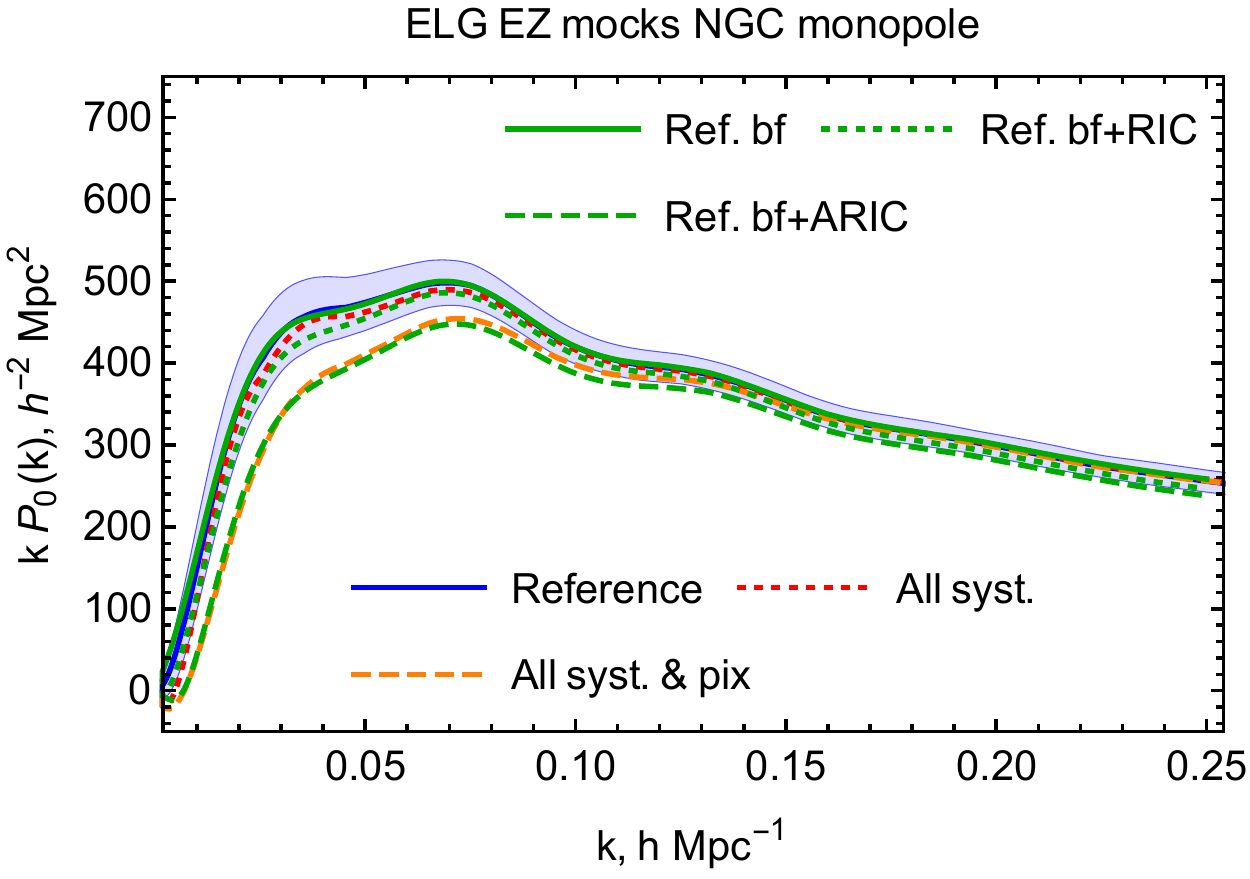}
\includegraphics[width=0.49\textwidth]{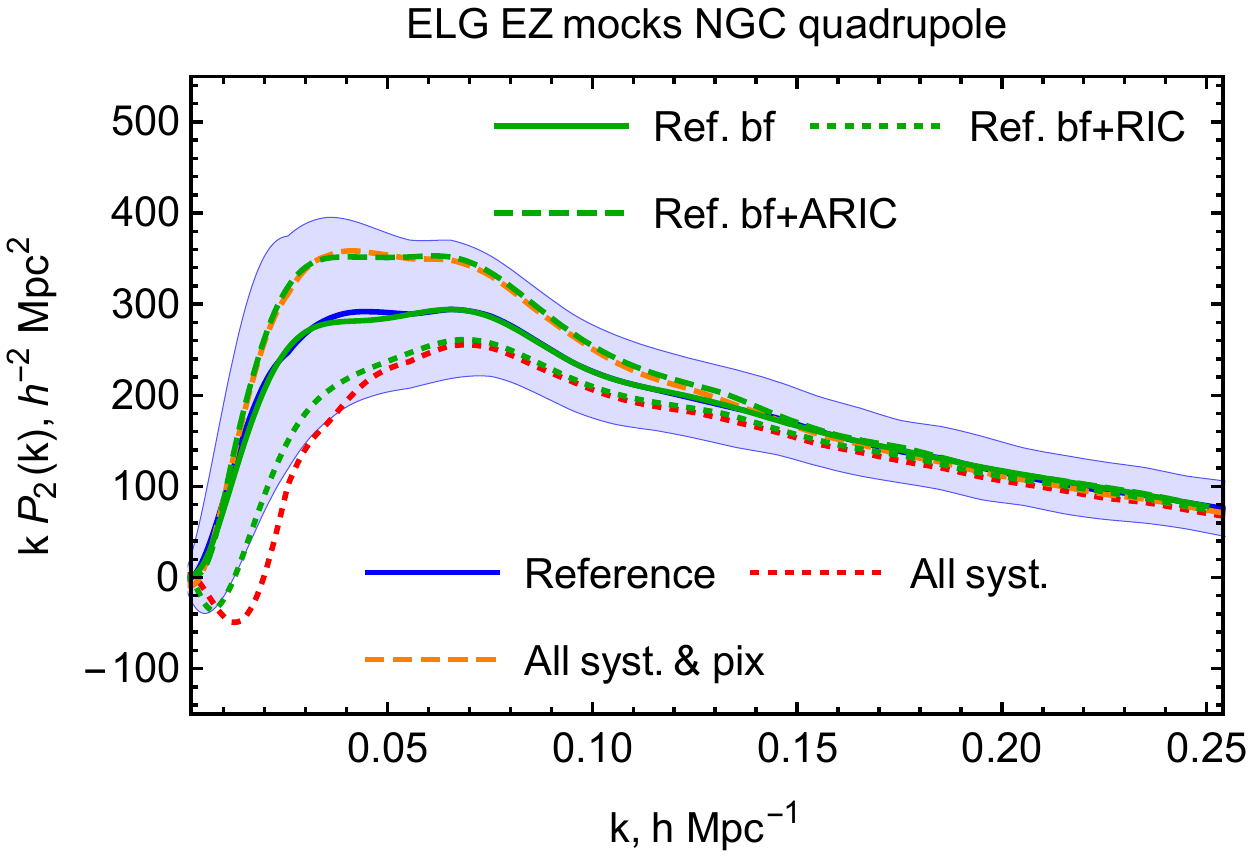}
\includegraphics[width=0.49\textwidth]{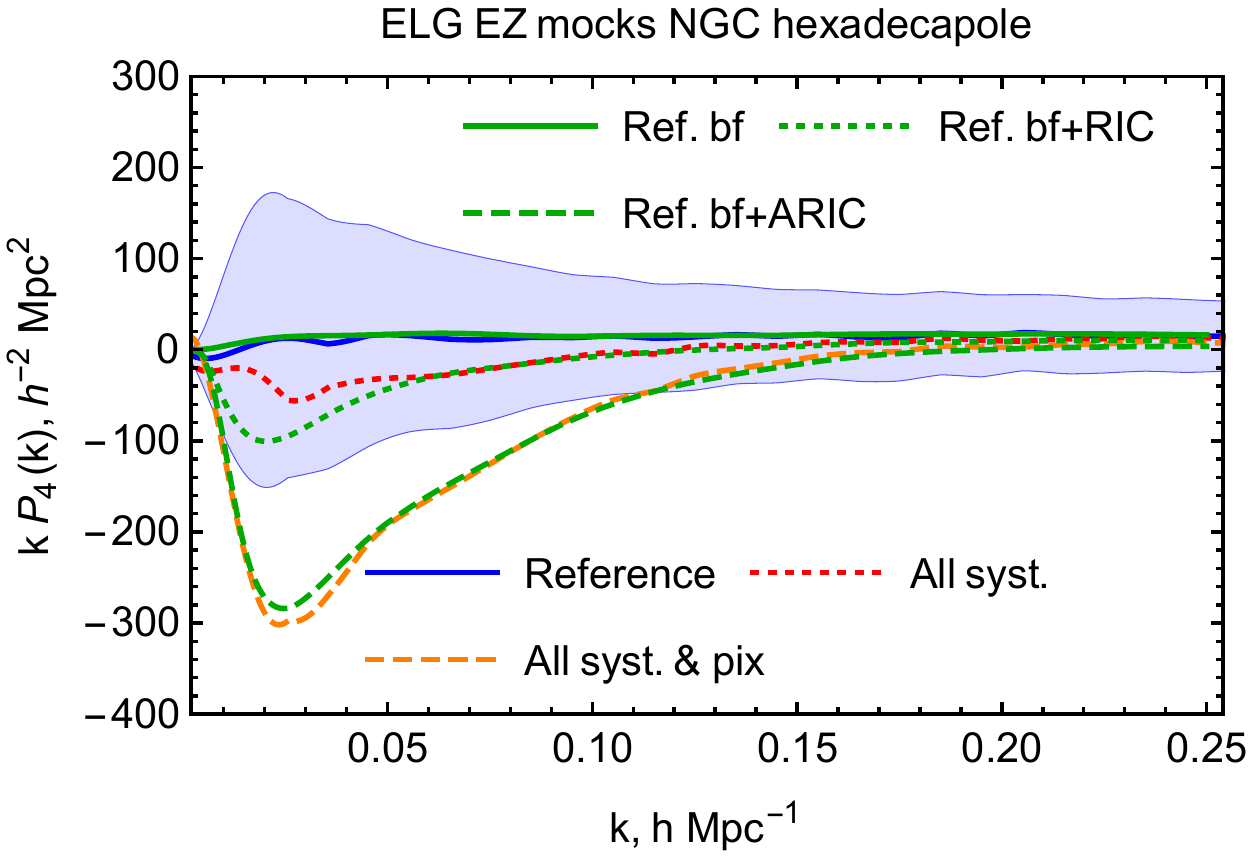}
\end{center}
\caption{
Effects of various systematic corrections on the monopole (upper panel),
quadrupole (middle panel), and hexadecapole (lower panel)
moments of the mock power spectra simulating the eBOSS NGC ELG sample.
We show mocks without systematics (``reference,'' solid blue curves),
with systematics (dashed red) and with systematics and pixelation (dashed orange).
Green curves of similar styles display appropriate modifications of the reference best-fit model.
\label{fig:arics} } 
\end{figure}

We can see that the contaminated mocks 
exhibit a clear suppression in the quadrupole,
and this suppression can be partly taken into account 
means of the radial integral constraint.
In contrast, after applying the pixelation scheme, 
the quadrupole becomes enhanced. This  enhancement can be taken into account by means of the radial and angular integral constraints. 
In order to illustrate that, in Fig.~\ref{fig:arics}
we overplot the fiducial theory templates curves on top
of curves from Fig.~\ref{fig:syst}. To obtain the theory predictions
we took the best-fit parameters of the reference mocks (``Ref. bf'')
and applied global+radial integral constraint to model the unpixelated 
data (``Ref. bf + RIC'') and the global+radial+angular integral constraint
to model the pixelated measurements (``Ref. bf + ARIC'').
We see that in the unpixelated case (``All syst'')
the agreement between the model and the mocks is not
perfect for the quadrupole and hexadecapole 
on large scales $k\lesssim 0.05~\hMpc$. This is a result of unaccounted for
angular systematics. Once the angular systematics
is mitigated by means of the pixelated scheme, 
the agreement between the mocks and the theory model
becomes restored essentially on all scales.

\begin{figure}[ht!]
\begin{center}
\includegraphics[width=0.49\textwidth]{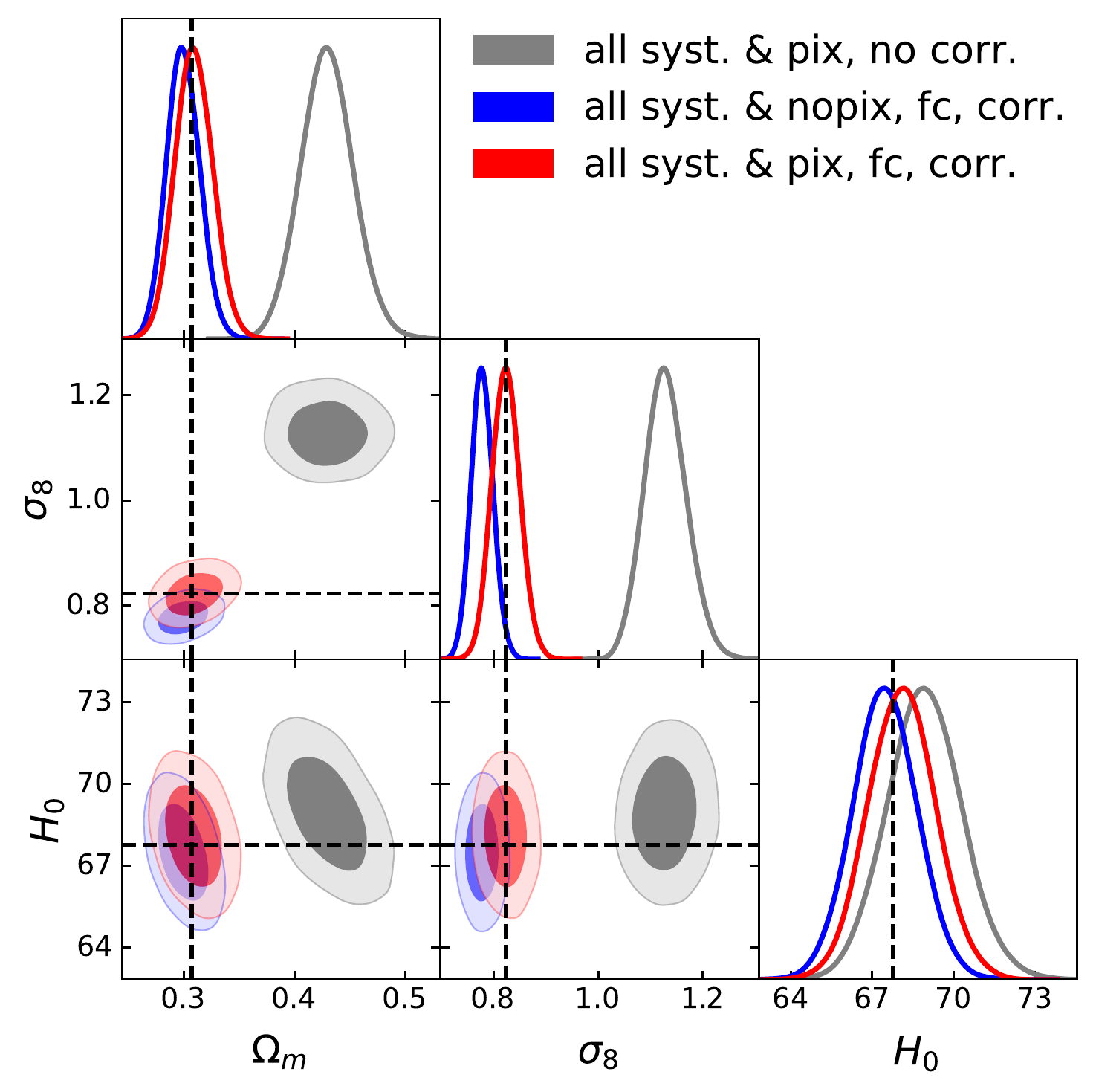}
\end{center}
\caption{
2d and 1d marginalized posteriors
from the analysis of the contaminated EZ mocks
with the partial (``all syst. \& pix, fc'') and full 
(``all syst. \& pix, fc, corr.'')
correction of systematic effects.
\label{fig:ezcont} } 
\end{figure}

If not accounted for, the scale-dependent modulation of the power spectrum due 
to radial and angular systematics would be a very strong 
source of systematics in our full-shape analysis.
As an example, in Fig.~\ref{fig:ezcont} and table~\ref{table4}
we present the results of the analysis 
of the pixelated spectra without 
implementing integral constraint corrections (``all syst. \& pix, no corr.''). 
We see that $\Omega_m$
and $\sigma_8$ are strongly biased in that case.
As a second test, we present the analysis 
of the non-pixelated spectra with all systematics injected
and with full modeling including the integral constraints 
and fiber collisions
(``all syst. \& nopix, corr.'').
We see that parameter recovery is 
satisfactory in this case, but 
the unaccounted for angular systematics 
leads to a $1\sigma$ bias on $\sigma_8$.
This bias originates due to
the additional suppression of the quadrupole on large scales
clearly seen in Fig.~\ref{fig:arics}.
As anticipated, cosmological parameter recovery 
is unbiased after implementing the pixelated scheme,
which is accurately modeled by means of the angular
integral constraint, see ``all syst. \& pix, corr.'' in Fig.~\ref{fig:ezcont}.

\begin{table}[!ht!]
  \begin{tabular}{|c||c|c|c|} \hline
   \diagbox{ { Type.}}{\small Param}  
   &  $\Omega_m$
   & $H_0$
   & $\sigma_8$
      \\ [0.2cm]
\hline
pix, no corr.
& $0.4301_{-0.024}^{+0.024}$
& $68.96_{-1.3}^{+1.5}$
& $1.129_{-0.043}^{+0.038}$
\\ 
\hline
nopix, fc, corr.    &  
$0.2997_{-0.016}^{+0.015}$
& $67.48_{-1.2}^{+1.2}$
  & $0.778_{-0.022}^{+0.021}$
\\ \hline
pix, fc, corr.    &  
$0.3093_{-0.018}^{+0.017}$
& $68.12_{-1.3}^{+1.2}$
  & $0.8232_{-0.028}^{+0.026}$
\\ \hline
\end{tabular}
\caption{1d marginalized intervals of parameters $\Omega_m,H_0$ (in units km/s/Mpc), and $\sigma_8$ extracted 
from the contaminated EZ mocks.
In all analyses the covariance was divided by 27.
The true values are $\Omega_m=0.307115$, $H_0=67.77$ km/s/Mpc,
$\sigma_8=0.8225$.
}
\label{table4}
\end{table}

\subsection{Post-reconstructed BAO data}

Now let us discuss the post-reconstructed power spectra, 
which we analyze using the method proposed in~\cite{Philcox:2020vvt}. We have found that the eBOSS ELG data 
cannot constrain the anisotropic BAO signal, and hence we focus
only on the isotropic part, which is extracted from the 
post-reconstructed monopole power spectrum moment. The angle-averaged
BAO signal is encapsulated by a parameter $\alpha$,
\be
\alpha = \frac{D_V(z_{\rm eff})}{r_s}
\left(\frac{r_s}{D_V(z_{\rm eff})}\right)\Bigg|_{\rm fid.}\,, 
\ee
where we introduced the effective spherically averaged comoving distance
\be
D_V(z)\equiv [(1+z)^2 z D^2_A(z) H^{-1}(z)]^{1/3}\,,
\ee
$D_A$ is the comoving angular diameter distance, 
$H$ is the Hubble parameter,
$r_s$ is the comoving sound horizon at the drag epoch,
and ``fid.'' refers to quantities computed in
reference cosmology used to generate the fixed power spectrum template.
We choose the same fiducial cosmology as Ref.~\cite{deMattia:2020fkb}.
We stress here that unlike the full-shape measurements,
the use of a fixed template 
is accurate in the BAO context. 
Our post-reconstructed power spectrum model is the same 
as in Ref.~\cite{Philcox:2020vvt}. We account for
non-linear effects on the broadband part 
by means of the theoretical error approach~\cite{Baldauf:2016sjb,Chudaykin:2020hbf}, which 
provides an efficient way to marginalize over 
the shape uncertainties.

\begin{figure}[ht]
\begin{center}
\includegraphics[width=0.49\textwidth]{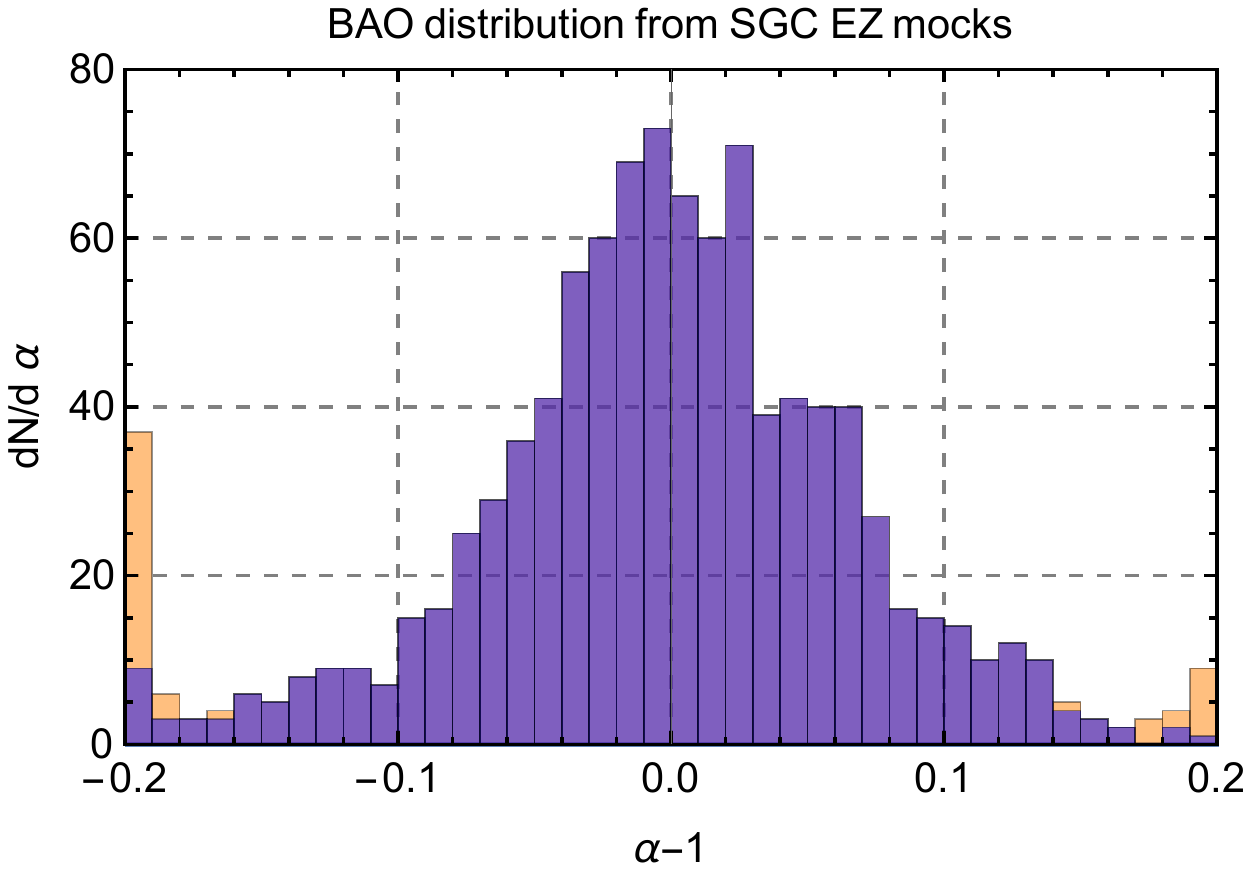}
\end{center}
\vspace{-0.5cm}
\caption{
Best fit distribution of the BAO parameter $\alpha-1$
from 1000 realizations of SGC EZ mocks. Orange histogram contains all
mocks, the violet histogram only those where the BAO is 
detected.
\label{fig:baodist} } 
\end{figure}

As a first step, we measure the isotropic BAO signal
from 1000 contaminated SGC EZ mocks. We use the mocks that span 
the full redshift range $0.6<z<1.1$, because the photometric 
systematics discussed above does not bias the BAO 
signal recovery. For each mock we fit
the parameters $\Sigma_{\rm NL}$ 
(post-reconstructed BAO smearing scale), $b_1$
and $\alpha$. The latter is constrained to a prior
range $[0.8,1.2]$~\cite{deMattia:2020fkb}.
The binned histogram of extracted best-fit values of 
$\alpha$ are shown in Fig.~\ref{fig:baodist}.
Apparent enhancements 
at the prior boundaries correspond to mocks 
where the BAO feature is not detected. 
The robust BAO detection can be characterized by 
requirements that the best-fit lies within $68\%$ confidence interval 
and this interval does not touch the prior boundaries,
\be 
\alpha_{\text{upper}~68\%~\text{CL}} < 1.2\,,\quad 
\alpha_{\text{lower}~68\%~\text{CL}} > 0.8\,,
\ee
where $\alpha_{\text{upper/lower}~68\%~\text{CL}}$
is the upper/lower boundary of the 68\% confidence interval computed using 
the two-tailed method.

Overall, we have not 
detected the BAO in 46 mocks out of 1000.
We discard them in the further analysis. 
The mean 
value of $\alpha$ from the remaining set 
matches the true value to $0.01\%$ accuracy,
\be
\frac{\langle \alpha \rangle}{\alpha_{\rm true.}}=
\frac{0.99895}{0.99937}=0.99958\,.
\ee
The rms standard deviation from the ``good'' mocks is $0.077$.
Using the best-fit value of $\alpha$ from each SGC mock
we compute the covariance between the 
pre-reconstructed SGC spectra $[P_{0},P_2,P_4]$
and $\alpha$ with the usual empirical estimator.
We have also repeated the same analysis for the NGC chunk,
but unfortunately we were not able to detect the BAO feature 
in the actual NGC data. 
This was also the case 
in the official eBOSS ELG analysis~\cite{deMattia:2020fkb}.
Therefore, for this footprint 
we proceed with the pre-reconstructed data only.

\subsection{Joint FS+BAO analysis}

Now we jointly fit the full-shape and BAO EZ mock data 
using the same pipeline and similar analysis choices as in 
the actual data analysis. We will vary now all relevant 
cosmological parameters assuming the priors \eqref{eq:priorcosm}.
As before, we will fit the mean of mocks and rescale the 
covariance by a factor 27. As a first step we analyze the full-shape 
ELG data with $\kmax=0.25~\hMpc$. As a second step we 
add the BAO data and fit the joint ELG FS+BAO likelihood.
As before, we use the FS-BAO EZ mock data covariance rescaled by 
27 and fit the mean of the BAO parameter $\alpha$ extracted from SGC EZ mocks.
For the NGC part we use only the full-shape measurements.

\begin{figure}[ht!]
\begin{center}
\includegraphics[width=0.49\textwidth]{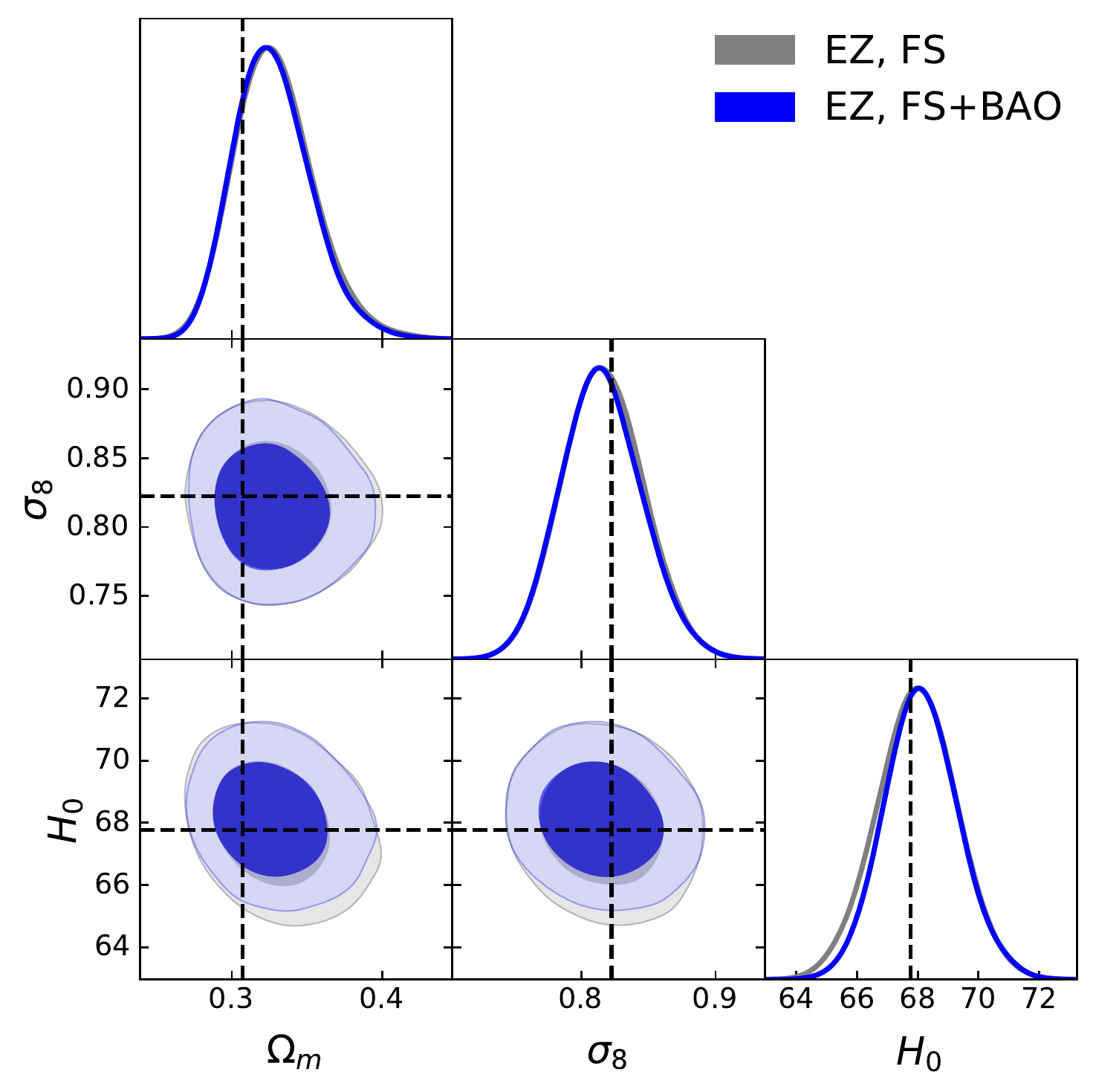}
\end{center}
\caption{
2d and 1d marginalized posteriors
of the full-shape 
and BAO data from the
contaminated EZ mocks obtained 
after marginalyzing
all $\L$CDM cosmological parameters within the priors
\eqref{eq:priorcosm}.
\label{fig:ezbao} } 
\end{figure}
\begin{table}[!ht!]
  \begin{tabular}{|c||c|c|c|} \hline
   \diagbox{ { Set}}{\small Param}  
   &  $\Omega_m$
   & $H_0$
   & $\sigma_8$
      \\ [0.2cm]
\hline
FS  
& $0.3285_{-0.03}^{+0.023}$
& $67.98_{-1.3}^{+1.4}$
& $0.8163_{-0.032}^{+0.03}$
\\ 
\hline
FS+BAO    &  
$0.3272_{-0.029}^{+0.022}$
& $68.13_{-1.3}^{+1.2}$
  & $0.8152_{-0.032}^{+0.029}$
\\ \hline
\end{tabular}
\caption{1d marginalized intervals of parameters $\Omega_m,H_0$ (in units km/s/Mpc), and $\sigma_8$ extracted 
from full-shape and BAO data from the mean 
of EZ mocks.
In all analyses the covariance was divided by 27.
The true values are $\Omega_m=0.307115$, $H_0=67.77$ km/s/Mpc,
$\sigma_8=0.8225$.
}
\label{table5}
\end{table}

Our results are presented in Fig.~\ref{fig:ezbao} and Table~\ref{table5}.
We observe that the contours 
have more non-Gaussian shape now because of correlations 
between the measured parameters $\Omega_m,\sigma_8,H_0$
and parameters $n_s,M_{\rm tot}$ that are determined by 
priors. This also explains some apparent shifts 
in the posteriors, which are driven by prior volume effects. 
In particular, $\sigma_8$ is shifted down w.r.t.
the reference analysis in table~\ref{table3} because 
of marginalization over the neutrino mass.
Finally, we see that the addition of 
the post-reconstructed BAO data improves 
the parameter constraints only marginally.
In particular, the posterior for $H_0$
sharpens only by $\sim 10\%$.

\section{Results}
\label{sec:res}

\textbf{ELG full-shape data alone.}
Let us start with the baseline analysis 
of the ELG FS data alone. 
We jointly fit the SGC+NGC monopole, 
quadrupole and hexadecapole power spectra
with $k_{\rm min}=0.03~\hMpc$
(following the analysis of Ref.~\cite{deMattia:2020fkb})
and for three choices of 
$\kmax=0.2$, $0.25$ and $0.3$ $\hMpc$.

In all analyses, we have found that the $H_0$ posterior is not sufficiently
narrower than the prior. Hence, we do not present its measurements. 
As we have explained before, this 
happens because of a relatively small effective 
volume and a particular shape 
of the ELG window function.

\begin{figure}[!thb]
\centering
\includegraphics[width=0.49\textwidth]{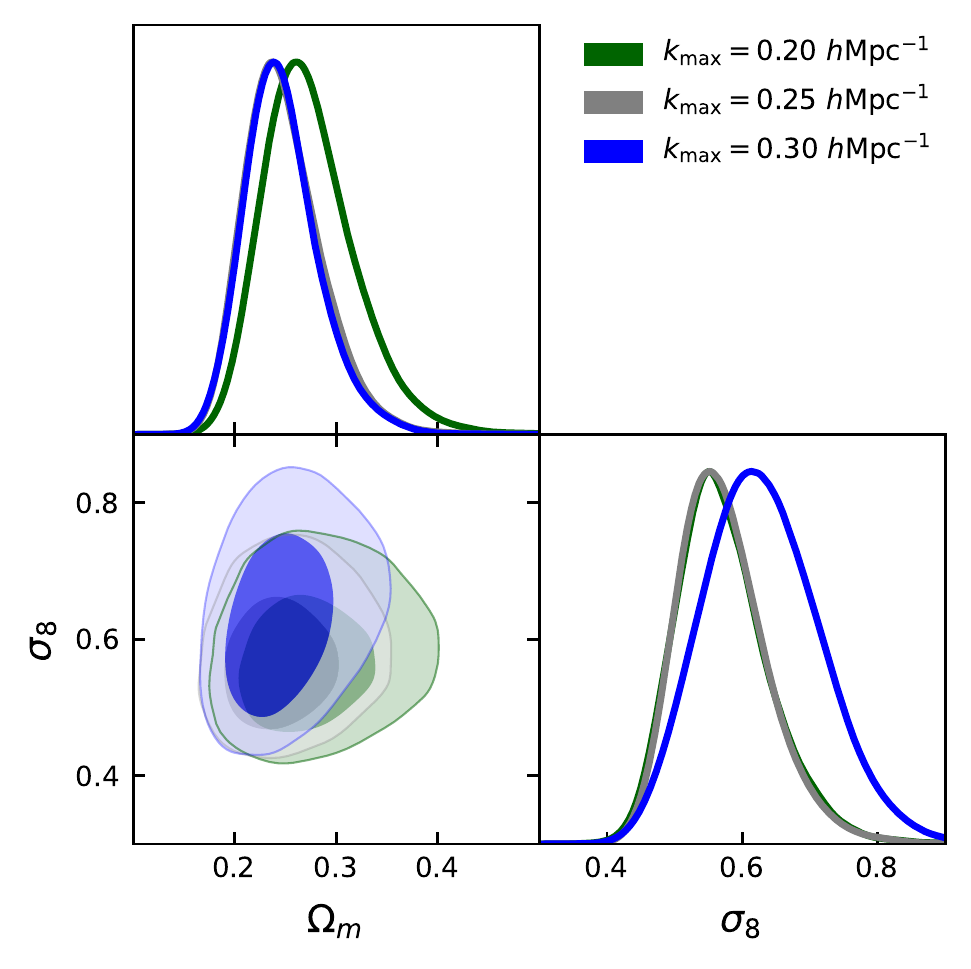}
\caption{Marginalized constraints (68\% and 95\% confidence intervals) on $\Omega_m$ and $\sigma_8$
in the flat $\L$CDM from the eBOSS ELG full-shape data
for three choices of the data cut $\kmax$.
\label{fig:eBOSSkmax} } 
\end{figure}

Due to a limited effective volume,
we cannot significantly constrain other parameters 
like $n_s$ and $M_{\rm tot}$.
Because of this, we present the results 
for $\Omega_m$ and $\sigma_8$ only. 
The posteriors for the three different 
data cuts are shown in Fig.~\ref{fig:eBOSSkmax} and 
in table~\ref{table6}.
The posterior is significantly 
non-Gaussian. 
We see that results at $\kmax=0.2~\hMpc$
and $\kmax=0.25~\hMpc$ are very consistent with one another, with the contours being somewhat narrower in the latter case.
However, the 1d marginalized 
$\sigma_8$ posterior shifts upward by 1$\sigma$ when $\kmax$
is increased up to $0.3~\hMpc$. This shift is very similar 
to the typical biases induced by two-loop corrections that we have 
seen in the simulation data analysis. 
Another interesting feature is the increase 
of the 1d marginalized error on $\sigma_8$ 
at $\kmax=0.3~\hMpc$. This may indicate 
residual systematics in the data or
be an artifact of the one-loop EFT, 
which is expected to break down at
these scales.
All in all, we select 
$\kmax=0.25~\hMpc$ as our baseline.

As an additional test, we repeat our baseline 
analysis without the fiber collision corrections, 
see table~\ref{table6}. In this case 
we find almost identical optimal values 
of relevant cosmological parameters. The
only noticeable effect is 
that the variance of $\sigma_8$ is 
underestimated by $\sim 10\%$
if 
the fiber collision corrections 
are completely ignored.

\textbf{Addition of the ELG BAO data.} 
Applying our BAO analysis pipeline to the SGC data we have 
obtained the following measurement of the isotropic BAO 
parameter,
\be 
\text{SGC:}\quad \alpha = 0.9889~(0.9926)_{-0.043}^{+0.048}\,,
\ee
where in the parenthesis we quote the best-fit value.
As an additional test, we fix $\Sigma_{\rm NL}=4~\hMpc$
as in Ref.~\cite{deMattia:2019vdg} and obtain essentially
the same result,
\be 
\text{SGC:}\quad \alpha = 0.9897~(0.9927)_{-0.045}^{+0.049}\,.
\ee

\begin{table*}[!ht!]
  \begin{tabular}{|c||c|c|c|} \hline
   \diagbox{ { Set}}{\small Param}  
   &  $\Omega_m$
   & $H_0$
   & $\sigma_8$
      \\ [0.2cm]
\hline
FS ($\kmax=0.20\hMpc$)  
& $0.2742_{-0.061}^{+0.043}$
& $>76.23~(95\%$ CL)
& $0.5725_{-0.08}^{+0.055}$
\\ 
\hline
FS ($\kmax=0.25\hMpc$)  
& $0.2486_{-0.047}^{+0.032}$
& $>75.17~(95\%$ CL)
& $0.5724_{-0.078}^{+0.053}$
\\ 
\hline
FS ($\kmax=0.30\hMpc$)  
& $0.2474_{-0.043}^{+0.031}$
& $>75.66~(95\%$ CL)
& $0.6292_{-0.097}^{+0.08}$
\\ 
\hline\hline
FS ($\kmax=0.25\hMpc$), no fc 
& $0.2426_{-0.05}^{+0.035}$
& $>75.19~(95\%$ CL)
& $0.5698_{-0.07}^{+0.048}$
\\  
\hline\hline
FS ($\kmax=0.25\hMpc$)+BAO    &  
$0.2565_{-0.045}^{+0.031}$
& $84.5_{-7}^{+5.8}$
  & $0.5713_{-0.076}^{+0.052}$
\\ \hline
\end{tabular}
\caption{1d marginalized intervals of parameters $\Omega_m,H_0$ (in units km/s/Mpc), and $\sigma_8$ extracted 
from the eBOSS ELG FS and BAO data. We use $\kmax=0.25~\hMpc$.
}
\label{table6}
\end{table*}

The NGC chunk does not carry any detectable BAO signal~\cite{deMattia:2020fkb}.
Now we perform a joint ELG FS+BAO fit, whose results
are summarized in Table~\ref{table6} and in Fig.~\ref{fig:money}. 
The addition of the BAO allows us to 
measure the Hubble constant, although 
with large errors. Indeed, unlike the case of the FS data alone, 
the $95\%$ confidence 
limit on $H_0$ from the ELG FS+BAO analysis 
is within the prior, $H_0=84.5^{+13.1}_{-11.6}$ km/s/Mpc
(95\% CL).

To put our measurements in context, in Fig.~\ref{fig:money}
we plot our posteriors along with the CMB \textit{Planck} 2018 
and FS+BAO+BBN measurements for the same parameters.
Overall, we observe a satisfactory agreement between 
the FS+BAO measurements from BOSS and eBOSS ELG.
At face value, the agreement with \textit{Planck} is worse, 
but the interpretation here
is complicated by the fact
that ELG contours are significantly non-Gaussian. 
In particular, the $\sigma_8$ posterior exhibits 
a massive tail towards large values. 
Regardless of this issue, the ELG posterior overlaps with 
the 
\textit{Planck} one within 99$\%$CL,
which is still consistent with a hypothesis 
that both data vectors are drawn from the same 
parameter set.

\textbf{Comparison with previous work.} 
Let us compare our results with 
the official ELG BOSS analysis from Ref.~\cite{deMattia:2020fkb}. 
Note that the RSD/AP parameters 
\mbox{$f\sigma_8(z=0.85)$}, 
\mbox{$D_H(z=0.85)/r_{\rm drag}$}
and $D_M(z=0.85)/r_{\rm drag}$ are\footnote{The latter are defined as 
	$D_H \equiv c/H(z)$, ($c$ is the speed of light) 
	$D_M \equiv (1+z)D_A(z)$.
}
derived parameters in our FS+BAO analysis,
and also they are subject to the $\L$CDM model priors.
The values extracted from 
our MCMC chains read: 
\be 
\begin{split}
& f\sigma_8(z=0.85)= 
0.309^{+0.029}_{-0.041}
\,,\\
& D_H(z=0.85)/r_{\rm drag} = 16.67^{+0.58}_{-0.52}    \,,\\
& D_M(z=0.85)/r_{\rm drag} = 17.89\pm 0.83  \,.  
\end{split}
\ee
Our measurements agree within $95\%$ confidence intervals with the
baseline RSD/AP measurements from the eBOSS ELG power spectrum reported in Ref.~\cite{deMattia:2020fkb}:
\be 
\begin{split}
& f\sigma_8(z=0.85)=0.289^{+0.085}_{-0.096}
\,,\\
& D_H(z=0.85)/r_{\rm drag} = 20.0^{+2.4}_{-2.2}\\
& D_M(z=0.85)/r_{\rm drag} = 19.17\pm 0.99\,.
\end{split}
\ee

\textbf{Joint analysis of ELG and LRG FS and BAO data.} 
Given a satisfactory agreement between the BOSS FS+BAO+BBN
dataset and the ELG FS+BAO data in the $\Omega_m-H_0-\sigma_8$ plains, we can combine them. 
The joint BOSS FS+BAO+BBN + ELG FS+BAO likelihood 
provides reasonable constraints on all cosmological parameters,
including the neutrino mass and the spectral index.
The results 
for this analysis, along with the \textit{Planck} CMB contours for the 
same $\nu\L$CDM model are displayed in Fig.~\ref{fig:money2}.
We see that the addition of the ELG data 
slightly tightens the LSS parameter constraints.
The maximal improvement can 
be seen in the $\ln(10^{10}A_s)-M_{\rm tot}$
plane.

A standard output of RSD measurements is the value of the rms 
velocity fluctuations $f\sigma_8$ at effective redshifts of given samples. These are derived parameters in our joint analysis,
which are given by
\be
\begin{split}
& f\sigma_8(z=0.38)=0.406\pm 0.024  \,,\\
& f\sigma_8(z=0.61)=0.399\pm 0.024  \,.
\end{split} 
\ee
Using the following tension metric between two measurements~\cite{Nunes:2021ipq},
\be
T\equiv \frac{|\text{mean}_1-\text{mean}_2|}{\sqrt{
	\text{variance}_1+\text{variance}_2
}} \,,
\ee
we estimate that our growth rates
from BOSS and eBOSS are in $2.8\sigma$ tension with the \textit{Planck} 2018 
results for the same $\nu\L$CDM model\footnote{\url{https://wiki.cosmos.esa.int/planck-legacy-archive/images/b/be/Baseline\_params\_table\_2018\_68pc.pdf}
}, $f\sigma_8(0.61)=0.4689^{+0.0070}_{-0.0045}$, 
$f\sigma_8(0.38)=0.4766^{+0.0062}_{-0.0053}$.

In order to check whether this conclusion is sensitive
to Bayesian marginalization effect, we do the following test.
First, we analyze the BOSS+ELG FS+BAO data fixing all cosmological parameters to the \textit{Planck} 2018 (P18) baseline best-fit values. 
All nuisance parameters are allowed to vary.
The total neutrino mass is fixed to it minimal value $0.06~$eV.
Second, we repeat the same analysis but additionally
vary the primordial 
spectral amplitude $A_s$ in our MCMC chains.
We find the constraints,
\be
\begin{split}
& \sigma_8=0.705~(0.711)_{-0.036}^{+0.035}\,,\\ 
&\ln (10^{10}A_{s })= 
2.761~(2.781)_{-0.1}^{+0.1}\,.
\end{split}
\ee
The effective $\chi^2$ difference between these two cases is 
\be 
\Delta\chi^2_{\rm eff}=\chi^2_{\rm eff}(\text{P18 fix.}+A_s)-\chi^2_{\rm eff}(\text{P18 fix.})=-5.4\,,
\ee
which corresponds to a $2.3\sigma$ tension. 
We conclude that the 
tension is present even from the 
frequentest point of view
but the
impact of 
Bayesian volume effects
is not negligible; it accounts for 
about $0.5\sigma$ of the ``Bayesian'' tension.

Another cosmological parameter important in the context 
of the weak lensing measurements is $S_8\equiv \sigma_8 \sqrt{\Omega_m/0.3}$,
whose value from our MCMC chains is given by
\be 
S_8 = 0.720\pm 0.042  \,.
\ee
This value is broadly consistent with the DES year 3 
weak lensing measurement $S_8 = 0.776^{+0.017}_{-0.017}$~\cite{Abbott:2021bzy} ($T=1.2$ in this case), but it is still in some 
tension with the \textit{Planck} CMB~\cite{Aghanim:2018eyx} $S_8=0.832 \pm 0.013
$, $T= 2.5$.

On the one hand, our analysis
confirms the trend of lower structure clustering
observed in other large-scale structure datasets, known
as ``$S_8$ tension,'' see Ref.~\cite{DiValentino:2020vvd} for a recent review. On the other hand, the statistical significance
is still not very significant, which suggests that this ``tension''
can still be treated as a statistical fluctuation.
Systematic errors in the RSD measurements, such as the contamination 
by selection bias~\cite{2009MNRAS.399.1074H,Obuljen:2020ypy}, 
are also not excluded.

\textbf{Bias parameters.} 
The posteriors for nuisance parameters 
are not significantly narrower than the priors,
except for the linear bias. The values extracted from the 
most constraining joint FS+BAO+BBN analysis yield
\be
b^{\rm NGC}_1=1.53_{-0.14}^{+0.14}\,,\quad 
b^{\rm SGC}_1=1.64_{-0.14}^{+0.14} \,,
\ee
which are broadly consistent with expected values
for the ELGs. For completeness, we present  
measurements of nuisance parameters, even though their 
posteriors are mostly determined by the priors (see Appendix~\ref{app}
for the full table including the BOSS samples),
\be
\begin{split}
&b^{\rm NGC}_2=0.87_{-1.9}^{+2}\,,\quad 
b^{\rm SGC}_2=0.13_{-2.1}^{+1.9} \,,\\
&b^{\rm NGC}_{\mathcal{G}_2}=-0.02_{-0.85}^{+0.94}\,,\quad 
b^{\rm SGC}_{\mathcal{G}_2}=0.63_{-0.85}^{+0.89} \,.
\end{split}
\ee

\section{Discussion}
\label{sec:disc}

In this work we have carried out a full-shape 
analysis of the eBOSS emission line galaxy sample. Our main
results are:
\begin{itemize}
\item We determined the data cut for the one-loop EFT description 
of the ELG power spectrum at $z\simeq 0.86$ to be $\kmax \simeq (0.23- 0.25)~\hMpc$, depending on the target theoretical error tolerance.
\item The ELG power spectrum model requires the next-to-leading 
order finger-of-God operator 
$k^4 \mu^4 P_{\rm lin}(k)$ just like the power spectrum of LRGs.  
However, the momentum cutoff scale of this operator $k_{\rm NL}^r$ is $\sim 20\%$ higher
for ELGs
than that of LRGs, and hence the EFT analysis could be pushed to somewhat smaller scales.
\item The current eBOSS ELG power spectrum and BAO data alone constrain 
$\Omega_m=0.257_{-0.045}^{+0.031}$, 
$\sigma_8=0.571_{-0.076}^{+0.052}$, and 
$h=0.845_{-0.07}^{+0.058}$.
These results are in a $\sim 3\sigma$ tension with the \textit{Planck} $\L$CDM model,
but this discrepancy is not statistically significant. 
Our measurements are broadly consistent
with other CMB and large-scale structure datasets.
\end{itemize}

Forecasts of Refs.~\cite{Chudaykin:2019ock,Sailer:2021yzm} suggest
that the full-shape analysis of power 
spectra from Euclid/DESI-like surveys will be able to 
strongly constrain our Universe's parameters and detect
neutrino masses. 
These surveys will mainly target emission line galaxies,
whose observational properties are not fully understood. 
In this work we have started 
exploring these galaxies 
in the context of the full-shape analysis
using the 
largest to date 
eBOSS catalog.
We believe that our analysis will initiate 
a systematic 
study of the ELG clustering statistics within 
the effective field theory framework, 
which will be important for cosmological parameter 
measurements from 
future spectroscopic surveys like DESI and Euclid.

\vspace{1cm}
\section*{Acknowledgments}

We are grateful to Giovanni Cabass, Arnaud de Mattia, 
Oliver Philcox,
Marko Simonovi\'c, Martin White 
and Matias Zaldarriaga for valuable comments on the draft. 
We thank Jay Wadekar for his help with the non-Gaussian covariance calculation. 
We also thank Santiago Avila for his help with the Outer Rim mocks.
We thank Arnaud de Mattia for sharing 
with us the eBOSS ELG EZ mock $P(k)$ measurements.

Parameter estimates presented in this paper have been obtained with the \texttt{CLASS-PT} 
Boltzmann code \cite{Chudaykin:2020aoj}\footnote{\url{https://github.com/Michalychforever/CLASS-PT}, \\\url{https://github.com/Michalychforever/lss_montepython}.
} 
(see also \cite{Blas:2011rf}) interfaced with the \texttt{Montepython} MCMC sampler \cite{Audren:2012wb,Brinckmann:2018cvx}\footnote{\url{https://github.com/brinckmann/montepython_public}.
}. 
The triangle plots are generated with the \texttt{getdist} package\footnote{\href{https://getdist.readthedocs.io/en/latest/}{
\textcolor{blue}{https://getdist.readthedocs.io/en/latest/}}
}~\cite{Lewis:2019xzd},
which is part of the \texttt{CosmoMC} code \cite{Lewis:2002ah,Lewis:2013hha}. 

\appendix

\section{Constraints table}
\label{app}

\begin{table}[!ht!]
\begin{tabular}{|l|c|c|c|c|}
 \hline
Param & best-fit & mean$\pm\sigma$ & 95\% lower & 95\% upper \\ \hline
$\omega_{cdm }$ &$0.1542$ & $0.1538_{-0.028}^{+0.021}$ & $0.1056$ & $0.2072$ \\
$n_{s }$ &$0.9187$ & $0.9496_{-0.08}^{+0.026}$ & $0.87$ & $1.05$ \\
$h$ &$0.8361$ & $0.845_{-0.07}^{+0.058}$ & $0.7295$ & $0.9764$ \\
$ln10^{10}A_{s }$ &$2.321$ & $2.155_{-0.31}^{+0.29}$ & $1.549$ & $2.767$ \\
$M_{tot }$ &$0.8151$ & $0.55^{+0.42}_{-0.19}$ & $0.$ & $1.$ \\
$b^{({\rm NGC})}_{1 }$ &$1.924$ & $2.044_{-0.31}^{+0.28}$ & $1.46$ & $2.656$ \\
$b^{({\rm NGC})}_{2 }$ &$0.004065$ & $0.6738_{-1.9}^{+1.9}$ & $-3.06$ & $4.447$ \\
$b^{({\rm NGC})}_{{\mathcal{G}_2} }$ &$-1.312$ & $-1.243_{-1.2}^{+1.4}$ & $-3.954$ & $1.259$ \\
$b^{({\rm SGC})}_{1 }$ &$1.983$ & $2.135_{-0.31}^{+0.28}$ & $1.542$ & $2.742$ \\
$b^{({\rm SGC})}_{2 }$ &$-0.1093$ & $0.1966_{-2}^{+1.9}$ & $-3.642$ & $4.082$ \\
$b^{({\rm SGC})}_{{\mathcal{G}_2} }$ &$0.4071$ & $0.2771_{-1.1}^{+1.3}$ & $-2.223$ & $2.67$ \\
$\Omega_{m }$ &$0.2651$ & $0.2565_{-0.045}^{+0.031}$ & $0.1821$ & $0.3374$ \\
$H_0$ &$83.61$ & $84.5_{-7}^{+5.8}$ & $72.95$ & $97.64$ \\
$\sigma_8$ &$0.581$ & $0.5713_{-0.076}^{+0.052}$ & $0.4442$ & $0.7106$ \\
\hline
 \end{tabular}
 \caption{
 \label{tab:elgonly}
 1d marginalized 
limits 
for cosmological and galaxy bias parameters 
of the ELG NGC and SGC FS+BAO samples
analyzed independently from other probes.
 } 
 \end{table}

\begin{table}[!ht!]
\begin{tabular}{|l|c|c|c|c|}
 \hline
Param & best-fit & mean$\pm\sigma$ & 95\% lower & 95\% upper \\ \hline
$100~\omega_{b }$ &$2.246$ & $2.267_{-0.039}^{+0.038}$ & $2.19$ & $2.346$ \\
$\omega_{cdm }$ &$0.1316$ & $0.1286_{-0.0097}^{+0.0081}$ & $0.1111$ & $0.1469$ \\
$h$ &$0.6882$ & $0.6858_{-0.011}^{+0.01}$ & $0.6649$ & $0.7068$ \\
$ln10^{10}A_{s }$ &$2.721$ & $2.733_{-0.19}^{+0.15}$ & $2.406$ & $3.086$ \\
$n_{s }$ &$0.9174$ & $0.9339_{-0.068}^{+0.063}$ & $0.8048$ & $1.066$ \\
$M_{tot }$ &$0.08057$ & $0.2258_{-0.23}^{+0.051}$ & $0.$ & $0.6272$ \\
$b^{(1)}_{1 }$ &$2.302$ & $2.375_{-0.16}^{+0.15}$ & $2.064$ & $2.686$ \\
$b^{(1)}_{2 }$ &$-1.321$ & $-1.1_{-1}^{+1}$ & $-3.147$ & $0.9087$ \\
$b^{(1)}_{{G 2} }$ &$-0.2543$ & $-0.1792_{-0.5}^{+0.53}$ & $-1.243$ & $0.8593$ \\
$b^{(2)}_{1 }$ &$2.5$ & $2.53_{-0.16}^{+0.15}$ & $2.225$ & $2.842$ \\
$b^{(2)}_{2 }$ &$-0.3781$ & $0.001504_{-0.99}^{+0.94}$ & $-1.896$ & $1.912$ \\
$b^{(2)}_{{\mathcal{G}_2} }$ &$-0.2396$ & $-0.2865_{-0.51}^{+0.53}$ & $-1.331$ & $0.7363$ \\
$b^{(3)}_{1 }$ &$2.221$ & $2.273_{-0.14}^{+0.14}$ & $2$ & $2.549$ \\
$b^{(3)}_{2 }$ &$-0.5626$ & $-0.3479_{-0.91}^{+0.85}$ & $-2.045$ & $1.419$ \\
$b^{(3)}_{{\mathcal{G}_2} }$ &$-0.4498$ & $-0.2702_{-0.42}^{+0.43}$ & $-1.114$ & $0.5783$ \\
$b^{(4)}_{1 }$ &$2.217$ & $2.299_{-0.14}^{+0.14}$ & $2.021$ & $2.578$ \\
$b^{(4)}_{2 }$ &$-0.4374$ & $-0.1263_{-0.94}^{+0.88}$ & $-1.914$ & $1.699$ \\
$b^{(4)}_{{\mathcal{G}_2} }$ &$0.04727$ & $0.1894_{-0.44}^{+0.44}$ & $-0.6876$ & $1.071$ \\
$b^{(5)}_{1 }$ &$1.482$ & $1.529_{-0.14}^{+0.14}$ & $1.248$ & $1.81$ \\
$b^{(5)}_{2 }$ &$0.4475$ & $0.87_{-1.9}^{+2}$ & $-2.916$ & $4.631$ \\
$b^{(5)}_{{\mathcal{G}_2} }$ &$-0.4962$ & $-0.01638_{-0.85}^{+0.94}$ & $-1.859$ & $1.79$ \\
$b^{(6)}_{1 }$ &$1.55$ & $1.643_{-0.14}^{+0.14}$ & $1.361$ & $1.928$ \\
$b^{(6)}_{2 }$ &$0.1653$ & $0.1282_{-2.1}^{+1.9}$ & $-3.745$ & $4.077$ \\
$b^{(6)}_{{\mathcal{G}_2} }$ &$0.8926$ & $0.6258_{-0.85}^{+0.89}$ & $-1.15$ & $2.386$ \\
$\Omega_{m }$ &$0.3271$ & $0.3267_{-0.016}^{+0.014}$ & $0.2986$ & $0.3561$ \\
$H_0$ &$68.82$ & $68.58_{-1.1}^{+1}$ & $66.49$ & $70.68$ \\
$\sigma_8$ &$0.7183$ & $0.6896_{-0.045}^{+0.038}$ & $0.6077$ & $0.7742$ \\
\hline
 \end{tabular} \\
 \caption{
\label{tab:big}
1d marginalized 
limits 
for cosmological and galaxy bias parameters 
from the joint analysis of the full-shape and BAO data from BOSS
and eBOSS ELG sample combined with the BBN prior on $\omega_b$.
The bias parameters are labeled as explained in Eq.~\eqref{eq:numbers}.
}
 \end{table}

In this appendix we present the full parameter constraints 
tables \ref{tab:elgonly}, \ref{tab:big} from our ELG FS+BAO
and 
BOSS+ELG FS+BAO+BBN analyses. We use the following 
convention for the upper indices of bias parameters:
\be 
\label{eq:numbers}
\begin{split}
&\text{BOSS NGC high-z}~(z_{\rm eff}=0.61)=(1)\,, \\
&\text{BOSS SGC high-z}~(z_{\rm eff}=0.61)=(2)\,, \\
&\text{BOSS NGC low-z}~(z_{\rm eff}=0.38)=(3)\,, \\
&\text{BOSS SGC low-z}~(z_{\rm eff}=0.38)=(4)\,, \\
&\text{eBOSS ELG NGC}~(z_{\rm eff}=0.86)=(5)\,, \\
&\text{eBOSS ELG SGC}~(z_{\rm eff}=0.86)=(6)\,. \\
\end{split}
\ee
The other nuisance parameters are marginalyzed over in
our likelihood by means of the exact analytic 
marginalization. 



\newpage 

\bibliography{short.bib}

\begin{thebibliography}{128}%
\makeatletter
\providecommand \@ifxundefined [1]{%
 \@ifx{#1\undefined}
}%
\providecommand \@ifnum [1]{%
 \ifnum #1\expandafter \@firstoftwo
 \else \expandafter \@secondoftwo
 \fi
}%
\providecommand \@ifx [1]{%
 \ifx #1\expandafter \@firstoftwo
 \else \expandafter \@secondoftwo
 \fi
}%
\providecommand \natexlab [1]{#1}%
\providecommand \enquote  [1]{``#1''}%
\providecommand \bibnamefont  [1]{#1}%
\providecommand \bibfnamefont [1]{#1}%
\providecommand \citenamefont [1]{#1}%
\providecommand \href@noop [0]{\@secondoftwo}%
\providecommand \href [0]{\begingroup \@sanitize@url \@href}%
\providecommand \@href[1]{\@@startlink{#1}\@@href}%
\providecommand \@@href[1]{\endgroup#1\@@endlink}%
\providecommand \@sanitize@url [0]{\catcode `\\12\catcode `\$12\catcode
  `\&12\catcode `\#12\catcode `\^12\catcode `\_12\catcode `\%12\relax}%
\providecommand \@@startlink[1]{}%
\providecommand \@@endlink[0]{}%
\providecommand \url  [0]{\begingroup\@sanitize@url \@url }%
\providecommand \@url [1]{\endgroup\@href {#1}{\urlprefix }}%
\providecommand \urlprefix  [0]{URL }%
\providecommand \Eprint [0]{\href }%
\providecommand \doibase [0]{http://dx.doi.org/}%
\providecommand \selectlanguage [0]{\@gobble}%
\providecommand \bibinfo  [0]{\@secondoftwo}%
\providecommand \bibfield  [0]{\@secondoftwo}%
\providecommand \translation [1]{[#1]}%
\providecommand \BibitemOpen [0]{}%
\providecommand \bibitemStop [0]{}%
\providecommand \bibitemNoStop [0]{.\EOS\space}%
\providecommand \EOS [0]{\spacefactor3000\relax}%
\providecommand \BibitemShut  [1]{\csname bibitem#1\endcsname}%
\let\auto@bib@innerbib\@empty
\bibitem [{\citenamefont {{Peebles}}(1980)}]{1980lssu.book.....P}%
  \BibitemOpen
  \bibfield  {author} {\bibinfo {author} {\bibfnamefont {P.~J.~E.}\
  \bibnamefont {{Peebles}}},\ }\href@noop {} {\emph {\bibinfo {title} {{The
  large-scale structure of the universe}}}}\ (\bibinfo {year}
  {1980})\BibitemShut {NoStop}%
\bibitem [{\citenamefont {Percival}\ \emph {et~al.}(2001)\citenamefont
  {Percival} \emph {et~al.}}]{Percival:2001hw}%
  \BibitemOpen
  \bibfield  {author} {\bibinfo {author} {\bibfnamefont {W.~J.}\ \bibnamefont
  {Percival}} \emph {et~al.} (\bibinfo {collaboration} {2dFGRS}),\ }\href
  {\doibase 10.1046/j.1365-8711.2001.04827.x} {\bibfield  {journal} {\bibinfo
  {journal} {Mon. Not. Roy. Astron. Soc.}\ }\textbf {\bibinfo {volume} {327}},\
  \bibinfo {pages} {1297} (\bibinfo {year} {2001})},\ \Eprint
  {http://arxiv.org/abs/astro-ph/0105252} {arXiv:astro-ph/0105252} \BibitemShut
  {NoStop}%
\bibitem [{\citenamefont {Aghanim}\ \emph {et~al.}(2018)\citenamefont {Aghanim}
  \emph {et~al.}}]{Aghanim:2018eyx}%
  \BibitemOpen
  \bibfield  {author} {\bibinfo {author} {\bibfnamefont {N.}~\bibnamefont
  {Aghanim}} \emph {et~al.} (\bibinfo {collaboration} {Planck}),\ }\href@noop
  {} {\  (\bibinfo {year} {2018})},\ \Eprint {http://arxiv.org/abs/1807.06209}
  {arXiv:1807.06209 [astro-ph.CO]} \BibitemShut {NoStop}%
\bibitem [{\citenamefont {Aiola}\ \emph {et~al.}(2020)\citenamefont {Aiola}
  \emph {et~al.}}]{Aiola:2020azj}%
  \BibitemOpen
  \bibfield  {author} {\bibinfo {author} {\bibfnamefont {S.}~\bibnamefont
  {Aiola}} \emph {et~al.} (\bibinfo {collaboration} {ACT}),\ }\href {\doibase
  10.1088/1475-7516/2020/12/047} {\bibfield  {journal} {\bibinfo  {journal}
  {JCAP}\ }\textbf {\bibinfo {volume} {12}},\ \bibinfo {pages} {047} (\bibinfo
  {year} {2020})},\ \Eprint {http://arxiv.org/abs/2007.07288} {arXiv:2007.07288
  [astro-ph.CO]} \BibitemShut {NoStop}%
\bibitem [{\citenamefont {Aylor}\ \emph {et~al.}(2017)\citenamefont {Aylor}
  \emph {et~al.}}]{Aylor:2017haa}%
  \BibitemOpen
  \bibfield  {author} {\bibinfo {author} {\bibfnamefont {K.}~\bibnamefont
  {Aylor}} \emph {et~al.} (\bibinfo {collaboration} {SPT}),\ }\href {\doibase
  10.3847/1538-4357/aa947b} {\bibfield  {journal} {\bibinfo  {journal}
  {Astrophys. J.}\ }\textbf {\bibinfo {volume} {850}},\ \bibinfo {pages} {101}
  (\bibinfo {year} {2017})},\ \Eprint {http://arxiv.org/abs/1706.10286}
  {arXiv:1706.10286 [astro-ph.CO]} \BibitemShut {NoStop}%
\bibitem [{\citenamefont {Bianchini}\ \emph {et~al.}(2020)\citenamefont
  {Bianchini} \emph {et~al.}}]{Bianchini:2019vxp}%
  \BibitemOpen
  \bibfield  {author} {\bibinfo {author} {\bibfnamefont {F.}~\bibnamefont
  {Bianchini}} \emph {et~al.} (\bibinfo {collaboration} {SPT}),\ }\href
  {\doibase 10.3847/1538-4357/ab6082} {\bibfield  {journal} {\bibinfo
  {journal} {Astrophys. J.}\ }\textbf {\bibinfo {volume} {888}},\ \bibinfo
  {pages} {119} (\bibinfo {year} {2020})},\ \Eprint
  {http://arxiv.org/abs/1910.07157} {arXiv:1910.07157 [astro-ph.CO]}
  \BibitemShut {NoStop}%
\bibitem [{\citenamefont {Asgari}\ \emph {et~al.}(2021)\citenamefont {Asgari}
  \emph {et~al.}}]{Asgari:2020wuj}%
  \BibitemOpen
  \bibfield  {author} {\bibinfo {author} {\bibfnamefont {M.}~\bibnamefont
  {Asgari}} \emph {et~al.} (\bibinfo {collaboration} {KiDS}),\ }\href {\doibase
  10.1051/0004-6361/202039070} {\bibfield  {journal} {\bibinfo  {journal}
  {Astron. Astrophys.}\ }\textbf {\bibinfo {volume} {645}},\ \bibinfo {pages}
  {A104} (\bibinfo {year} {2021})},\ \Eprint {http://arxiv.org/abs/2007.15633}
  {arXiv:2007.15633 [astro-ph.CO]} \BibitemShut {NoStop}%
\bibitem [{\citenamefont {Abbott}\ \emph {et~al.}(2018)\citenamefont {Abbott}
  \emph {et~al.}}]{Abbott:2017wau}%
  \BibitemOpen
  \bibfield  {author} {\bibinfo {author} {\bibfnamefont {T.~M.~C.}\
  \bibnamefont {Abbott}} \emph {et~al.} (\bibinfo {collaboration} {DES}),\
  }\href {\doibase 10.1103/PhysRevD.98.043526} {\bibfield  {journal} {\bibinfo
  {journal} {Phys. Rev.}\ }\textbf {\bibinfo {volume} {D98}},\ \bibinfo {pages}
  {043526} (\bibinfo {year} {2018})},\ \Eprint
  {http://arxiv.org/abs/1708.01530} {arXiv:1708.01530 [astro-ph.CO]}
  \BibitemShut {NoStop}%
\bibitem [{\citenamefont {Abbott}\ \emph {et~al.}(2021)\citenamefont {Abbott}
  \emph {et~al.}}]{Abbott:2021bzy}%
  \BibitemOpen
  \bibfield  {author} {\bibinfo {author} {\bibfnamefont {T.~M.~C.}\
  \bibnamefont {Abbott}} \emph {et~al.} (\bibinfo {collaboration} {DES}),\
  }\href@noop {} {\  (\bibinfo {year} {2021})},\ \Eprint
  {http://arxiv.org/abs/2105.13549} {arXiv:2105.13549 [astro-ph.CO]}
  \BibitemShut {NoStop}%
\bibitem [{\citenamefont {Alam}\ \emph {et~al.}(2017)\citenamefont {Alam} \emph
  {et~al.}}]{Alam:2016hwk}%
  \BibitemOpen
  \bibfield  {author} {\bibinfo {author} {\bibfnamefont {S.}~\bibnamefont
  {Alam}} \emph {et~al.} (\bibinfo {collaboration} {BOSS}),\ }\href {\doibase
  10.1093/mnras/stx721} {\bibfield  {journal} {\bibinfo  {journal} {Mon. Not.
  Roy. Astron. Soc.}\ }\textbf {\bibinfo {volume} {470}},\ \bibinfo {pages}
  {2617} (\bibinfo {year} {2017})},\ \Eprint {http://arxiv.org/abs/1607.03155}
  {arXiv:1607.03155 [astro-ph.CO]} \BibitemShut {NoStop}%
\bibitem [{\citenamefont {Alam}\ \emph {et~al.}(2021)\citenamefont {Alam} \emph
  {et~al.}}]{Alam:2020sor}%
  \BibitemOpen
  \bibfield  {author} {\bibinfo {author} {\bibfnamefont {S.}~\bibnamefont
  {Alam}} \emph {et~al.} (\bibinfo {collaboration} {eBOSS}),\ }\href {\doibase
  10.1103/PhysRevD.103.083533} {\bibfield  {journal} {\bibinfo  {journal}
  {Phys. Rev. D}\ }\textbf {\bibinfo {volume} {103}},\ \bibinfo {pages}
  {083533} (\bibinfo {year} {2021})},\ \Eprint
  {http://arxiv.org/abs/2007.08991} {arXiv:2007.08991 [astro-ph.CO]}
  \BibitemShut {NoStop}%
\bibitem [{\citenamefont {Tegmark}\ \emph {et~al.}(2002)\citenamefont {Tegmark}
  \emph {et~al.}}]{Tegmark:2001xb}%
  \BibitemOpen
  \bibfield  {author} {\bibinfo {author} {\bibfnamefont {M.}~\bibnamefont
  {Tegmark}} \emph {et~al.} (\bibinfo {collaboration} {SDSS}),\ }\href
  {\doibase 10.1086/339894} {\bibfield  {journal} {\bibinfo  {journal}
  {Astrophys. J.}\ }\textbf {\bibinfo {volume} {571}},\ \bibinfo {pages} {191}
  (\bibinfo {year} {2002})},\ \Eprint {http://arxiv.org/abs/astro-ph/0107418}
  {arXiv:astro-ph/0107418} \BibitemShut {NoStop}%
\bibitem [{\citenamefont {Tegmark}\ \emph {et~al.}(2004)\citenamefont {Tegmark}
  \emph {et~al.}}]{Tegmark:2003uf}%
  \BibitemOpen
  \bibfield  {author} {\bibinfo {author} {\bibfnamefont {M.}~\bibnamefont
  {Tegmark}} \emph {et~al.} (\bibinfo {collaboration} {SDSS}),\ }\href
  {\doibase 10.1086/382125} {\bibfield  {journal} {\bibinfo  {journal}
  {Astrophys. J.}\ }\textbf {\bibinfo {volume} {606}},\ \bibinfo {pages} {702}
  (\bibinfo {year} {2004})},\ \Eprint {http://arxiv.org/abs/astro-ph/0310725}
  {arXiv:astro-ph/0310725} \BibitemShut {NoStop}%
\bibitem [{\citenamefont {Cole}\ \emph {et~al.}(2005)\citenamefont {Cole} \emph
  {et~al.}}]{Cole:2005sx}%
  \BibitemOpen
  \bibfield  {author} {\bibinfo {author} {\bibfnamefont {S.}~\bibnamefont
  {Cole}} \emph {et~al.} (\bibinfo {collaboration} {2dFGRS}),\ }\href {\doibase
  10.1111/j.1365-2966.2005.09318.x} {\bibfield  {journal} {\bibinfo  {journal}
  {Mon. Not. Roy. Astron. Soc.}\ }\textbf {\bibinfo {volume} {362}},\ \bibinfo
  {pages} {505} (\bibinfo {year} {2005})},\ \Eprint
  {http://arxiv.org/abs/astro-ph/0501174} {arXiv:astro-ph/0501174} \BibitemShut
  {NoStop}%
\bibitem [{\citenamefont {Percival}\ \emph
  {et~al.}(2007{\natexlab{a}})\citenamefont {Percival} \emph
  {et~al.}}]{Percival:2006gt}%
  \BibitemOpen
  \bibfield  {author} {\bibinfo {author} {\bibfnamefont {W.~J.}\ \bibnamefont
  {Percival}} \emph {et~al.},\ }\href {\doibase 10.1086/510615} {\bibfield
  {journal} {\bibinfo  {journal} {Astrophys. J.}\ }\textbf {\bibinfo {volume}
  {657}},\ \bibinfo {pages} {645} (\bibinfo {year} {2007}{\natexlab{a}})},\
  \Eprint {http://arxiv.org/abs/astro-ph/0608636} {arXiv:astro-ph/0608636}
  \BibitemShut {NoStop}%
\bibitem [{\citenamefont {Smith}\ \emph {et~al.}(2003)\citenamefont {Smith},
  \citenamefont {Peacock}, \citenamefont {Jenkins}, \citenamefont {White},
  \citenamefont {Frenk}, \citenamefont {Pearce}, \citenamefont {Thomas},
  \citenamefont {Efstathiou},\ and\ \citenamefont {Couchmann}}]{Smith:2002dz}%
  \BibitemOpen
  \bibfield  {author} {\bibinfo {author} {\bibfnamefont {R.~E.}\ \bibnamefont
  {Smith}}, \bibinfo {author} {\bibfnamefont {J.~A.}\ \bibnamefont {Peacock}},
  \bibinfo {author} {\bibfnamefont {A.}~\bibnamefont {Jenkins}}, \bibinfo
  {author} {\bibfnamefont {S.~D.~M.}\ \bibnamefont {White}}, \bibinfo {author}
  {\bibfnamefont {C.~S.}\ \bibnamefont {Frenk}}, \bibinfo {author}
  {\bibfnamefont {F.~R.}\ \bibnamefont {Pearce}}, \bibinfo {author}
  {\bibfnamefont {P.~A.}\ \bibnamefont {Thomas}}, \bibinfo {author}
  {\bibfnamefont {G.}~\bibnamefont {Efstathiou}}, \ and\ \bibinfo {author}
  {\bibfnamefont {H.~M.~P.}\ \bibnamefont {Couchmann}} (\bibinfo
  {collaboration} {VIRGO Consortium}),\ }\href {\doibase
  10.1046/j.1365-8711.2003.06503.x} {\bibfield  {journal} {\bibinfo  {journal}
  {Mon. Not. Roy. Astron. Soc.}\ }\textbf {\bibinfo {volume} {341}},\ \bibinfo
  {pages} {1311} (\bibinfo {year} {2003})},\ \Eprint
  {http://arxiv.org/abs/astro-ph/0207664} {arXiv:astro-ph/0207664} \BibitemShut
  {NoStop}%
\bibitem [{\citenamefont {Percival}\ \emph
  {et~al.}(2007{\natexlab{b}})\citenamefont {Percival}, \citenamefont {Cole},
  \citenamefont {Eisenstein}, \citenamefont {Nichol}, \citenamefont {Peacock},
  \citenamefont {Pope},\ and\ \citenamefont {Szalay}}]{Percival:2007yw}%
  \BibitemOpen
  \bibfield  {author} {\bibinfo {author} {\bibfnamefont {W.~J.}\ \bibnamefont
  {Percival}}, \bibinfo {author} {\bibfnamefont {S.}~\bibnamefont {Cole}},
  \bibinfo {author} {\bibfnamefont {D.~J.}\ \bibnamefont {Eisenstein}},
  \bibinfo {author} {\bibfnamefont {R.~C.}\ \bibnamefont {Nichol}}, \bibinfo
  {author} {\bibfnamefont {J.~A.}\ \bibnamefont {Peacock}}, \bibinfo {author}
  {\bibfnamefont {A.~C.}\ \bibnamefont {Pope}}, \ and\ \bibinfo {author}
  {\bibfnamefont {A.~S.}\ \bibnamefont {Szalay}},\ }\href {\doibase
  10.1111/j.1365-2966.2007.12268.x} {\bibfield  {journal} {\bibinfo  {journal}
  {Mon. Not. Roy. Astron. Soc.}\ }\textbf {\bibinfo {volume} {381}},\ \bibinfo
  {pages} {1053} (\bibinfo {year} {2007}{\natexlab{b}})},\ \Eprint
  {http://arxiv.org/abs/0705.3323} {arXiv:0705.3323 [astro-ph]} \BibitemShut
  {NoStop}%
\bibitem [{\citenamefont {Alcock}\ and\ \citenamefont
  {Paczynski}(1979)}]{Alcock:1979mp}%
  \BibitemOpen
  \bibfield  {author} {\bibinfo {author} {\bibfnamefont {C.}~\bibnamefont
  {Alcock}}\ and\ \bibinfo {author} {\bibfnamefont {B.}~\bibnamefont
  {Paczynski}},\ }\href {\doibase 10.1038/281358a0} {\bibfield  {journal}
  {\bibinfo  {journal} {Nature}\ }\textbf {\bibinfo {volume} {281}},\ \bibinfo
  {pages} {358} (\bibinfo {year} {1979})}\BibitemShut {NoStop}%
\bibitem [{\citenamefont {Cabre}\ and\ \citenamefont
  {Gaztanaga}(2009)}]{Cabre:2008sz}%
  \BibitemOpen
  \bibfield  {author} {\bibinfo {author} {\bibfnamefont {A.}~\bibnamefont
  {Cabre}}\ and\ \bibinfo {author} {\bibfnamefont {E.}~\bibnamefont
  {Gaztanaga}},\ }\href {\doibase 10.1111/j.1365-2966.2008.14281.x} {\bibfield
  {journal} {\bibinfo  {journal} {Mon. Not. Roy. Astron. Soc.}\ }\textbf
  {\bibinfo {volume} {393}},\ \bibinfo {pages} {1183} (\bibinfo {year}
  {2009})},\ \Eprint {http://arxiv.org/abs/0807.2460} {arXiv:0807.2460
  [astro-ph]} \BibitemShut {NoStop}%
\bibitem [{\citenamefont {Blake}\ \emph {et~al.}(2011)\citenamefont {Blake}
  \emph {et~al.}}]{Blake:2011ep}%
  \BibitemOpen
  \bibfield  {author} {\bibinfo {author} {\bibfnamefont {C.}~\bibnamefont
  {Blake}} \emph {et~al.},\ }\href {\doibase 10.1111/j.1365-2966.2011.19606.x}
  {\bibfield  {journal} {\bibinfo  {journal} {Mon. Not. Roy. Astron. Soc.}\
  }\textbf {\bibinfo {volume} {418}},\ \bibinfo {pages} {1725} (\bibinfo {year}
  {2011})},\ \Eprint {http://arxiv.org/abs/1108.2637} {arXiv:1108.2637
  [astro-ph.CO]} \BibitemShut {NoStop}%
\bibitem [{\citenamefont {Samushia}\ \emph {et~al.}(2012)\citenamefont
  {Samushia}, \citenamefont {Percival},\ and\ \citenamefont
  {Raccanelli}}]{Samushia:2011cs}%
  \BibitemOpen
  \bibfield  {author} {\bibinfo {author} {\bibfnamefont {L.}~\bibnamefont
  {Samushia}}, \bibinfo {author} {\bibfnamefont {W.~J.}\ \bibnamefont
  {Percival}}, \ and\ \bibinfo {author} {\bibfnamefont {A.}~\bibnamefont
  {Raccanelli}},\ }\href {\doibase 10.1111/j.1365-2966.2011.20169.x} {\bibfield
   {journal} {\bibinfo  {journal} {Mon. Not. Roy. Astron. Soc.}\ }\textbf
  {\bibinfo {volume} {420}},\ \bibinfo {pages} {2102} (\bibinfo {year}
  {2012})},\ \Eprint {http://arxiv.org/abs/1102.1014} {arXiv:1102.1014
  [astro-ph.CO]} \BibitemShut {NoStop}%
\bibitem [{\citenamefont {Chuang}\ and\ \citenamefont
  {Wang}(2012)}]{Chuang:2011fy}%
  \BibitemOpen
  \bibfield  {author} {\bibinfo {author} {\bibfnamefont {C.-H.}\ \bibnamefont
  {Chuang}}\ and\ \bibinfo {author} {\bibfnamefont {Y.}~\bibnamefont {Wang}},\
  }\href {\doibase 10.1111/j.1365-2966.2012.21565.x} {\bibfield  {journal}
  {\bibinfo  {journal} {Mon. Not. Roy. Astron. Soc.}\ }\textbf {\bibinfo
  {volume} {426}},\ \bibinfo {pages} {226} (\bibinfo {year} {2012})},\ \Eprint
  {http://arxiv.org/abs/1102.2251} {arXiv:1102.2251 [astro-ph.CO]} \BibitemShut
  {NoStop}%
\bibitem [{\citenamefont {de~Mattia}\ \emph {et~al.}(2021)\citenamefont
  {de~Mattia} \emph {et~al.}}]{deMattia:2020fkb}%
  \BibitemOpen
  \bibfield  {author} {\bibinfo {author} {\bibfnamefont {A.}~\bibnamefont
  {de~Mattia}} \emph {et~al.},\ }\href {\doibase 10.1093/mnras/staa3891}
  {\bibfield  {journal} {\bibinfo  {journal} {Mon. Not. Roy. Astron. Soc.}\
  }\textbf {\bibinfo {volume} {501}},\ \bibinfo {pages} {5616} (\bibinfo {year}
  {2021})},\ \Eprint {http://arxiv.org/abs/2007.09008} {arXiv:2007.09008
  [astro-ph.CO]} \BibitemShut {NoStop}%
\bibitem [{\citenamefont {Smith}\ \emph
  {et~al.}(2020{\natexlab{a}})\citenamefont {Smith} \emph
  {et~al.}}]{Smith:2020stf}%
  \BibitemOpen
  \bibfield  {author} {\bibinfo {author} {\bibfnamefont {A.}~\bibnamefont
  {Smith}} \emph {et~al.},\ }\href {\doibase 10.1093/mnras/staa2825} {\bibfield
   {journal} {\bibinfo  {journal} {Mon. Not. Roy. Astron. Soc.}\ }\textbf
  {\bibinfo {volume} {499}},\ \bibinfo {pages} {269} (\bibinfo {year}
  {2020}{\natexlab{a}})},\ \Eprint {http://arxiv.org/abs/2007.09003}
  {arXiv:2007.09003 [astro-ph.CO]} \BibitemShut {NoStop}%
\bibitem [{\citenamefont {Aghamousa}\ \emph {et~al.}(2016)\citenamefont
  {Aghamousa} \emph {et~al.}}]{Aghamousa:2016zmz}%
  \BibitemOpen
  \bibfield  {author} {\bibinfo {author} {\bibfnamefont {A.}~\bibnamefont
  {Aghamousa}} \emph {et~al.} (\bibinfo {collaboration} {DESI}),\ }\href@noop
  {} {\  (\bibinfo {year} {2016})},\ \Eprint {http://arxiv.org/abs/1611.00036}
  {arXiv:1611.00036 [astro-ph.IM]} \BibitemShut {NoStop}%
\bibitem [{\citenamefont {Laureijs}\ \emph {et~al.}(2011)\citenamefont
  {Laureijs} \emph {et~al.}}]{Laureijs:2011gra}%
  \BibitemOpen
  \bibfield  {author} {\bibinfo {author} {\bibfnamefont {R.}~\bibnamefont
  {Laureijs}} \emph {et~al.} (\bibinfo {collaboration} {EUCLID}),\ }\href@noop
  {} {\  (\bibinfo {year} {2011})},\ \Eprint {http://arxiv.org/abs/1110.3193}
  {arXiv:1110.3193 [astro-ph.CO]} \BibitemShut {NoStop}%
\bibitem [{\citenamefont {Lesgourgues}\ and\ \citenamefont
  {Pastor}(2006)}]{Lesgourgues:2006nd}%
  \BibitemOpen
  \bibfield  {author} {\bibinfo {author} {\bibfnamefont {J.}~\bibnamefont
  {Lesgourgues}}\ and\ \bibinfo {author} {\bibfnamefont {S.}~\bibnamefont
  {Pastor}},\ }\href {\doibase 10.1016/j.physrep.2006.04.001} {\bibfield
  {journal} {\bibinfo  {journal} {Phys. Rept.}\ }\textbf {\bibinfo {volume}
  {429}},\ \bibinfo {pages} {307} (\bibinfo {year} {2006})},\ \Eprint
  {http://arxiv.org/abs/astro-ph/0603494} {arXiv:astro-ph/0603494 [astro-ph]}
  \BibitemShut {NoStop}%
\bibitem [{\citenamefont {Poulin}\ \emph {et~al.}(2019)\citenamefont {Poulin},
  \citenamefont {Smith}, \citenamefont {Karwal},\ and\ \citenamefont
  {Kamionkowski}}]{Poulin:2018cxd}%
  \BibitemOpen
  \bibfield  {author} {\bibinfo {author} {\bibfnamefont {V.}~\bibnamefont
  {Poulin}}, \bibinfo {author} {\bibfnamefont {T.~L.}\ \bibnamefont {Smith}},
  \bibinfo {author} {\bibfnamefont {T.}~\bibnamefont {Karwal}}, \ and\ \bibinfo
  {author} {\bibfnamefont {M.}~\bibnamefont {Kamionkowski}},\ }\href {\doibase
  10.1103/PhysRevLett.122.221301} {\bibfield  {journal} {\bibinfo  {journal}
  {Phys. Rev. Lett.}\ }\textbf {\bibinfo {volume} {122}},\ \bibinfo {pages}
  {221301} (\bibinfo {year} {2019})},\ \Eprint
  {http://arxiv.org/abs/1811.04083} {arXiv:1811.04083 [astro-ph.CO]}
  \BibitemShut {NoStop}%
\bibitem [{\citenamefont {Hill}\ \emph {et~al.}(2020)\citenamefont {Hill},
  \citenamefont {McDonough}, \citenamefont {Toomey},\ and\ \citenamefont
  {Alexander}}]{Hill:2020osr}%
  \BibitemOpen
  \bibfield  {author} {\bibinfo {author} {\bibfnamefont {J.~C.}\ \bibnamefont
  {Hill}}, \bibinfo {author} {\bibfnamefont {E.}~\bibnamefont {McDonough}},
  \bibinfo {author} {\bibfnamefont {M.~W.}\ \bibnamefont {Toomey}}, \ and\
  \bibinfo {author} {\bibfnamefont {S.}~\bibnamefont {Alexander}},\ }\href
  {\doibase 10.1103/PhysRevD.102.043507} {\bibfield  {journal} {\bibinfo
  {journal} {Phys. Rev. D}\ }\textbf {\bibinfo {volume} {102}},\ \bibinfo
  {pages} {043507} (\bibinfo {year} {2020})},\ \Eprint
  {http://arxiv.org/abs/2003.07355} {arXiv:2003.07355 [astro-ph.CO]}
  \BibitemShut {NoStop}%
\bibitem [{\citenamefont {Ivanov}\ \emph
  {et~al.}(2020{\natexlab{a}})\citenamefont {Ivanov}, \citenamefont
  {McDonough}, \citenamefont {Hill}, \citenamefont {Simonovi\'c}, \citenamefont
  {Toomey}, \citenamefont {Alexander},\ and\ \citenamefont
  {Zaldarriaga}}]{Ivanov:2020ril}%
  \BibitemOpen
  \bibfield  {author} {\bibinfo {author} {\bibfnamefont {M.~M.}\ \bibnamefont
  {Ivanov}}, \bibinfo {author} {\bibfnamefont {E.}~\bibnamefont {McDonough}},
  \bibinfo {author} {\bibfnamefont {J.~C.}\ \bibnamefont {Hill}}, \bibinfo
  {author} {\bibfnamefont {M.}~\bibnamefont {Simonovi\'c}}, \bibinfo {author}
  {\bibfnamefont {M.~W.}\ \bibnamefont {Toomey}}, \bibinfo {author}
  {\bibfnamefont {S.}~\bibnamefont {Alexander}}, \ and\ \bibinfo {author}
  {\bibfnamefont {M.}~\bibnamefont {Zaldarriaga}},\ }\href {\doibase
  10.1103/PhysRevD.102.103502} {\bibfield  {journal} {\bibinfo  {journal}
  {Phys. Rev. D}\ }\textbf {\bibinfo {volume} {102}},\ \bibinfo {pages}
  {103502} (\bibinfo {year} {2020}{\natexlab{a}})},\ \Eprint
  {http://arxiv.org/abs/2006.11235} {arXiv:2006.11235 [astro-ph.CO]}
  \BibitemShut {NoStop}%
\bibitem [{\citenamefont {D'Amico}\ \emph
  {et~al.}(2020{\natexlab{a}})\citenamefont {D'Amico}, \citenamefont
  {Senatore}, \citenamefont {Zhang},\ and\ \citenamefont
  {Zheng}}]{DAmico:2020ods}%
  \BibitemOpen
  \bibfield  {author} {\bibinfo {author} {\bibfnamefont {G.}~\bibnamefont
  {D'Amico}}, \bibinfo {author} {\bibfnamefont {L.}~\bibnamefont {Senatore}},
  \bibinfo {author} {\bibfnamefont {P.}~\bibnamefont {Zhang}}, \ and\ \bibinfo
  {author} {\bibfnamefont {H.}~\bibnamefont {Zheng}},\ }\href@noop {} {\
  (\bibinfo {year} {2020}{\natexlab{a}})},\ \Eprint
  {http://arxiv.org/abs/2006.12420} {arXiv:2006.12420 [astro-ph.CO]}
  \BibitemShut {NoStop}%
\bibitem [{\citenamefont {Baumann}\ \emph {et~al.}(2012)\citenamefont
  {Baumann}, \citenamefont {Nicolis}, \citenamefont {Senatore},\ and\
  \citenamefont {Zaldarriaga}}]{Baumann:2010tm}%
  \BibitemOpen
  \bibfield  {author} {\bibinfo {author} {\bibfnamefont {D.}~\bibnamefont
  {Baumann}}, \bibinfo {author} {\bibfnamefont {A.}~\bibnamefont {Nicolis}},
  \bibinfo {author} {\bibfnamefont {L.}~\bibnamefont {Senatore}}, \ and\
  \bibinfo {author} {\bibfnamefont {M.}~\bibnamefont {Zaldarriaga}},\ }\href
  {\doibase 10.1088/1475-7516/2012/07/051} {\bibfield  {journal} {\bibinfo
  {journal} {JCAP}\ }\textbf {\bibinfo {volume} {1207}},\ \bibinfo {pages}
  {051} (\bibinfo {year} {2012})},\ \Eprint {http://arxiv.org/abs/1004.2488}
  {arXiv:1004.2488 [astro-ph.CO]} \BibitemShut {NoStop}%
\bibitem [{\citenamefont {Carrasco}\ \emph {et~al.}(2012)\citenamefont
  {Carrasco}, \citenamefont {Hertzberg},\ and\ \citenamefont
  {Senatore}}]{Carrasco:2012cv}%
  \BibitemOpen
  \bibfield  {author} {\bibinfo {author} {\bibfnamefont {J.~J.~M.}\
  \bibnamefont {Carrasco}}, \bibinfo {author} {\bibfnamefont {M.~P.}\
  \bibnamefont {Hertzberg}}, \ and\ \bibinfo {author} {\bibfnamefont
  {L.}~\bibnamefont {Senatore}},\ }\href {\doibase 10.1007/JHEP09(2012)082}
  {\bibfield  {journal} {\bibinfo  {journal} {JHEP}\ }\textbf {\bibinfo
  {volume} {09}},\ \bibinfo {pages} {082} (\bibinfo {year} {2012})},\ \Eprint
  {http://arxiv.org/abs/1206.2926} {arXiv:1206.2926 [astro-ph.CO]} \BibitemShut
  {NoStop}%
\bibitem [{\citenamefont {Porto}\ \emph {et~al.}(2014)\citenamefont {Porto},
  \citenamefont {Senatore},\ and\ \citenamefont {Zaldarriaga}}]{Porto:2013qua}%
  \BibitemOpen
  \bibfield  {author} {\bibinfo {author} {\bibfnamefont {R.~A.}\ \bibnamefont
  {Porto}}, \bibinfo {author} {\bibfnamefont {L.}~\bibnamefont {Senatore}}, \
  and\ \bibinfo {author} {\bibfnamefont {M.}~\bibnamefont {Zaldarriaga}},\
  }\href {\doibase 10.1088/1475-7516/2014/05/022} {\bibfield  {journal}
  {\bibinfo  {journal} {JCAP}\ }\textbf {\bibinfo {volume} {1405}},\ \bibinfo
  {pages} {022} (\bibinfo {year} {2014})},\ \Eprint
  {http://arxiv.org/abs/1311.2168} {arXiv:1311.2168 [astro-ph.CO]} \BibitemShut
  {NoStop}%
\bibitem [{\citenamefont {Vlah}\ \emph {et~al.}(2015)\citenamefont {Vlah},
  \citenamefont {White},\ and\ \citenamefont {Aviles}}]{Vlah:2015sea}%
  \BibitemOpen
  \bibfield  {author} {\bibinfo {author} {\bibfnamefont {Z.}~\bibnamefont
  {Vlah}}, \bibinfo {author} {\bibfnamefont {M.}~\bibnamefont {White}}, \ and\
  \bibinfo {author} {\bibfnamefont {A.}~\bibnamefont {Aviles}},\ }\href
  {\doibase 10.1088/1475-7516/2015/09/014} {\bibfield  {journal} {\bibinfo
  {journal} {JCAP}\ }\textbf {\bibinfo {volume} {09}},\ \bibinfo {pages} {014}
  (\bibinfo {year} {2015})},\ \Eprint {http://arxiv.org/abs/1506.05264}
  {arXiv:1506.05264 [astro-ph.CO]} \BibitemShut {NoStop}%
\bibitem [{\citenamefont {Blas}\ \emph
  {et~al.}(2016{\natexlab{a}})\citenamefont {Blas}, \citenamefont {Garny},
  \citenamefont {Ivanov},\ and\ \citenamefont {Sibiryakov}}]{Blas:2015qsi}%
  \BibitemOpen
  \bibfield  {author} {\bibinfo {author} {\bibfnamefont {D.}~\bibnamefont
  {Blas}}, \bibinfo {author} {\bibfnamefont {M.}~\bibnamefont {Garny}},
  \bibinfo {author} {\bibfnamefont {M.~M.}\ \bibnamefont {Ivanov}}, \ and\
  \bibinfo {author} {\bibfnamefont {S.}~\bibnamefont {Sibiryakov}},\ }\href
  {\doibase 10.1088/1475-7516/2016/07/052} {\bibfield  {journal} {\bibinfo
  {journal} {JCAP}\ }\textbf {\bibinfo {volume} {1607}},\ \bibinfo {pages}
  {052} (\bibinfo {year} {2016}{\natexlab{a}})},\ \Eprint
  {http://arxiv.org/abs/1512.05807} {arXiv:1512.05807 [astro-ph.CO]}
  \BibitemShut {NoStop}%
\bibitem [{\citenamefont {Chen}\ \emph {et~al.}(2021)\citenamefont {Chen},
  \citenamefont {Vlah}, \citenamefont {Castorina},\ and\ \citenamefont
  {White}}]{Chen:2020zjt}%
  \BibitemOpen
  \bibfield  {author} {\bibinfo {author} {\bibfnamefont {S.-F.}\ \bibnamefont
  {Chen}}, \bibinfo {author} {\bibfnamefont {Z.}~\bibnamefont {Vlah}}, \bibinfo
  {author} {\bibfnamefont {E.}~\bibnamefont {Castorina}}, \ and\ \bibinfo
  {author} {\bibfnamefont {M.}~\bibnamefont {White}},\ }\href {\doibase
  10.1088/1475-7516/2021/03/100} {\bibfield  {journal} {\bibinfo  {journal}
  {JCAP}\ }\textbf {\bibinfo {volume} {03}},\ \bibinfo {pages} {100} (\bibinfo
  {year} {2021})},\ \Eprint {http://arxiv.org/abs/2012.04636} {arXiv:2012.04636
  [astro-ph.CO]} \BibitemShut {NoStop}%
\bibitem [{\citenamefont {Carrasco}\ \emph
  {et~al.}(2014{\natexlab{a}})\citenamefont {Carrasco}, \citenamefont
  {Foreman}, \citenamefont {Green},\ and\ \citenamefont
  {Senatore}}]{Carrasco:2013sva}%
  \BibitemOpen
  \bibfield  {author} {\bibinfo {author} {\bibfnamefont {J.~J.~M.}\
  \bibnamefont {Carrasco}}, \bibinfo {author} {\bibfnamefont {S.}~\bibnamefont
  {Foreman}}, \bibinfo {author} {\bibfnamefont {D.}~\bibnamefont {Green}}, \
  and\ \bibinfo {author} {\bibfnamefont {L.}~\bibnamefont {Senatore}},\ }\href
  {\doibase 10.1088/1475-7516/2014/07/056} {\bibfield  {journal} {\bibinfo
  {journal} {JCAP}\ }\textbf {\bibinfo {volume} {07}},\ \bibinfo {pages} {056}
  (\bibinfo {year} {2014}{\natexlab{a}})},\ \Eprint
  {http://arxiv.org/abs/1304.4946} {arXiv:1304.4946 [astro-ph.CO]} \BibitemShut
  {NoStop}%
\bibitem [{\citenamefont {Carrasco}\ \emph
  {et~al.}(2014{\natexlab{b}})\citenamefont {Carrasco}, \citenamefont
  {Foreman}, \citenamefont {Green},\ and\ \citenamefont
  {Senatore}}]{Carrasco:2013mua}%
  \BibitemOpen
  \bibfield  {author} {\bibinfo {author} {\bibfnamefont {J.~J.~M.}\
  \bibnamefont {Carrasco}}, \bibinfo {author} {\bibfnamefont {S.}~\bibnamefont
  {Foreman}}, \bibinfo {author} {\bibfnamefont {D.}~\bibnamefont {Green}}, \
  and\ \bibinfo {author} {\bibfnamefont {L.}~\bibnamefont {Senatore}},\ }\href
  {\doibase 10.1088/1475-7516/2014/07/057} {\bibfield  {journal} {\bibinfo
  {journal} {JCAP}\ }\textbf {\bibinfo {volume} {07}},\ \bibinfo {pages} {057}
  (\bibinfo {year} {2014}{\natexlab{b}})},\ \Eprint
  {http://arxiv.org/abs/1310.0464} {arXiv:1310.0464 [astro-ph.CO]} \BibitemShut
  {NoStop}%
\bibitem [{\citenamefont {Foreman}\ \emph {et~al.}(2016)\citenamefont
  {Foreman}, \citenamefont {Perrier},\ and\ \citenamefont
  {Senatore}}]{Foreman:2015lca}%
  \BibitemOpen
  \bibfield  {author} {\bibinfo {author} {\bibfnamefont {S.}~\bibnamefont
  {Foreman}}, \bibinfo {author} {\bibfnamefont {H.}~\bibnamefont {Perrier}}, \
  and\ \bibinfo {author} {\bibfnamefont {L.}~\bibnamefont {Senatore}},\ }\href
  {\doibase 10.1088/1475-7516/2016/05/027} {\bibfield  {journal} {\bibinfo
  {journal} {JCAP}\ }\textbf {\bibinfo {volume} {05}},\ \bibinfo {pages} {027}
  (\bibinfo {year} {2016})},\ \Eprint {http://arxiv.org/abs/1507.05326}
  {arXiv:1507.05326 [astro-ph.CO]} \BibitemShut {NoStop}%
\bibitem [{\citenamefont {Baldauf}\ \emph
  {et~al.}(2015{\natexlab{a}})\citenamefont {Baldauf}, \citenamefont
  {Mercolli},\ and\ \citenamefont {Zaldarriaga}}]{Baldauf:2015aha}%
  \BibitemOpen
  \bibfield  {author} {\bibinfo {author} {\bibfnamefont {T.}~\bibnamefont
  {Baldauf}}, \bibinfo {author} {\bibfnamefont {L.}~\bibnamefont {Mercolli}}, \
  and\ \bibinfo {author} {\bibfnamefont {M.}~\bibnamefont {Zaldarriaga}},\
  }\href {\doibase 10.1103/PhysRevD.92.123007} {\bibfield  {journal} {\bibinfo
  {journal} {Phys. Rev.}\ }\textbf {\bibinfo {volume} {D92}},\ \bibinfo {pages}
  {123007} (\bibinfo {year} {2015}{\natexlab{a}})},\ \Eprint
  {http://arxiv.org/abs/1507.02256} {arXiv:1507.02256 [astro-ph.CO]}
  \BibitemShut {NoStop}%
\bibitem [{\citenamefont {Baldauf}\ \emph
  {et~al.}(2016{\natexlab{a}})\citenamefont {Baldauf}, \citenamefont {Schaan},\
  and\ \citenamefont {Zaldarriaga}}]{Baldauf:2015zga}%
  \BibitemOpen
  \bibfield  {author} {\bibinfo {author} {\bibfnamefont {T.}~\bibnamefont
  {Baldauf}}, \bibinfo {author} {\bibfnamefont {E.}~\bibnamefont {Schaan}}, \
  and\ \bibinfo {author} {\bibfnamefont {M.}~\bibnamefont {Zaldarriaga}},\
  }\href {\doibase 10.1088/1475-7516/2016/03/007} {\bibfield  {journal}
  {\bibinfo  {journal} {JCAP}\ }\textbf {\bibinfo {volume} {03}},\ \bibinfo
  {pages} {007} (\bibinfo {year} {2016}{\natexlab{a}})},\ \Eprint
  {http://arxiv.org/abs/1507.02255} {arXiv:1507.02255 [astro-ph.CO]}
  \BibitemShut {NoStop}%
\bibitem [{\citenamefont {Baldauf}\ \emph
  {et~al.}(2016{\natexlab{b}})\citenamefont {Baldauf}, \citenamefont {Schaan},\
  and\ \citenamefont {Zaldarriaga}}]{Baldauf:2015tla}%
  \BibitemOpen
  \bibfield  {author} {\bibinfo {author} {\bibfnamefont {T.}~\bibnamefont
  {Baldauf}}, \bibinfo {author} {\bibfnamefont {E.}~\bibnamefont {Schaan}}, \
  and\ \bibinfo {author} {\bibfnamefont {M.}~\bibnamefont {Zaldarriaga}},\
  }\href {\doibase 10.1088/1475-7516/2016/03/017} {\bibfield  {journal}
  {\bibinfo  {journal} {JCAP}\ }\textbf {\bibinfo {volume} {03}},\ \bibinfo
  {pages} {017} (\bibinfo {year} {2016}{\natexlab{b}})},\ \Eprint
  {http://arxiv.org/abs/1505.07098} {arXiv:1505.07098 [astro-ph.CO]}
  \BibitemShut {NoStop}%
\bibitem [{\citenamefont {Konstandin}\ \emph {et~al.}(2019)\citenamefont
  {Konstandin}, \citenamefont {Porto},\ and\ \citenamefont
  {Rubira}}]{Konstandin:2019bay}%
  \BibitemOpen
  \bibfield  {author} {\bibinfo {author} {\bibfnamefont {T.}~\bibnamefont
  {Konstandin}}, \bibinfo {author} {\bibfnamefont {R.~A.}\ \bibnamefont
  {Porto}}, \ and\ \bibinfo {author} {\bibfnamefont {H.}~\bibnamefont
  {Rubira}},\ }\href {\doibase 10.1088/1475-7516/2019/11/027} {\bibfield
  {journal} {\bibinfo  {journal} {JCAP}\ }\textbf {\bibinfo {volume} {11}},\
  \bibinfo {pages} {027} (\bibinfo {year} {2019})},\ \Eprint
  {http://arxiv.org/abs/1906.00997} {arXiv:1906.00997 [astro-ph.CO]}
  \BibitemShut {NoStop}%
\bibitem [{\citenamefont {Senatore}(2015)}]{Senatore:2014eva}%
  \BibitemOpen
  \bibfield  {author} {\bibinfo {author} {\bibfnamefont {L.}~\bibnamefont
  {Senatore}},\ }\href {\doibase 10.1088/1475-7516/2015/11/007} {\bibfield
  {journal} {\bibinfo  {journal} {JCAP}\ }\textbf {\bibinfo {volume} {1511}},\
  \bibinfo {pages} {007} (\bibinfo {year} {2015})},\ \Eprint
  {http://arxiv.org/abs/1406.7843} {arXiv:1406.7843 [astro-ph.CO]} \BibitemShut
  {NoStop}%
\bibitem [{\citenamefont {Angulo}\ \emph {et~al.}(2015)\citenamefont {Angulo},
  \citenamefont {Fasiello}, \citenamefont {Senatore},\ and\ \citenamefont
  {Vlah}}]{Angulo:2015eqa}%
  \BibitemOpen
  \bibfield  {author} {\bibinfo {author} {\bibfnamefont {R.}~\bibnamefont
  {Angulo}}, \bibinfo {author} {\bibfnamefont {M.}~\bibnamefont {Fasiello}},
  \bibinfo {author} {\bibfnamefont {L.}~\bibnamefont {Senatore}}, \ and\
  \bibinfo {author} {\bibfnamefont {Z.}~\bibnamefont {Vlah}},\ }\href {\doibase
  10.1088/1475-7516/2015/09/029, 10.1088/1475-7516/2015/9/029} {\bibfield
  {journal} {\bibinfo  {journal} {JCAP}\ }\textbf {\bibinfo {volume} {1509}},\
  \bibinfo {pages} {029} (\bibinfo {year} {2015})},\ \Eprint
  {http://arxiv.org/abs/1503.08826} {arXiv:1503.08826 [astro-ph.CO]}
  \BibitemShut {NoStop}%
\bibitem [{\citenamefont {Mirbabayi}\ \emph {et~al.}(2015)\citenamefont
  {Mirbabayi}, \citenamefont {Schmidt},\ and\ \citenamefont
  {Zaldarriaga}}]{Mirbabayi:2014zca}%
  \BibitemOpen
  \bibfield  {author} {\bibinfo {author} {\bibfnamefont {M.}~\bibnamefont
  {Mirbabayi}}, \bibinfo {author} {\bibfnamefont {F.}~\bibnamefont {Schmidt}},
  \ and\ \bibinfo {author} {\bibfnamefont {M.}~\bibnamefont {Zaldarriaga}},\
  }\href {\doibase 10.1088/1475-7516/2015/07/030} {\bibfield  {journal}
  {\bibinfo  {journal} {JCAP}\ }\textbf {\bibinfo {volume} {1507}},\ \bibinfo
  {pages} {030} (\bibinfo {year} {2015})},\ \Eprint
  {http://arxiv.org/abs/1412.5169} {arXiv:1412.5169 [astro-ph.CO]} \BibitemShut
  {NoStop}%
\bibitem [{\citenamefont {Assassi}\ \emph {et~al.}(2014)\citenamefont
  {Assassi}, \citenamefont {Baumann}, \citenamefont {Green},\ and\
  \citenamefont {Zaldarriaga}}]{Assassi:2014fva}%
  \BibitemOpen
  \bibfield  {author} {\bibinfo {author} {\bibfnamefont {V.}~\bibnamefont
  {Assassi}}, \bibinfo {author} {\bibfnamefont {D.}~\bibnamefont {Baumann}},
  \bibinfo {author} {\bibfnamefont {D.}~\bibnamefont {Green}}, \ and\ \bibinfo
  {author} {\bibfnamefont {M.}~\bibnamefont {Zaldarriaga}},\ }\href {\doibase
  10.1088/1475-7516/2014/08/056} {\bibfield  {journal} {\bibinfo  {journal}
  {JCAP}\ }\textbf {\bibinfo {volume} {1408}},\ \bibinfo {pages} {056}
  (\bibinfo {year} {2014})},\ \Eprint {http://arxiv.org/abs/1402.5916}
  {arXiv:1402.5916 [astro-ph.CO]} \BibitemShut {NoStop}%
\bibitem [{\citenamefont {Desjacques}\ \emph {et~al.}(2018)\citenamefont
  {Desjacques}, \citenamefont {Jeong},\ and\ \citenamefont
  {Schmidt}}]{Desjacques:2016bnm}%
  \BibitemOpen
  \bibfield  {author} {\bibinfo {author} {\bibfnamefont {V.}~\bibnamefont
  {Desjacques}}, \bibinfo {author} {\bibfnamefont {D.}~\bibnamefont {Jeong}}, \
  and\ \bibinfo {author} {\bibfnamefont {F.}~\bibnamefont {Schmidt}},\ }\href
  {\doibase 10.1016/j.physrep.2017.12.002} {\bibfield  {journal} {\bibinfo
  {journal} {Phys. Rept.}\ }\textbf {\bibinfo {volume} {733}},\ \bibinfo
  {pages} {1} (\bibinfo {year} {2018})},\ \Eprint
  {http://arxiv.org/abs/1611.09787} {arXiv:1611.09787 [astro-ph.CO]}
  \BibitemShut {NoStop}%
\bibitem [{\citenamefont {Senatore}\ and\ \citenamefont
  {Zaldarriaga}(2014)}]{Senatore:2014vja}%
  \BibitemOpen
  \bibfield  {author} {\bibinfo {author} {\bibfnamefont {L.}~\bibnamefont
  {Senatore}}\ and\ \bibinfo {author} {\bibfnamefont {M.}~\bibnamefont
  {Zaldarriaga}},\ }\href@noop {} {\  (\bibinfo {year} {2014})},\ \Eprint
  {http://arxiv.org/abs/1409.1225} {arXiv:1409.1225 [astro-ph.CO]} \BibitemShut
  {NoStop}%
\bibitem [{\citenamefont {Lewandowski}\ \emph {et~al.}(2018)\citenamefont
  {Lewandowski}, \citenamefont {Senatore}, \citenamefont {Prada}, \citenamefont
  {Zhao},\ and\ \citenamefont {Chuang}}]{Lewandowski:2015ziq}%
  \BibitemOpen
  \bibfield  {author} {\bibinfo {author} {\bibfnamefont {M.}~\bibnamefont
  {Lewandowski}}, \bibinfo {author} {\bibfnamefont {L.}~\bibnamefont
  {Senatore}}, \bibinfo {author} {\bibfnamefont {F.}~\bibnamefont {Prada}},
  \bibinfo {author} {\bibfnamefont {C.}~\bibnamefont {Zhao}}, \ and\ \bibinfo
  {author} {\bibfnamefont {C.-H.}\ \bibnamefont {Chuang}},\ }\href {\doibase
  10.1103/PhysRevD.97.063526} {\bibfield  {journal} {\bibinfo  {journal} {Phys.
  Rev. D}\ }\textbf {\bibinfo {volume} {97}},\ \bibinfo {pages} {063526}
  (\bibinfo {year} {2018})},\ \Eprint {http://arxiv.org/abs/1512.06831}
  {arXiv:1512.06831 [astro-ph.CO]} \BibitemShut {NoStop}%
\bibitem [{\citenamefont {Perko}\ \emph {et~al.}(2016)\citenamefont {Perko},
  \citenamefont {Senatore}, \citenamefont {Jennings},\ and\ \citenamefont
  {Wechsler}}]{Perko:2016puo}%
  \BibitemOpen
  \bibfield  {author} {\bibinfo {author} {\bibfnamefont {A.}~\bibnamefont
  {Perko}}, \bibinfo {author} {\bibfnamefont {L.}~\bibnamefont {Senatore}},
  \bibinfo {author} {\bibfnamefont {E.}~\bibnamefont {Jennings}}, \ and\
  \bibinfo {author} {\bibfnamefont {R.~H.}\ \bibnamefont {Wechsler}},\
  }\href@noop {} {\  (\bibinfo {year} {2016})},\ \Eprint
  {http://arxiv.org/abs/1610.09321} {arXiv:1610.09321 [astro-ph.CO]}
  \BibitemShut {NoStop}%
\bibitem [{\citenamefont {Ivanov}\ and\ \citenamefont
  {Sibiryakov}(2018)}]{Ivanov:2018gjr}%
  \BibitemOpen
  \bibfield  {author} {\bibinfo {author} {\bibfnamefont {M.~M.}\ \bibnamefont
  {Ivanov}}\ and\ \bibinfo {author} {\bibfnamefont {S.}~\bibnamefont
  {Sibiryakov}},\ }\href {\doibase 10.1088/1475-7516/2018/07/053} {\bibfield
  {journal} {\bibinfo  {journal} {JCAP}\ }\textbf {\bibinfo {volume} {1807}},\
  \bibinfo {pages} {053} (\bibinfo {year} {2018})},\ \Eprint
  {http://arxiv.org/abs/1804.05080} {arXiv:1804.05080 [astro-ph.CO]}
  \BibitemShut {NoStop}%
\bibitem [{\citenamefont {Vlah}\ and\ \citenamefont
  {White}(2019)}]{Vlah:2018ygt}%
  \BibitemOpen
  \bibfield  {author} {\bibinfo {author} {\bibfnamefont {Z.}~\bibnamefont
  {Vlah}}\ and\ \bibinfo {author} {\bibfnamefont {M.}~\bibnamefont {White}},\
  }\href {\doibase 10.1088/1475-7516/2019/03/007} {\bibfield  {journal}
  {\bibinfo  {journal} {JCAP}\ }\textbf {\bibinfo {volume} {1903}},\ \bibinfo
  {pages} {007} (\bibinfo {year} {2019})},\ \Eprint
  {http://arxiv.org/abs/1812.02775} {arXiv:1812.02775 [astro-ph.CO]}
  \BibitemShut {NoStop}%
\bibitem [{\citenamefont {Lewandowski}\ \emph {et~al.}(2015)\citenamefont
  {Lewandowski}, \citenamefont {Perko},\ and\ \citenamefont
  {Senatore}}]{Lewandowski:2014rca}%
  \BibitemOpen
  \bibfield  {author} {\bibinfo {author} {\bibfnamefont {M.}~\bibnamefont
  {Lewandowski}}, \bibinfo {author} {\bibfnamefont {A.}~\bibnamefont {Perko}},
  \ and\ \bibinfo {author} {\bibfnamefont {L.}~\bibnamefont {Senatore}},\
  }\href {\doibase 10.1088/1475-7516/2015/05/019} {\bibfield  {journal}
  {\bibinfo  {journal} {JCAP}\ }\textbf {\bibinfo {volume} {1505}},\ \bibinfo
  {pages} {019} (\bibinfo {year} {2015})},\ \Eprint
  {http://arxiv.org/abs/1412.5049} {arXiv:1412.5049 [astro-ph.CO]} \BibitemShut
  {NoStop}%
\bibitem [{\citenamefont {Senatore}\ and\ \citenamefont
  {Zaldarriaga}(2017)}]{Senatore:2017hyk}%
  \BibitemOpen
  \bibfield  {author} {\bibinfo {author} {\bibfnamefont {L.}~\bibnamefont
  {Senatore}}\ and\ \bibinfo {author} {\bibfnamefont {M.}~\bibnamefont
  {Zaldarriaga}},\ }\href@noop {} {\  (\bibinfo {year} {2017})},\ \Eprint
  {http://arxiv.org/abs/1707.04698} {arXiv:1707.04698 [astro-ph.CO]}
  \BibitemShut {NoStop}%
\bibitem [{\citenamefont {Senatore}\ and\ \citenamefont
  {Zaldarriaga}(2015)}]{Senatore:2014via}%
  \BibitemOpen
  \bibfield  {author} {\bibinfo {author} {\bibfnamefont {L.}~\bibnamefont
  {Senatore}}\ and\ \bibinfo {author} {\bibfnamefont {M.}~\bibnamefont
  {Zaldarriaga}},\ }\href {\doibase 10.1088/1475-7516/2015/02/013} {\bibfield
  {journal} {\bibinfo  {journal} {JCAP}\ }\textbf {\bibinfo {volume} {1502}},\
  \bibinfo {pages} {013} (\bibinfo {year} {2015})},\ \Eprint
  {http://arxiv.org/abs/1404.5954} {arXiv:1404.5954 [astro-ph.CO]} \BibitemShut
  {NoStop}%
\bibitem [{\citenamefont {Vlah}\ \emph {et~al.}(2016)\citenamefont {Vlah},
  \citenamefont {Seljak}, \citenamefont {Chu},\ and\ \citenamefont
  {Feng}}]{Vlah:2015zda}%
  \BibitemOpen
  \bibfield  {author} {\bibinfo {author} {\bibfnamefont {Z.}~\bibnamefont
  {Vlah}}, \bibinfo {author} {\bibfnamefont {U.}~\bibnamefont {Seljak}},
  \bibinfo {author} {\bibfnamefont {M.~Y.}\ \bibnamefont {Chu}}, \ and\
  \bibinfo {author} {\bibfnamefont {Y.}~\bibnamefont {Feng}},\ }\href {\doibase
  10.1088/1475-7516/2016/03/057} {\bibfield  {journal} {\bibinfo  {journal}
  {JCAP}\ }\textbf {\bibinfo {volume} {1603}},\ \bibinfo {pages} {057}
  (\bibinfo {year} {2016})},\ \Eprint {http://arxiv.org/abs/1509.02120}
  {arXiv:1509.02120 [astro-ph.CO]} \BibitemShut {NoStop}%
\bibitem [{\citenamefont {Blas}\ \emph
  {et~al.}(2016{\natexlab{b}})\citenamefont {Blas}, \citenamefont {Garny},
  \citenamefont {Ivanov},\ and\ \citenamefont {Sibiryakov}}]{Blas:2016sfa}%
  \BibitemOpen
  \bibfield  {author} {\bibinfo {author} {\bibfnamefont {D.}~\bibnamefont
  {Blas}}, \bibinfo {author} {\bibfnamefont {M.}~\bibnamefont {Garny}},
  \bibinfo {author} {\bibfnamefont {M.~M.}\ \bibnamefont {Ivanov}}, \ and\
  \bibinfo {author} {\bibfnamefont {S.}~\bibnamefont {Sibiryakov}},\ }\href
  {\doibase 10.1088/1475-7516/2016/07/028} {\bibfield  {journal} {\bibinfo
  {journal} {JCAP}\ }\textbf {\bibinfo {volume} {1607}},\ \bibinfo {pages}
  {028} (\bibinfo {year} {2016}{\natexlab{b}})},\ \Eprint
  {http://arxiv.org/abs/1605.02149} {arXiv:1605.02149 [astro-ph.CO]}
  \BibitemShut {NoStop}%
\bibitem [{\citenamefont {Senatore}\ and\ \citenamefont
  {Trevisan}(2018)}]{Senatore:2017pbn}%
  \BibitemOpen
  \bibfield  {author} {\bibinfo {author} {\bibfnamefont {L.}~\bibnamefont
  {Senatore}}\ and\ \bibinfo {author} {\bibfnamefont {G.}~\bibnamefont
  {Trevisan}},\ }\href {\doibase 10.1088/1475-7516/2018/05/019} {\bibfield
  {journal} {\bibinfo  {journal} {JCAP}\ }\textbf {\bibinfo {volume} {1805}},\
  \bibinfo {pages} {019} (\bibinfo {year} {2018})},\ \Eprint
  {http://arxiv.org/abs/1710.02178} {arXiv:1710.02178 [astro-ph.CO]}
  \BibitemShut {NoStop}%
\bibitem [{\citenamefont {Baldauf}\ \emph
  {et~al.}(2015{\natexlab{b}})\citenamefont {Baldauf}, \citenamefont
  {Mirbabayi}, \citenamefont {Simonović},\ and\ \citenamefont
  {Zaldarriaga}}]{Baldauf:2015xfa}%
  \BibitemOpen
  \bibfield  {author} {\bibinfo {author} {\bibfnamefont {T.}~\bibnamefont
  {Baldauf}}, \bibinfo {author} {\bibfnamefont {M.}~\bibnamefont {Mirbabayi}},
  \bibinfo {author} {\bibfnamefont {M.}~\bibnamefont {Simonović}}, \ and\
  \bibinfo {author} {\bibfnamefont {M.}~\bibnamefont {Zaldarriaga}},\ }\href
  {\doibase 10.1103/PhysRevD.92.043514} {\bibfield  {journal} {\bibinfo
  {journal} {Phys. Rev.}\ }\textbf {\bibinfo {volume} {D92}},\ \bibinfo {pages}
  {043514} (\bibinfo {year} {2015}{\natexlab{b}})},\ \Eprint
  {http://arxiv.org/abs/1504.04366} {arXiv:1504.04366 [astro-ph.CO]}
  \BibitemShut {NoStop}%
\bibitem [{\citenamefont {Baldauf}\ \emph
  {et~al.}(2016{\natexlab{c}})\citenamefont {Baldauf}, \citenamefont
  {Mirbabayi}, \citenamefont {Simonović},\ and\ \citenamefont
  {Zaldarriaga}}]{Baldauf:2016sjb}%
  \BibitemOpen
  \bibfield  {author} {\bibinfo {author} {\bibfnamefont {T.}~\bibnamefont
  {Baldauf}}, \bibinfo {author} {\bibfnamefont {M.}~\bibnamefont {Mirbabayi}},
  \bibinfo {author} {\bibfnamefont {M.}~\bibnamefont {Simonović}}, \ and\
  \bibinfo {author} {\bibfnamefont {M.}~\bibnamefont {Zaldarriaga}},\
  }\href@noop {} {\  (\bibinfo {year} {2016}{\natexlab{c}})},\ \Eprint
  {http://arxiv.org/abs/1602.00674} {arXiv:1602.00674 [astro-ph.CO]}
  \BibitemShut {NoStop}%
\bibitem [{\citenamefont {Chudaykin}\ \emph
  {et~al.}(2021{\natexlab{a}})\citenamefont {Chudaykin}, \citenamefont
  {Ivanov},\ and\ \citenamefont {Simonovi\'c}}]{Chudaykin:2020hbf}%
  \BibitemOpen
  \bibfield  {author} {\bibinfo {author} {\bibfnamefont {A.}~\bibnamefont
  {Chudaykin}}, \bibinfo {author} {\bibfnamefont {M.~M.}\ \bibnamefont
  {Ivanov}}, \ and\ \bibinfo {author} {\bibfnamefont {M.}~\bibnamefont
  {Simonovi\'c}},\ }\href {\doibase 10.1103/PhysRevD.103.043525} {\bibfield
  {journal} {\bibinfo  {journal} {Phys. Rev. D}\ }\textbf {\bibinfo {volume}
  {103}},\ \bibinfo {pages} {043525} (\bibinfo {year} {2021}{\natexlab{a}})},\
  \Eprint {http://arxiv.org/abs/2009.10724} {arXiv:2009.10724 [astro-ph.CO]}
  \BibitemShut {NoStop}%
\bibitem [{\citenamefont {Ivanov}\ \emph
  {et~al.}(2020{\natexlab{b}})\citenamefont {Ivanov}, \citenamefont
  {Simonovi\'c},\ and\ \citenamefont {Zaldarriaga}}]{Ivanov:2019pdj}%
  \BibitemOpen
  \bibfield  {author} {\bibinfo {author} {\bibfnamefont {M.~M.}\ \bibnamefont
  {Ivanov}}, \bibinfo {author} {\bibfnamefont {M.}~\bibnamefont {Simonovi\'c}},
  \ and\ \bibinfo {author} {\bibfnamefont {M.}~\bibnamefont {Zaldarriaga}},\
  }\href {\doibase 10.1088/1475-7516/2020/05/042} {\bibfield  {journal}
  {\bibinfo  {journal} {JCAP}\ }\textbf {\bibinfo {volume} {05}},\ \bibinfo
  {pages} {042} (\bibinfo {year} {2020}{\natexlab{b}})},\ \Eprint
  {http://arxiv.org/abs/1909.05277} {arXiv:1909.05277 [astro-ph.CO]}
  \BibitemShut {NoStop}%
\bibitem [{\citenamefont {D'Amico}\ \emph {et~al.}(2019)\citenamefont
  {D'Amico}, \citenamefont {Gleyzes}, \citenamefont {Kokron}, \citenamefont
  {Markovic}, \citenamefont {Senatore}, \citenamefont {Zhang}, \citenamefont
  {Beutler},\ and\ \citenamefont {Gil-Marín}}]{DAmico:2019fhj}%
  \BibitemOpen
  \bibfield  {author} {\bibinfo {author} {\bibfnamefont {G.}~\bibnamefont
  {D'Amico}}, \bibinfo {author} {\bibfnamefont {J.}~\bibnamefont {Gleyzes}},
  \bibinfo {author} {\bibfnamefont {N.}~\bibnamefont {Kokron}}, \bibinfo
  {author} {\bibfnamefont {D.}~\bibnamefont {Markovic}}, \bibinfo {author}
  {\bibfnamefont {L.}~\bibnamefont {Senatore}}, \bibinfo {author}
  {\bibfnamefont {P.}~\bibnamefont {Zhang}}, \bibinfo {author} {\bibfnamefont
  {F.}~\bibnamefont {Beutler}}, \ and\ \bibinfo {author} {\bibfnamefont
  {H.}~\bibnamefont {Gil-Marín}},\ }\href@noop {} {\  (\bibinfo {year}
  {2019})},\ \Eprint {http://arxiv.org/abs/1909.05271} {arXiv:1909.05271
  [astro-ph.CO]} \BibitemShut {NoStop}%
\bibitem [{\citenamefont {Philcox}\ \emph {et~al.}(2020)\citenamefont
  {Philcox}, \citenamefont {Ivanov}, \citenamefont {Simonovi\'c},\ and\
  \citenamefont {Zaldarriaga}}]{Philcox:2020vvt}%
  \BibitemOpen
  \bibfield  {author} {\bibinfo {author} {\bibfnamefont {O.~H.~E.}\
  \bibnamefont {Philcox}}, \bibinfo {author} {\bibfnamefont {M.~M.}\
  \bibnamefont {Ivanov}}, \bibinfo {author} {\bibfnamefont {M.}~\bibnamefont
  {Simonovi\'c}}, \ and\ \bibinfo {author} {\bibfnamefont {M.}~\bibnamefont
  {Zaldarriaga}},\ }\href {\doibase 10.1088/1475-7516/2020/05/032} {\bibfield
  {journal} {\bibinfo  {journal} {JCAP}\ }\textbf {\bibinfo {volume} {05}},\
  \bibinfo {pages} {032} (\bibinfo {year} {2020})},\ \Eprint
  {http://arxiv.org/abs/2002.04035} {arXiv:2002.04035 [astro-ph.CO]}
  \BibitemShut {NoStop}%
\bibitem [{\citenamefont {Philcox}\ \emph
  {et~al.}(2021{\natexlab{a}})\citenamefont {Philcox}, \citenamefont {Sherwin},
  \citenamefont {Farren},\ and\ \citenamefont {Baxter}}]{Philcox:2020xbv}%
  \BibitemOpen
  \bibfield  {author} {\bibinfo {author} {\bibfnamefont {O.~H.~E.}\
  \bibnamefont {Philcox}}, \bibinfo {author} {\bibfnamefont {B.~D.}\
  \bibnamefont {Sherwin}}, \bibinfo {author} {\bibfnamefont {G.~S.}\
  \bibnamefont {Farren}}, \ and\ \bibinfo {author} {\bibfnamefont {E.~J.}\
  \bibnamefont {Baxter}},\ }\href {\doibase 10.1103/PhysRevD.103.023538}
  {\bibfield  {journal} {\bibinfo  {journal} {Phys. Rev. D}\ }\textbf {\bibinfo
  {volume} {103}},\ \bibinfo {pages} {023538} (\bibinfo {year}
  {2021}{\natexlab{a}})},\ \Eprint {http://arxiv.org/abs/2008.08084}
  {arXiv:2008.08084 [astro-ph.CO]} \BibitemShut {NoStop}%
\bibitem [{\citenamefont {D'Amico}\ \emph {et~al.}(2021)\citenamefont
  {D'Amico}, \citenamefont {Senatore},\ and\ \citenamefont
  {Zhang}}]{DAmico:2020kxu}%
  \BibitemOpen
  \bibfield  {author} {\bibinfo {author} {\bibfnamefont {G.}~\bibnamefont
  {D'Amico}}, \bibinfo {author} {\bibfnamefont {L.}~\bibnamefont {Senatore}}, \
  and\ \bibinfo {author} {\bibfnamefont {P.}~\bibnamefont {Zhang}},\ }\href
  {\doibase 10.1088/1475-7516/2021/01/006} {\bibfield  {journal} {\bibinfo
  {journal} {JCAP}\ }\textbf {\bibinfo {volume} {01}},\ \bibinfo {pages} {006}
  (\bibinfo {year} {2021})},\ \Eprint {http://arxiv.org/abs/2003.07956}
  {arXiv:2003.07956 [astro-ph.CO]} \BibitemShut {NoStop}%
\bibitem [{\citenamefont {D'Amico}\ \emph
  {et~al.}(2020{\natexlab{b}})\citenamefont {D'Amico}, \citenamefont {Donath},
  \citenamefont {Senatore},\ and\ \citenamefont {Zhang}}]{DAmico:2020tty}%
  \BibitemOpen
  \bibfield  {author} {\bibinfo {author} {\bibfnamefont {G.}~\bibnamefont
  {D'Amico}}, \bibinfo {author} {\bibfnamefont {Y.}~\bibnamefont {Donath}},
  \bibinfo {author} {\bibfnamefont {L.}~\bibnamefont {Senatore}}, \ and\
  \bibinfo {author} {\bibfnamefont {P.}~\bibnamefont {Zhang}},\ }\href@noop {}
  {\  (\bibinfo {year} {2020}{\natexlab{b}})},\ \Eprint
  {http://arxiv.org/abs/2012.07554} {arXiv:2012.07554 [astro-ph.CO]}
  \BibitemShut {NoStop}%
\bibitem [{\citenamefont {Chudaykin}\ \emph
  {et~al.}(2021{\natexlab{b}})\citenamefont {Chudaykin}, \citenamefont
  {Dolgikh},\ and\ \citenamefont {Ivanov}}]{Chudaykin:2020ghx}%
  \BibitemOpen
  \bibfield  {author} {\bibinfo {author} {\bibfnamefont {A.}~\bibnamefont
  {Chudaykin}}, \bibinfo {author} {\bibfnamefont {K.}~\bibnamefont {Dolgikh}},
  \ and\ \bibinfo {author} {\bibfnamefont {M.~M.}\ \bibnamefont {Ivanov}},\
  }\href {\doibase 10.1103/PhysRevD.103.023507} {\bibfield  {journal} {\bibinfo
   {journal} {Phys. Rev. D}\ }\textbf {\bibinfo {volume} {103}},\ \bibinfo
  {pages} {023507} (\bibinfo {year} {2021}{\natexlab{b}})},\ \Eprint
  {http://arxiv.org/abs/2009.10106} {arXiv:2009.10106 [astro-ph.CO]}
  \BibitemShut {NoStop}%
\bibitem [{\citenamefont {Lagu\"e}\ \emph {et~al.}(2021)\citenamefont
  {Lagu\"e}, \citenamefont {Bond}, \citenamefont {Hlo\v{z}ek}, \citenamefont
  {Rogers}, \citenamefont {Marsh},\ and\ \citenamefont {Grin}}]{Lague:2021frh}%
  \BibitemOpen
  \bibfield  {author} {\bibinfo {author} {\bibfnamefont {A.}~\bibnamefont
  {Lagu\"e}}, \bibinfo {author} {\bibfnamefont {J.~R.}\ \bibnamefont {Bond}},
  \bibinfo {author} {\bibfnamefont {R.}~\bibnamefont {Hlo\v{z}ek}}, \bibinfo
  {author} {\bibfnamefont {K.~K.}\ \bibnamefont {Rogers}}, \bibinfo {author}
  {\bibfnamefont {D.~J.~E.}\ \bibnamefont {Marsh}}, \ and\ \bibinfo {author}
  {\bibfnamefont {D.}~\bibnamefont {Grin}},\ }\href@noop {} {\  (\bibinfo
  {year} {2021})},\ \Eprint {http://arxiv.org/abs/2104.07802} {arXiv:2104.07802
  [astro-ph.CO]} \BibitemShut {NoStop}%
\bibitem [{\citenamefont {Ivanov}\ \emph
  {et~al.}(2020{\natexlab{c}})\citenamefont {Ivanov}, \citenamefont
  {Simonovi\'c},\ and\ \citenamefont {Zaldarriaga}}]{Ivanov:2019hqk}%
  \BibitemOpen
  \bibfield  {author} {\bibinfo {author} {\bibfnamefont {M.~M.}\ \bibnamefont
  {Ivanov}}, \bibinfo {author} {\bibfnamefont {M.}~\bibnamefont {Simonovi\'c}},
  \ and\ \bibinfo {author} {\bibfnamefont {M.}~\bibnamefont {Zaldarriaga}},\
  }\href {\doibase 10.1103/PhysRevD.101.083504} {\bibfield  {journal} {\bibinfo
   {journal} {Phys. Rev. D}\ }\textbf {\bibinfo {volume} {101}},\ \bibinfo
  {pages} {083504} (\bibinfo {year} {2020}{\natexlab{c}})},\ \Eprint
  {http://arxiv.org/abs/1912.08208} {arXiv:1912.08208 [astro-ph.CO]}
  \BibitemShut {NoStop}%
\bibitem [{\citenamefont {Chudaykin}\ and\ \citenamefont
  {Ivanov}(2019)}]{Chudaykin:2019ock}%
  \BibitemOpen
  \bibfield  {author} {\bibinfo {author} {\bibfnamefont {A.}~\bibnamefont
  {Chudaykin}}\ and\ \bibinfo {author} {\bibfnamefont {M.~M.}\ \bibnamefont
  {Ivanov}},\ }\href {\doibase 10.1088/1475-7516/2019/11/034} {\bibfield
  {journal} {\bibinfo  {journal} {JCAP}\ }\textbf {\bibinfo {volume} {11}},\
  \bibinfo {pages} {034} (\bibinfo {year} {2019})},\ \Eprint
  {http://arxiv.org/abs/1907.06666} {arXiv:1907.06666 [astro-ph.CO]}
  \BibitemShut {NoStop}%
\bibitem [{\citenamefont {Sailer}\ \emph {et~al.}(2021)\citenamefont {Sailer},
  \citenamefont {Castorina}, \citenamefont {Ferraro},\ and\ \citenamefont
  {White}}]{Sailer:2021yzm}%
  \BibitemOpen
  \bibfield  {author} {\bibinfo {author} {\bibfnamefont {N.}~\bibnamefont
  {Sailer}}, \bibinfo {author} {\bibfnamefont {E.}~\bibnamefont {Castorina}},
  \bibinfo {author} {\bibfnamefont {S.}~\bibnamefont {Ferraro}}, \ and\
  \bibinfo {author} {\bibfnamefont {M.}~\bibnamefont {White}},\ }\href@noop {}
  {\  (\bibinfo {year} {2021})},\ \Eprint {http://arxiv.org/abs/2106.09713}
  {arXiv:2106.09713 [astro-ph.CO]} \BibitemShut {NoStop}%
\bibitem [{\citenamefont {Avila}\ \emph {et~al.}(2020)\citenamefont {Avila}
  \emph {et~al.}}]{Avila:2020rmp}%
  \BibitemOpen
  \bibfield  {author} {\bibinfo {author} {\bibfnamefont {S.}~\bibnamefont
  {Avila}} \emph {et~al.},\ }\href {\doibase 10.1093/mnras/staa2951} {\bibfield
   {journal} {\bibinfo  {journal} {Mon. Not. Roy. Astron. Soc.}\ }\textbf
  {\bibinfo {volume} {499}},\ \bibinfo {pages} {5486} (\bibinfo {year}
  {2020})},\ \Eprint {http://arxiv.org/abs/2007.09012} {arXiv:2007.09012
  [astro-ph.CO]} \BibitemShut {NoStop}%
\bibitem [{\citenamefont {Orsi}\ \emph {et~al.}(2010)\citenamefont {Orsi},
  \citenamefont {Baugh}, \citenamefont {Lacey}, \citenamefont {Cimatti},
  \citenamefont {Wang},\ and\ \citenamefont {Zamorani}}]{Orsi:2009mj}%
  \BibitemOpen
  \bibfield  {author} {\bibinfo {author} {\bibfnamefont {A.}~\bibnamefont
  {Orsi}}, \bibinfo {author} {\bibfnamefont {C.~M.}\ \bibnamefont {Baugh}},
  \bibinfo {author} {\bibfnamefont {C.~G.}\ \bibnamefont {Lacey}}, \bibinfo
  {author} {\bibfnamefont {A.}~\bibnamefont {Cimatti}}, \bibinfo {author}
  {\bibfnamefont {Y.}~\bibnamefont {Wang}}, \ and\ \bibinfo {author}
  {\bibfnamefont {G.}~\bibnamefont {Zamorani}},\ }\href {\doibase
  10.1111/j.1365-2966.2010.16585.x} {\bibfield  {journal} {\bibinfo  {journal}
  {Mon. Not. Roy. Astron. Soc.}\ }\textbf {\bibinfo {volume} {405}},\ \bibinfo
  {pages} {1006} (\bibinfo {year} {2010})},\ \Eprint
  {http://arxiv.org/abs/0911.0669} {arXiv:0911.0669 [astro-ph.CO]} \BibitemShut
  {NoStop}%
\bibitem [{\citenamefont {Okumura}\ \emph {et~al.}(2016)\citenamefont {Okumura}
  \emph {et~al.}}]{Okumura:2015lvp}%
  \BibitemOpen
  \bibfield  {author} {\bibinfo {author} {\bibfnamefont {T.}~\bibnamefont
  {Okumura}} \emph {et~al.},\ }\href {\doibase 10.1093/pasj/psw029} {\bibfield
  {journal} {\bibinfo  {journal} {Publ. Astron. Soc. Jap.}\ }\textbf {\bibinfo
  {volume} {68}},\ \bibinfo {pages} {38} (\bibinfo {year} {2016})},\ \Eprint
  {http://arxiv.org/abs/1511.08083} {arXiv:1511.08083 [astro-ph.CO]}
  \BibitemShut {NoStop}%
\bibitem [{\citenamefont {Orsi}\ and\ \citenamefont
  {Angulo}(2018)}]{Orsi:2017ggf}%
  \BibitemOpen
  \bibfield  {author} {\bibinfo {author} {\bibfnamefont {A.~A.}\ \bibnamefont
  {Orsi}}\ and\ \bibinfo {author} {\bibfnamefont {R.~E.}\ \bibnamefont
  {Angulo}},\ }\href {\doibase 10.1093/mnras/stx3349} {\bibfield  {journal}
  {\bibinfo  {journal} {Mon. Not. Roy. Astron. Soc.}\ }\textbf {\bibinfo
  {volume} {475}},\ \bibinfo {pages} {2530} (\bibinfo {year} {2018})},\ \Eprint
  {http://arxiv.org/abs/1708.00956} {arXiv:1708.00956 [astro-ph.CO]}
  \BibitemShut {NoStop}%
\bibitem [{\citenamefont {Ross}\ \emph {et~al.}(2020)\citenamefont {Ross} \emph
  {et~al.}}]{Ross:2020lqz}%
  \BibitemOpen
  \bibfield  {author} {\bibinfo {author} {\bibfnamefont {A.~J.}\ \bibnamefont
  {Ross}} \emph {et~al.},\ }\href {\doibase 10.1093/mnras/staa2416} {\bibfield
  {journal} {\bibinfo  {journal} {Mon. Not. Roy. Astron. Soc.}\ }\textbf
  {\bibinfo {volume} {498}},\ \bibinfo {pages} {2354} (\bibinfo {year}
  {2020})},\ \Eprint {http://arxiv.org/abs/2007.09000} {arXiv:2007.09000
  [astro-ph.CO]} \BibitemShut {NoStop}%
\bibitem [{\citenamefont {Raichoor}\ \emph {et~al.}(2020)\citenamefont
  {Raichoor} \emph {et~al.}}]{Raichoor:2020vio}%
  \BibitemOpen
  \bibfield  {author} {\bibinfo {author} {\bibfnamefont {A.}~\bibnamefont
  {Raichoor}} \emph {et~al.},\ }\href {\doibase 10.1093/mnras/staa3336}
  {\bibfield  {journal} {\bibinfo  {journal} {Mon. Not. Roy. Astron. Soc.}\
  }\textbf {\bibinfo {volume} {500}},\ \bibinfo {pages} {3254} (\bibinfo {year}
  {2020})},\ \Eprint {http://arxiv.org/abs/2007.09007} {arXiv:2007.09007
  [astro-ph.CO]} \BibitemShut {NoStop}%
\bibitem [{\citenamefont {Raichoor}\ \emph {et~al.}(2017)\citenamefont
  {Raichoor} \emph {et~al.}}]{Raichoor:2017nuz}%
  \BibitemOpen
  \bibfield  {author} {\bibinfo {author} {\bibfnamefont {A.}~\bibnamefont
  {Raichoor}} \emph {et~al.},\ }\href {\doibase 10.1093/mnras/stx1790}
  {\bibfield  {journal} {\bibinfo  {journal} {Mon. Not. Roy. Astron. Soc.}\
  }\textbf {\bibinfo {volume} {471}},\ \bibinfo {pages} {3955} (\bibinfo {year}
  {2017})},\ \Eprint {http://arxiv.org/abs/1704.00338} {arXiv:1704.00338
  [astro-ph.CO]} \BibitemShut {NoStop}%
\bibitem [{\citenamefont {Karim}\ \emph {et~al.}(2020)\citenamefont {Karim},
  \citenamefont {Lee}, \citenamefont {Eisenstein}, \citenamefont {Burtin},
  \citenamefont {Moustakas}, \citenamefont {Raichoor},\ and\ \citenamefont
  {Y\`eche}}]{Karim:2020qao}%
  \BibitemOpen
  \bibfield  {author} {\bibinfo {author} {\bibfnamefont {T.}~\bibnamefont
  {Karim}}, \bibinfo {author} {\bibfnamefont {J.~H.}\ \bibnamefont {Lee}},
  \bibinfo {author} {\bibfnamefont {D.~J.}\ \bibnamefont {Eisenstein}},
  \bibinfo {author} {\bibfnamefont {E.}~\bibnamefont {Burtin}}, \bibinfo
  {author} {\bibfnamefont {J.}~\bibnamefont {Moustakas}}, \bibinfo {author}
  {\bibfnamefont {A.}~\bibnamefont {Raichoor}}, \ and\ \bibinfo {author}
  {\bibfnamefont {C.}~\bibnamefont {Y\`eche}},\ }\href {\doibase
  10.1093/mnras/staa2270} {\bibfield  {journal} {\bibinfo  {journal} {Mon. Not.
  Roy. Astron. Soc.}\ }\textbf {\bibinfo {volume} {497}},\ \bibinfo {pages}
  {4587} (\bibinfo {year} {2020})},\ \Eprint {http://arxiv.org/abs/2007.14484}
  {arXiv:2007.14484 [astro-ph.GA]} \BibitemShut {NoStop}%
\bibitem [{\citenamefont {de~Mattia}\ and\ \citenamefont
  {Ruhlmann-Kleider}(2019)}]{deMattia:2019vdg}%
  \BibitemOpen
  \bibfield  {author} {\bibinfo {author} {\bibfnamefont {A.}~\bibnamefont
  {de~Mattia}}\ and\ \bibinfo {author} {\bibfnamefont {V.}~\bibnamefont
  {Ruhlmann-Kleider}},\ }\href {\doibase 10.1088/1475-7516/2019/08/036}
  {\bibfield  {journal} {\bibinfo  {journal} {JCAP}\ }\textbf {\bibinfo
  {volume} {08}},\ \bibinfo {pages} {036} (\bibinfo {year} {2019})},\ \Eprint
  {http://arxiv.org/abs/1904.08851} {arXiv:1904.08851 [astro-ph.CO]}
  \BibitemShut {NoStop}%
\bibitem [{\citenamefont {Tamone}\ \emph {et~al.}(2020)\citenamefont {Tamone}
  \emph {et~al.}}]{Tamone:2020qrl}%
  \BibitemOpen
  \bibfield  {author} {\bibinfo {author} {\bibfnamefont {A.}~\bibnamefont
  {Tamone}} \emph {et~al.},\ }\href {\doibase 10.1093/mnras/staa3050}
  {\bibfield  {journal} {\bibinfo  {journal} {Mon. Not. Roy. Astron. Soc.}\
  }\textbf {\bibinfo {volume} {499}},\ \bibinfo {pages} {5527} (\bibinfo {year}
  {2020})},\ \Eprint {http://arxiv.org/abs/2007.09009} {arXiv:2007.09009
  [astro-ph.CO]} \BibitemShut {NoStop}%
\bibitem [{\citenamefont {Eisenstein}\ \emph {et~al.}(2007)\citenamefont
  {Eisenstein}, \citenamefont {Seo}, \citenamefont {Sirko},\ and\ \citenamefont
  {Spergel}}]{Eisenstein:2006nk}%
  \BibitemOpen
  \bibfield  {author} {\bibinfo {author} {\bibfnamefont {D.~J.}\ \bibnamefont
  {Eisenstein}}, \bibinfo {author} {\bibfnamefont {H.-j.}\ \bibnamefont {Seo}},
  \bibinfo {author} {\bibfnamefont {E.}~\bibnamefont {Sirko}}, \ and\ \bibinfo
  {author} {\bibfnamefont {D.}~\bibnamefont {Spergel}},\ }\href {\doibase
  10.1086/518712} {\bibfield  {journal} {\bibinfo  {journal} {Astrophys. J.}\
  }\textbf {\bibinfo {volume} {664}},\ \bibinfo {pages} {675} (\bibinfo {year}
  {2007})},\ \Eprint {http://arxiv.org/abs/astro-ph/0604362}
  {arXiv:astro-ph/0604362 [astro-ph]} \BibitemShut {NoStop}%
\bibitem [{\citenamefont {Feldman}\ \emph {et~al.}(1994)\citenamefont
  {Feldman}, \citenamefont {Kaiser},\ and\ \citenamefont
  {Peacock}}]{Feldman:1993ky}%
  \BibitemOpen
  \bibfield  {author} {\bibinfo {author} {\bibfnamefont {H.~A.}\ \bibnamefont
  {Feldman}}, \bibinfo {author} {\bibfnamefont {N.}~\bibnamefont {Kaiser}}, \
  and\ \bibinfo {author} {\bibfnamefont {J.~A.}\ \bibnamefont {Peacock}},\
  }\href {\doibase 10.1086/174036} {\bibfield  {journal} {\bibinfo  {journal}
  {Astrophys. J.}\ }\textbf {\bibinfo {volume} {426}},\ \bibinfo {pages} {23}
  (\bibinfo {year} {1994})},\ \Eprint {http://arxiv.org/abs/astro-ph/9304022}
  {arXiv:astro-ph/9304022} \BibitemShut {NoStop}%
\bibitem [{\citenamefont {Yamamoto}\ \emph {et~al.}(2006)\citenamefont
  {Yamamoto}, \citenamefont {Nakamichi}, \citenamefont {Kamino}, \citenamefont
  {Bassett},\ and\ \citenamefont {Nishioka}}]{Yamamoto:2005dz}%
  \BibitemOpen
  \bibfield  {author} {\bibinfo {author} {\bibfnamefont {K.}~\bibnamefont
  {Yamamoto}}, \bibinfo {author} {\bibfnamefont {M.}~\bibnamefont {Nakamichi}},
  \bibinfo {author} {\bibfnamefont {A.}~\bibnamefont {Kamino}}, \bibinfo
  {author} {\bibfnamefont {B.~A.}\ \bibnamefont {Bassett}}, \ and\ \bibinfo
  {author} {\bibfnamefont {H.}~\bibnamefont {Nishioka}},\ }\href {\doibase
  10.1093/pasj/58.1.93} {\bibfield  {journal} {\bibinfo  {journal} {Publ.
  Astron. Soc. Jap.}\ }\textbf {\bibinfo {volume} {58}},\ \bibinfo {pages} {93}
  (\bibinfo {year} {2006})},\ \Eprint {http://arxiv.org/abs/astro-ph/0505115}
  {arXiv:astro-ph/0505115} \BibitemShut {NoStop}%
\bibitem [{\citenamefont {Hand}\ \emph {et~al.}(2018)\citenamefont {Hand},
  \citenamefont {Feng}, \citenamefont {Beutler}, \citenamefont {Li},
  \citenamefont {Modi}, \citenamefont {Seljak},\ and\ \citenamefont
  {Slepian}}]{Hand:2017pqn}%
  \BibitemOpen
  \bibfield  {author} {\bibinfo {author} {\bibfnamefont {N.}~\bibnamefont
  {Hand}}, \bibinfo {author} {\bibfnamefont {Y.}~\bibnamefont {Feng}}, \bibinfo
  {author} {\bibfnamefont {F.}~\bibnamefont {Beutler}}, \bibinfo {author}
  {\bibfnamefont {Y.}~\bibnamefont {Li}}, \bibinfo {author} {\bibfnamefont
  {C.}~\bibnamefont {Modi}}, \bibinfo {author} {\bibfnamefont {U.}~\bibnamefont
  {Seljak}}, \ and\ \bibinfo {author} {\bibfnamefont {Z.}~\bibnamefont
  {Slepian}},\ }\href {\doibase 10.3847/1538-3881/aadae0} {\bibfield  {journal}
  {\bibinfo  {journal} {Astron. J.}\ }\textbf {\bibinfo {volume} {156}},\
  \bibinfo {pages} {160} (\bibinfo {year} {2018})},\ \Eprint
  {http://arxiv.org/abs/1712.05834} {arXiv:1712.05834 [astro-ph.IM]}
  \BibitemShut {NoStop}%
\bibitem [{\citenamefont {Chudaykin}\ \emph {et~al.}(2020)\citenamefont
  {Chudaykin}, \citenamefont {Ivanov}, \citenamefont {Philcox},\ and\
  \citenamefont {Simonovi\'c}}]{Chudaykin:2020aoj}%
  \BibitemOpen
  \bibfield  {author} {\bibinfo {author} {\bibfnamefont {A.}~\bibnamefont
  {Chudaykin}}, \bibinfo {author} {\bibfnamefont {M.~M.}\ \bibnamefont
  {Ivanov}}, \bibinfo {author} {\bibfnamefont {O.~H.~E.}\ \bibnamefont
  {Philcox}}, \ and\ \bibinfo {author} {\bibfnamefont {M.}~\bibnamefont
  {Simonovi\'c}},\ }\href {\doibase 10.1103/PhysRevD.102.063533} {\bibfield
  {journal} {\bibinfo  {journal} {Phys. Rev. D}\ }\textbf {\bibinfo {volume}
  {102}},\ \bibinfo {pages} {063533} (\bibinfo {year} {2020})},\ \Eprint
  {http://arxiv.org/abs/2004.10607} {arXiv:2004.10607 [astro-ph.CO]}
  \BibitemShut {NoStop}%
\bibitem [{\citenamefont {Wadekar}\ \emph {et~al.}(2020)\citenamefont
  {Wadekar}, \citenamefont {Ivanov},\ and\ \citenamefont
  {Scoccimarro}}]{Wadekar:2020hax}%
  \BibitemOpen
  \bibfield  {author} {\bibinfo {author} {\bibfnamefont {D.}~\bibnamefont
  {Wadekar}}, \bibinfo {author} {\bibfnamefont {M.~M.}\ \bibnamefont {Ivanov}},
  \ and\ \bibinfo {author} {\bibfnamefont {R.}~\bibnamefont {Scoccimarro}},\
  }\href {\doibase 10.1103/PhysRevD.102.123521} {\bibfield  {journal} {\bibinfo
   {journal} {Phys. Rev. D}\ }\textbf {\bibinfo {volume} {102}},\ \bibinfo
  {pages} {123521} (\bibinfo {year} {2020})},\ \Eprint
  {http://arxiv.org/abs/2009.00622} {arXiv:2009.00622 [astro-ph.CO]}
  \BibitemShut {NoStop}%
\bibitem [{\citenamefont {Philcox}(2021)}]{Philcox:2020vbm}%
  \BibitemOpen
  \bibfield  {author} {\bibinfo {author} {\bibfnamefont {O.~H.~E.}\
  \bibnamefont {Philcox}},\ }\href {\doibase 10.1103/PhysRevD.103.103504}
  {\bibfield  {journal} {\bibinfo  {journal} {Phys. Rev. D}\ }\textbf {\bibinfo
  {volume} {103}},\ \bibinfo {pages} {103504} (\bibinfo {year} {2021})},\
  \Eprint {http://arxiv.org/abs/2012.09389} {arXiv:2012.09389 [astro-ph.CO]}
  \BibitemShut {NoStop}%
\bibitem [{\citenamefont {Zhao}\ \emph
  {et~al.}(2021{\natexlab{a}})\citenamefont {Zhao} \emph
  {et~al.}}]{Zhao:2020tis}%
  \BibitemOpen
  \bibfield  {author} {\bibinfo {author} {\bibfnamefont {G.-B.}\ \bibnamefont
  {Zhao}} \emph {et~al.},\ }\href {\doibase 10.1093/mnras/stab849} {\bibfield
  {journal} {\bibinfo  {journal} {Mon. Not. Roy. Astron. Soc.}\ }\textbf
  {\bibinfo {volume} {504}},\ \bibinfo {pages} {33} (\bibinfo {year}
  {2021}{\natexlab{a}})},\ \Eprint {http://arxiv.org/abs/2007.09011}
  {arXiv:2007.09011 [astro-ph.CO]} \BibitemShut {NoStop}%
\bibitem [{\citenamefont {Beutler}\ \emph
  {et~al.}(2017{\natexlab{a}})\citenamefont {Beutler} \emph
  {et~al.}}]{Beutler:2016ixs}%
  \BibitemOpen
  \bibfield  {author} {\bibinfo {author} {\bibfnamefont {F.}~\bibnamefont
  {Beutler}} \emph {et~al.} (\bibinfo {collaboration} {BOSS}),\ }\href
  {\doibase 10.1093/mnras/stw2373} {\bibfield  {journal} {\bibinfo  {journal}
  {Mon. Not. Roy. Astron. Soc.}\ }\textbf {\bibinfo {volume} {464}},\ \bibinfo
  {pages} {3409} (\bibinfo {year} {2017}{\natexlab{a}})},\ \Eprint
  {http://arxiv.org/abs/1607.03149} {arXiv:1607.03149 [astro-ph.CO]}
  \BibitemShut {NoStop}%
\bibitem [{\citenamefont {Kitaura}\ \emph {et~al.}(2016)\citenamefont {Kitaura}
  \emph {et~al.}}]{Kitaura:2015uqa}%
  \BibitemOpen
  \bibfield  {author} {\bibinfo {author} {\bibfnamefont {F.-S.}\ \bibnamefont
  {Kitaura}} \emph {et~al.},\ }\href {\doibase 10.1093/mnras/stv2826}
  {\bibfield  {journal} {\bibinfo  {journal} {Mon. Not. Roy. Astron. Soc.}\
  }\textbf {\bibinfo {volume} {456}},\ \bibinfo {pages} {4156} (\bibinfo {year}
  {2016})},\ \Eprint {http://arxiv.org/abs/1509.06400} {arXiv:1509.06400
  [astro-ph.CO]} \BibitemShut {NoStop}%
\bibitem [{\citenamefont {Ross}\ \emph {et~al.}(2015)\citenamefont {Ross},
  \citenamefont {Samushia}, \citenamefont {Howlett}, \citenamefont {Percival},
  \citenamefont {Burden},\ and\ \citenamefont {Manera}}]{Ross:2014qpa}%
  \BibitemOpen
  \bibfield  {author} {\bibinfo {author} {\bibfnamefont {A.~J.}\ \bibnamefont
  {Ross}}, \bibinfo {author} {\bibfnamefont {L.}~\bibnamefont {Samushia}},
  \bibinfo {author} {\bibfnamefont {C.}~\bibnamefont {Howlett}}, \bibinfo
  {author} {\bibfnamefont {W.~J.}\ \bibnamefont {Percival}}, \bibinfo {author}
  {\bibfnamefont {A.}~\bibnamefont {Burden}}, \ and\ \bibinfo {author}
  {\bibfnamefont {M.}~\bibnamefont {Manera}},\ }\href {\doibase
  10.1093/mnras/stv154} {\bibfield  {journal} {\bibinfo  {journal} {Mon. Not.
  Roy. Astron. Soc.}\ }\textbf {\bibinfo {volume} {449}},\ \bibinfo {pages}
  {835} (\bibinfo {year} {2015})},\ \Eprint {http://arxiv.org/abs/1409.3242}
  {arXiv:1409.3242 [astro-ph.CO]} \BibitemShut {NoStop}%
\bibitem [{\citenamefont {Beutler}\ \emph {et~al.}(2011)\citenamefont
  {Beutler}, \citenamefont {Blake}, \citenamefont {Colless}, \citenamefont
  {Jones}, \citenamefont {Staveley-Smith}, \citenamefont {Campbell},
  \citenamefont {Parker}, \citenamefont {Saunders},\ and\ \citenamefont
  {Watson}}]{Beutler:2011hx}%
  \BibitemOpen
  \bibfield  {author} {\bibinfo {author} {\bibfnamefont {F.}~\bibnamefont
  {Beutler}}, \bibinfo {author} {\bibfnamefont {C.}~\bibnamefont {Blake}},
  \bibinfo {author} {\bibfnamefont {M.}~\bibnamefont {Colless}}, \bibinfo
  {author} {\bibfnamefont {D.}~\bibnamefont {Jones}}, \bibinfo {author}
  {\bibfnamefont {L.}~\bibnamefont {Staveley-Smith}}, \bibinfo {author}
  {\bibfnamefont {L.}~\bibnamefont {Campbell}}, \bibinfo {author}
  {\bibfnamefont {Q.}~\bibnamefont {Parker}}, \bibinfo {author} {\bibfnamefont
  {W.}~\bibnamefont {Saunders}}, \ and\ \bibinfo {author} {\bibfnamefont
  {F.}~\bibnamefont {Watson}},\ }\href {\doibase
  10.1111/j.1365-2966.2011.19250.x} {\bibfield  {journal} {\bibinfo  {journal}
  {Mon. Not. Roy. Astron. Soc.}\ }\textbf {\bibinfo {volume} {416}},\ \bibinfo
  {pages} {3017} (\bibinfo {year} {2011})},\ \Eprint
  {http://arxiv.org/abs/1106.3366} {arXiv:1106.3366 [astro-ph.CO]} \BibitemShut
  {NoStop}%
\bibitem [{\citenamefont {du~Mas~des Bourboux}\ \emph
  {et~al.}(2020)\citenamefont {du~Mas~des Bourboux} \emph
  {et~al.}}]{duMasdesBourboux:2020pck}%
  \BibitemOpen
  \bibfield  {author} {\bibinfo {author} {\bibfnamefont {H.}~\bibnamefont
  {du~Mas~des Bourboux}} \emph {et~al.},\ }\href@noop {} {\  (\bibinfo {year}
  {2020})},\ \Eprint {http://arxiv.org/abs/2007.08995} {arXiv:2007.08995
  [astro-ph.CO]} \BibitemShut {NoStop}%
\bibitem [{\citenamefont {Neveux}\ \emph {et~al.}(2020)\citenamefont {Neveux}
  \emph {et~al.}}]{Neveux:2020voa}%
  \BibitemOpen
  \bibfield  {author} {\bibinfo {author} {\bibfnamefont {R.}~\bibnamefont
  {Neveux}} \emph {et~al.},\ }\href@noop {} {\  (\bibinfo {year} {2020})},\
  \Eprint {http://arxiv.org/abs/2007.08999} {arXiv:2007.08999 [astro-ph.CO]}
  \BibitemShut {NoStop}%
\bibitem [{\citenamefont {Aver}\ \emph {et~al.}(2015)\citenamefont {Aver},
  \citenamefont {Olive},\ and\ \citenamefont {Skillman}}]{Aver:2015iza}%
  \BibitemOpen
  \bibfield  {author} {\bibinfo {author} {\bibfnamefont {E.}~\bibnamefont
  {Aver}}, \bibinfo {author} {\bibfnamefont {K.~A.}\ \bibnamefont {Olive}}, \
  and\ \bibinfo {author} {\bibfnamefont {E.~D.}\ \bibnamefont {Skillman}},\
  }\href {\doibase 10.1088/1475-7516/2015/07/011} {\bibfield  {journal}
  {\bibinfo  {journal} {JCAP}\ }\textbf {\bibinfo {volume} {07}},\ \bibinfo
  {pages} {011} (\bibinfo {year} {2015})},\ \Eprint
  {http://arxiv.org/abs/1503.08146} {arXiv:1503.08146 [astro-ph.CO]}
  \BibitemShut {NoStop}%
\bibitem [{\citenamefont {Cooke}\ \emph {et~al.}(2018)\citenamefont {Cooke},
  \citenamefont {Pettini},\ and\ \citenamefont {Steidel}}]{Cooke:2017cwo}%
  \BibitemOpen
  \bibfield  {author} {\bibinfo {author} {\bibfnamefont {R.~J.}\ \bibnamefont
  {Cooke}}, \bibinfo {author} {\bibfnamefont {M.}~\bibnamefont {Pettini}}, \
  and\ \bibinfo {author} {\bibfnamefont {C.~C.}\ \bibnamefont {Steidel}},\
  }\href {\doibase 10.3847/1538-4357/aaab53} {\bibfield  {journal} {\bibinfo
  {journal} {Astrophys. J.}\ }\textbf {\bibinfo {volume} {855}},\ \bibinfo
  {pages} {102} (\bibinfo {year} {2018})},\ \Eprint
  {http://arxiv.org/abs/1710.11129} {arXiv:1710.11129 [astro-ph.CO]}
  \BibitemShut {NoStop}%
\bibitem [{\citenamefont {Wilson}\ \emph {et~al.}(2017)\citenamefont {Wilson},
  \citenamefont {Peacock}, \citenamefont {Taylor},\ and\ \citenamefont {de~la
  Torre}}]{Wilson:2015lup}%
  \BibitemOpen
  \bibfield  {author} {\bibinfo {author} {\bibfnamefont {M.~J.}\ \bibnamefont
  {Wilson}}, \bibinfo {author} {\bibfnamefont {J.~A.}\ \bibnamefont {Peacock}},
  \bibinfo {author} {\bibfnamefont {A.~N.}\ \bibnamefont {Taylor}}, \ and\
  \bibinfo {author} {\bibfnamefont {S.}~\bibnamefont {de~la Torre}},\ }\href
  {\doibase 10.1093/mnras/stw2576} {\bibfield  {journal} {\bibinfo  {journal}
  {Mon. Not. Roy. Astron. Soc.}\ }\textbf {\bibinfo {volume} {464}},\ \bibinfo
  {pages} {3121} (\bibinfo {year} {2017})},\ \Eprint
  {http://arxiv.org/abs/1511.07799} {arXiv:1511.07799 [astro-ph.CO]}
  \BibitemShut {NoStop}%
\bibitem [{\citenamefont {Beutler}\ \emph
  {et~al.}(2017{\natexlab{b}})\citenamefont {Beutler} \emph
  {et~al.}}]{Beutler:2016arn}%
  \BibitemOpen
  \bibfield  {author} {\bibinfo {author} {\bibfnamefont {F.}~\bibnamefont
  {Beutler}} \emph {et~al.} (\bibinfo {collaboration} {BOSS}),\ }\href
  {\doibase 10.1093/mnras/stw3298} {\bibfield  {journal} {\bibinfo  {journal}
  {Mon. Not. Roy. Astron. Soc.}\ }\textbf {\bibinfo {volume} {466}},\ \bibinfo
  {pages} {2242} (\bibinfo {year} {2017}{\natexlab{b}})},\ \Eprint
  {http://arxiv.org/abs/1607.03150} {arXiv:1607.03150 [astro-ph.CO]}
  \BibitemShut {NoStop}%
\bibitem [{\citenamefont {Hahn}\ \emph {et~al.}(2017)\citenamefont {Hahn},
  \citenamefont {Scoccimarro}, \citenamefont {Blanton}, \citenamefont
  {Tinker},\ and\ \citenamefont {Rodr\'\i{}guez-Torres}}]{Hahn:2016kiy}%
  \BibitemOpen
  \bibfield  {author} {\bibinfo {author} {\bibfnamefont {C.}~\bibnamefont
  {Hahn}}, \bibinfo {author} {\bibfnamefont {R.}~\bibnamefont {Scoccimarro}},
  \bibinfo {author} {\bibfnamefont {M.~R.}\ \bibnamefont {Blanton}}, \bibinfo
  {author} {\bibfnamefont {J.~L.}\ \bibnamefont {Tinker}}, \ and\ \bibinfo
  {author} {\bibfnamefont {S.~A.}\ \bibnamefont {Rodr\'\i{}guez-Torres}},\
  }\href {\doibase 10.1093/mnras/stx185} {\bibfield  {journal} {\bibinfo
  {journal} {Mon. Not. Roy. Astron. Soc.}\ }\textbf {\bibinfo {volume} {467}},\
  \bibinfo {pages} {1940} (\bibinfo {year} {2017})},\ \Eprint
  {http://arxiv.org/abs/1609.01714} {arXiv:1609.01714 [astro-ph.CO]}
  \BibitemShut {NoStop}%
\bibitem [{\citenamefont {Eggemeier}\ \emph {et~al.}(2021)\citenamefont
  {Eggemeier}, \citenamefont {Scoccimarro}, \citenamefont {Smith},
  \citenamefont {Crocce}, \citenamefont {Pezzotta},\ and\ \citenamefont
  {S\'anchez}}]{Eggemeier:2021cam}%
  \BibitemOpen
  \bibfield  {author} {\bibinfo {author} {\bibfnamefont {A.}~\bibnamefont
  {Eggemeier}}, \bibinfo {author} {\bibfnamefont {R.}~\bibnamefont
  {Scoccimarro}}, \bibinfo {author} {\bibfnamefont {R.~E.}\ \bibnamefont
  {Smith}}, \bibinfo {author} {\bibfnamefont {M.}~\bibnamefont {Crocce}},
  \bibinfo {author} {\bibfnamefont {A.}~\bibnamefont {Pezzotta}}, \ and\
  \bibinfo {author} {\bibfnamefont {A.~G.}\ \bibnamefont {S\'anchez}},\
  }\href@noop {} {\  (\bibinfo {year} {2021})},\ \Eprint
  {http://arxiv.org/abs/2102.06902} {arXiv:2102.06902 [astro-ph.CO]}
  \BibitemShut {NoStop}%
\bibitem [{\citenamefont {Barreira}\ \emph {et~al.}(2021)\citenamefont
  {Barreira}, \citenamefont {Lazeyras},\ and\ \citenamefont
  {Schmidt}}]{Barreira:2021ukk}%
  \BibitemOpen
  \bibfield  {author} {\bibinfo {author} {\bibfnamefont {A.}~\bibnamefont
  {Barreira}}, \bibinfo {author} {\bibfnamefont {T.}~\bibnamefont {Lazeyras}},
  \ and\ \bibinfo {author} {\bibfnamefont {F.}~\bibnamefont {Schmidt}},\
  }\href@noop {} {\  (\bibinfo {year} {2021})},\ \Eprint
  {http://arxiv.org/abs/2105.02876} {arXiv:2105.02876 [astro-ph.CO]}
  \BibitemShut {NoStop}%
\bibitem [{\citenamefont {Schmittfull}\ \emph {et~al.}(2019)\citenamefont
  {Schmittfull}, \citenamefont {Simonović}, \citenamefont {Assassi},\ and\
  \citenamefont {Zaldarriaga}}]{Schmittfull:2018yuk}%
  \BibitemOpen
  \bibfield  {author} {\bibinfo {author} {\bibfnamefont {M.}~\bibnamefont
  {Schmittfull}}, \bibinfo {author} {\bibfnamefont {M.}~\bibnamefont
  {Simonović}}, \bibinfo {author} {\bibfnamefont {V.}~\bibnamefont {Assassi}},
  \ and\ \bibinfo {author} {\bibfnamefont {M.}~\bibnamefont {Zaldarriaga}},\
  }\href {\doibase 10.1103/PhysRevD.100.043514} {\bibfield  {journal} {\bibinfo
   {journal} {Phys.\ Rev.\ D}\ }\textbf {\bibinfo {volume} {100}},\ \bibinfo
  {pages} {043514} (\bibinfo {year} {2019})},\ \Eprint
  {http://arxiv.org/abs/1811.10640} {arXiv:1811.10640 [astro-ph.CO]}
  \BibitemShut {NoStop}%
\bibitem [{\citenamefont {Zhao}\ \emph
  {et~al.}(2021{\natexlab{b}})\citenamefont {Zhao} \emph
  {et~al.}}]{Zhao:2020bib}%
  \BibitemOpen
  \bibfield  {author} {\bibinfo {author} {\bibfnamefont {C.}~\bibnamefont
  {Zhao}} \emph {et~al.},\ }\href {\doibase 10.1093/mnras/stab510} {\bibfield
  {journal} {\bibinfo  {journal} {Mon. Not. Roy. Astron. Soc.}\ }\textbf
  {\bibinfo {volume} {503}},\ \bibinfo {pages} {1149} (\bibinfo {year}
  {2021}{\natexlab{b}})},\ \Eprint {http://arxiv.org/abs/2007.08997}
  {arXiv:2007.08997 [astro-ph.CO]} \BibitemShut {NoStop}%
\bibitem [{\citenamefont {Alam}\ \emph {et~al.}(2020)\citenamefont {Alam} \emph
  {et~al.}}]{Alam:2020jvh}%
  \BibitemOpen
  \bibfield  {author} {\bibinfo {author} {\bibfnamefont {S.}~\bibnamefont
  {Alam}} \emph {et~al.},\ }\href {\doibase 10.1093/mnras/stab1150} {\
  (\bibinfo {year} {2020}),\ 10.1093/mnras/stab1150},\ \Eprint
  {http://arxiv.org/abs/2007.09004} {arXiv:2007.09004 [astro-ph.CO]}
  \BibitemShut {NoStop}%
\bibitem [{\citenamefont {Rossi}\ \emph {et~al.}(2020)\citenamefont {Rossi}
  \emph {et~al.}}]{Rossi:2020wxx}%
  \BibitemOpen
  \bibfield  {author} {\bibinfo {author} {\bibfnamefont {G.}~\bibnamefont
  {Rossi}} \emph {et~al.},\ }\href {\doibase 10.1093/mnras/staa3955} {\
  (\bibinfo {year} {2020}),\ 10.1093/mnras/staa3955},\ \Eprint
  {http://arxiv.org/abs/2007.09002} {arXiv:2007.09002 [astro-ph.CO]}
  \BibitemShut {NoStop}%
\bibitem [{\citenamefont {Heitmann}\ \emph {et~al.}(2019)\citenamefont
  {Heitmann} \emph {et~al.}}]{Heitmann:2019ytn}%
  \BibitemOpen
  \bibfield  {author} {\bibinfo {author} {\bibfnamefont {K.}~\bibnamefont
  {Heitmann}} \emph {et~al.},\ }\href {\doibase 10.3847/1538-4365/ab4da1}
  {\bibfield  {journal} {\bibinfo  {journal} {Astrophys. J. Suppl.}\ }\textbf
  {\bibinfo {volume} {245}},\ \bibinfo {pages} {16} (\bibinfo {year} {2019})},\
  \Eprint {http://arxiv.org/abs/1904.11970} {arXiv:1904.11970 [astro-ph.CO]}
  \BibitemShut {NoStop}%
\bibitem [{\citenamefont {Hand}\ \emph {et~al.}(2017)\citenamefont {Hand},
  \citenamefont {Seljak}, \citenamefont {Beutler},\ and\ \citenamefont
  {Vlah}}]{Hand:2017ilm}%
  \BibitemOpen
  \bibfield  {author} {\bibinfo {author} {\bibfnamefont {N.}~\bibnamefont
  {Hand}}, \bibinfo {author} {\bibfnamefont {U.}~\bibnamefont {Seljak}},
  \bibinfo {author} {\bibfnamefont {F.}~\bibnamefont {Beutler}}, \ and\
  \bibinfo {author} {\bibfnamefont {Z.}~\bibnamefont {Vlah}},\ }\href {\doibase
  10.1088/1475-7516/2017/10/009} {\bibfield  {journal} {\bibinfo  {journal}
  {JCAP}\ }\textbf {\bibinfo {volume} {1710}},\ \bibinfo {pages} {009}
  (\bibinfo {year} {2017})},\ \Eprint {http://arxiv.org/abs/1706.02362}
  {arXiv:1706.02362 [astro-ph.CO]} \BibitemShut {NoStop}%
\bibitem [{\citenamefont {Smith}\ \emph
  {et~al.}(2020{\natexlab{b}})\citenamefont {Smith}, \citenamefont {de~Mattia},
  \citenamefont {Burtin}, \citenamefont {Chuang},\ and\ \citenamefont
  {Zhao}}]{Smith:2020dpo}%
  \BibitemOpen
  \bibfield  {author} {\bibinfo {author} {\bibfnamefont {A.}~\bibnamefont
  {Smith}}, \bibinfo {author} {\bibfnamefont {A.}~\bibnamefont {de~Mattia}},
  \bibinfo {author} {\bibfnamefont {E.}~\bibnamefont {Burtin}}, \bibinfo
  {author} {\bibfnamefont {C.-H.}\ \bibnamefont {Chuang}}, \ and\ \bibinfo
  {author} {\bibfnamefont {C.}~\bibnamefont {Zhao}},\ }\href {\doibase
  10.1093/mnras/staa3244} {\bibfield  {journal} {\bibinfo  {journal} {Mon. Not.
  Roy. Astron. Soc.}\ }\textbf {\bibinfo {volume} {500}},\ \bibinfo {pages}
  {259} (\bibinfo {year} {2020}{\natexlab{b}})},\ \Eprint
  {http://arxiv.org/abs/2007.11417} {arXiv:2007.11417 [astro-ph.CO]}
  \BibitemShut {NoStop}%
\bibitem [{\citenamefont {Nishimichi}\ \emph {et~al.}(2020)\citenamefont
  {Nishimichi}, \citenamefont {D'Amico}, \citenamefont {Ivanov}, \citenamefont
  {Senatore}, \citenamefont {Simonovi\'c}, \citenamefont {Takada},
  \citenamefont {Zaldarriaga},\ and\ \citenamefont
  {Zhang}}]{Nishimichi:2020tvu}%
  \BibitemOpen
  \bibfield  {author} {\bibinfo {author} {\bibfnamefont {T.}~\bibnamefont
  {Nishimichi}}, \bibinfo {author} {\bibfnamefont {G.}~\bibnamefont {D'Amico}},
  \bibinfo {author} {\bibfnamefont {M.~M.}\ \bibnamefont {Ivanov}}, \bibinfo
  {author} {\bibfnamefont {L.}~\bibnamefont {Senatore}}, \bibinfo {author}
  {\bibfnamefont {M.}~\bibnamefont {Simonovi\'c}}, \bibinfo {author}
  {\bibfnamefont {M.}~\bibnamefont {Takada}}, \bibinfo {author} {\bibfnamefont
  {M.}~\bibnamefont {Zaldarriaga}}, \ and\ \bibinfo {author} {\bibfnamefont
  {P.}~\bibnamefont {Zhang}},\ }\href {\doibase 10.1103/PhysRevD.102.123541}
  {\bibfield  {journal} {\bibinfo  {journal} {Phys. Rev. D}\ }\textbf {\bibinfo
  {volume} {102}},\ \bibinfo {pages} {123541} (\bibinfo {year} {2020})},\
  \Eprint {http://arxiv.org/abs/2003.08277} {arXiv:2003.08277 [astro-ph.CO]}
  \BibitemShut {NoStop}%
\bibitem [{\citenamefont {Wadekar}\ and\ \citenamefont
  {Scoccimarro}(2019)}]{Wadekar:2019rdu}%
  \BibitemOpen
  \bibfield  {author} {\bibinfo {author} {\bibfnamefont {D.}~\bibnamefont
  {Wadekar}}\ and\ \bibinfo {author} {\bibfnamefont {R.}~\bibnamefont
  {Scoccimarro}},\ }\href@noop {} {\  (\bibinfo {year} {2019})},\ \Eprint
  {http://arxiv.org/abs/1910.02914} {arXiv:1910.02914 [astro-ph.CO]}
  \BibitemShut {NoStop}%
\bibitem [{\citenamefont {de~Putter}\ \emph {et~al.}(2012)\citenamefont
  {de~Putter}, \citenamefont {Wagner}, \citenamefont {Mena}, \citenamefont
  {Verde},\ and\ \citenamefont {Percival}}]{dePutter:2011ah}%
  \BibitemOpen
  \bibfield  {author} {\bibinfo {author} {\bibfnamefont {R.}~\bibnamefont
  {de~Putter}}, \bibinfo {author} {\bibfnamefont {C.}~\bibnamefont {Wagner}},
  \bibinfo {author} {\bibfnamefont {O.}~\bibnamefont {Mena}}, \bibinfo {author}
  {\bibfnamefont {L.}~\bibnamefont {Verde}}, \ and\ \bibinfo {author}
  {\bibfnamefont {W.}~\bibnamefont {Percival}},\ }\href {\doibase
  10.1088/1475-7516/2012/04/019} {\bibfield  {journal} {\bibinfo  {journal}
  {JCAP}\ }\textbf {\bibinfo {volume} {04}},\ \bibinfo {pages} {019} (\bibinfo
  {year} {2012})},\ \Eprint {http://arxiv.org/abs/1111.6596} {arXiv:1111.6596
  [astro-ph.CO]} \BibitemShut {NoStop}%
\bibitem [{\citenamefont {Philcox}\ \emph
  {et~al.}(2021{\natexlab{b}})\citenamefont {Philcox}, \citenamefont {Ivanov},
  \citenamefont {Zaldarriaga}, \citenamefont {Simonovic},\ and\ \citenamefont
  {Schmittfull}}]{Philcox:2020zyp}%
  \BibitemOpen
  \bibfield  {author} {\bibinfo {author} {\bibfnamefont {O.~H.~E.}\
  \bibnamefont {Philcox}}, \bibinfo {author} {\bibfnamefont {M.~M.}\
  \bibnamefont {Ivanov}}, \bibinfo {author} {\bibfnamefont {M.}~\bibnamefont
  {Zaldarriaga}}, \bibinfo {author} {\bibfnamefont {M.}~\bibnamefont
  {Simonovic}}, \ and\ \bibinfo {author} {\bibfnamefont {M.}~\bibnamefont
  {Schmittfull}},\ }\href {\doibase 10.1103/PhysRevD.103.043508} {\bibfield
  {journal} {\bibinfo  {journal} {Phys. Rev. D}\ }\textbf {\bibinfo {volume}
  {103}},\ \bibinfo {pages} {043508} (\bibinfo {year} {2021}{\natexlab{b}})},\
  \Eprint {http://arxiv.org/abs/2009.03311} {arXiv:2009.03311 [astro-ph.CO]}
  \BibitemShut {NoStop}%
\bibitem [{\citenamefont {Jackson}(1972)}]{Jackson:2008yv}%
  \BibitemOpen
  \bibfield  {author} {\bibinfo {author} {\bibfnamefont {J.~C.}\ \bibnamefont
  {Jackson}},\ }\href {\doibase 10.1093/mnras/156.1.1P} {\bibfield  {journal}
  {\bibinfo  {journal} {Mon. Not. Roy. Astron. Soc.}\ }\textbf {\bibinfo
  {volume} {156}},\ \bibinfo {pages} {1P} (\bibinfo {year} {1972})},\ \Eprint
  {http://arxiv.org/abs/0810.3908} {arXiv:0810.3908 [astro-ph]} \BibitemShut
  {NoStop}%
\bibitem [{\citenamefont {Chen}\ \emph {et~al.}(2020)\citenamefont {Chen},
  \citenamefont {Vlah},\ and\ \citenamefont {White}}]{Chen:2020fxs}%
  \BibitemOpen
  \bibfield  {author} {\bibinfo {author} {\bibfnamefont {S.-F.}\ \bibnamefont
  {Chen}}, \bibinfo {author} {\bibfnamefont {Z.}~\bibnamefont {Vlah}}, \ and\
  \bibinfo {author} {\bibfnamefont {M.}~\bibnamefont {White}},\ }\href
  {\doibase 10.1088/1475-7516/2020/07/062} {\bibfield  {journal} {\bibinfo
  {journal} {JCAP}\ }\textbf {\bibinfo {volume} {07}},\ \bibinfo {pages} {062}
  (\bibinfo {year} {2020})},\ \Eprint {http://arxiv.org/abs/2005.00523}
  {arXiv:2005.00523 [astro-ph.CO]} \BibitemShut {NoStop}%
\bibitem [{\citenamefont {Nunes}\ and\ \citenamefont
  {Vagnozzi}(2021)}]{Nunes:2021ipq}%
  \BibitemOpen
  \bibfield  {author} {\bibinfo {author} {\bibfnamefont {R.~C.}\ \bibnamefont
  {Nunes}}\ and\ \bibinfo {author} {\bibfnamefont {S.}~\bibnamefont
  {Vagnozzi}},\ }\href {\doibase 10.1093/mnras/stab1613} {\bibfield  {journal}
  {\bibinfo  {journal} {Mon. Not. Roy. Astron. Soc.}\ }\textbf {\bibinfo
  {volume} {505}},\ \bibinfo {pages} {5427} (\bibinfo {year} {2021})},\ \Eprint
  {http://arxiv.org/abs/2106.01208} {arXiv:2106.01208 [astro-ph.CO]}
  \BibitemShut {NoStop}%
\bibitem [{\citenamefont {Di~Valentino}\ \emph {et~al.}(2021)\citenamefont
  {Di~Valentino} \emph {et~al.}}]{DiValentino:2020vvd}%
  \BibitemOpen
  \bibfield  {author} {\bibinfo {author} {\bibfnamefont {E.}~\bibnamefont
  {Di~Valentino}} \emph {et~al.},\ }\href {\doibase
  10.1016/j.astropartphys.2021.102604} {\bibfield  {journal} {\bibinfo
  {journal} {Astropart. Phys.}\ }\textbf {\bibinfo {volume} {131}},\ \bibinfo
  {pages} {102604} (\bibinfo {year} {2021})},\ \Eprint
  {http://arxiv.org/abs/2008.11285} {arXiv:2008.11285 [astro-ph.CO]}
  \BibitemShut {NoStop}%
\bibitem [{\citenamefont {{Hirata}}(2009)}]{2009MNRAS.399.1074H}%
  \BibitemOpen
  \bibfield  {author} {\bibinfo {author} {\bibfnamefont {C.~M.}\ \bibnamefont
  {{Hirata}}},\ }\href {\doibase 10.1111/j.1365-2966.2009.15353.x} {\bibfield
  {journal} {\bibinfo  {journal} {MNRAS}\ }\textbf {\bibinfo {volume} {399}},\
  \bibinfo {pages} {1074} (\bibinfo {year} {2009})},\ \Eprint
  {http://arxiv.org/abs/0903.4929} {arXiv:0903.4929 [astro-ph.CO]} \BibitemShut
  {NoStop}%
\bibitem [{\citenamefont {Obuljen}\ \emph {et~al.}(2020)\citenamefont
  {Obuljen}, \citenamefont {Percival},\ and\ \citenamefont
  {Dalal}}]{Obuljen:2020ypy}%
  \BibitemOpen
  \bibfield  {author} {\bibinfo {author} {\bibfnamefont {A.}~\bibnamefont
  {Obuljen}}, \bibinfo {author} {\bibfnamefont {W.~J.}\ \bibnamefont
  {Percival}}, \ and\ \bibinfo {author} {\bibfnamefont {N.}~\bibnamefont
  {Dalal}},\ }\href {\doibase 10.1088/1475-7516/2020/10/058} {\bibfield
  {journal} {\bibinfo  {journal} {JCAP}\ }\textbf {\bibinfo {volume} {10}},\
  \bibinfo {pages} {058} (\bibinfo {year} {2020})},\ \Eprint
  {http://arxiv.org/abs/2004.07240} {arXiv:2004.07240 [astro-ph.CO]}
  \BibitemShut {NoStop}%
\bibitem [{\citenamefont {Blas}\ \emph {et~al.}(2011)\citenamefont {Blas},
  \citenamefont {Lesgourgues},\ and\ \citenamefont {Tram}}]{Blas:2011rf}%
  \BibitemOpen
  \bibfield  {author} {\bibinfo {author} {\bibfnamefont {D.}~\bibnamefont
  {Blas}}, \bibinfo {author} {\bibfnamefont {J.}~\bibnamefont {Lesgourgues}}, \
  and\ \bibinfo {author} {\bibfnamefont {T.}~\bibnamefont {Tram}},\ }\href
  {\doibase 10.1088/1475-7516/2011/07/034} {\bibfield  {journal} {\bibinfo
  {journal} {JCAP}\ }\textbf {\bibinfo {volume} {1107}},\ \bibinfo {pages}
  {034} (\bibinfo {year} {2011})},\ \Eprint {http://arxiv.org/abs/1104.2933}
  {arXiv:1104.2933 [astro-ph.CO]} \BibitemShut {NoStop}%
\bibitem [{\citenamefont {Audren}\ \emph {et~al.}(2013)\citenamefont {Audren},
  \citenamefont {Lesgourgues}, \citenamefont {Benabed},\ and\ \citenamefont
  {Prunet}}]{Audren:2012wb}%
  \BibitemOpen
  \bibfield  {author} {\bibinfo {author} {\bibfnamefont {B.}~\bibnamefont
  {Audren}}, \bibinfo {author} {\bibfnamefont {J.}~\bibnamefont {Lesgourgues}},
  \bibinfo {author} {\bibfnamefont {K.}~\bibnamefont {Benabed}}, \ and\
  \bibinfo {author} {\bibfnamefont {S.}~\bibnamefont {Prunet}},\ }\href
  {\doibase 10.1088/1475-7516/2013/02/001} {\bibfield  {journal} {\bibinfo
  {journal} {JCAP}\ }\textbf {\bibinfo {volume} {1302}},\ \bibinfo {pages}
  {001} (\bibinfo {year} {2013})},\ \Eprint {http://arxiv.org/abs/1210.7183}
  {arXiv:1210.7183 [astro-ph.CO]} \BibitemShut {NoStop}%
\bibitem [{\citenamefont {Brinckmann}\ and\ \citenamefont
  {Lesgourgues}(2019)}]{Brinckmann:2018cvx}%
  \BibitemOpen
  \bibfield  {author} {\bibinfo {author} {\bibfnamefont {T.}~\bibnamefont
  {Brinckmann}}\ and\ \bibinfo {author} {\bibfnamefont {J.}~\bibnamefont
  {Lesgourgues}},\ }\href {\doibase 10.1016/j.dark.2018.100260} {\bibfield
  {journal} {\bibinfo  {journal} {Phys. Dark Univ.}\ }\textbf {\bibinfo
  {volume} {24}},\ \bibinfo {pages} {100260} (\bibinfo {year} {2019})},\
  \Eprint {http://arxiv.org/abs/1804.07261} {arXiv:1804.07261 [astro-ph.CO]}
  \BibitemShut {NoStop}%
\bibitem [{\citenamefont {Lewis}(2019)}]{Lewis:2019xzd}%
  \BibitemOpen
  \bibfield  {author} {\bibinfo {author} {\bibfnamefont {A.}~\bibnamefont
  {Lewis}},\ }\href@noop {} {\  (\bibinfo {year} {2019})},\ \Eprint
  {http://arxiv.org/abs/1910.13970} {arXiv:1910.13970 [astro-ph.IM]}
  \BibitemShut {NoStop}%
\bibitem [{\citenamefont {Lewis}\ and\ \citenamefont
  {Bridle}(2002)}]{Lewis:2002ah}%
  \BibitemOpen
  \bibfield  {author} {\bibinfo {author} {\bibfnamefont {A.}~\bibnamefont
  {Lewis}}\ and\ \bibinfo {author} {\bibfnamefont {S.}~\bibnamefont {Bridle}},\
  }\href {\doibase 10.1103/PhysRevD.66.103511} {\bibfield  {journal} {\bibinfo
  {journal} {Phys. Rev.}\ }\textbf {\bibinfo {volume} {D66}},\ \bibinfo {pages}
  {103511} (\bibinfo {year} {2002})},\ \Eprint
  {http://arxiv.org/abs/astro-ph/0205436} {arXiv:astro-ph/0205436 [astro-ph]}
  \BibitemShut {NoStop}%
\bibitem [{\citenamefont {Lewis}(2013)}]{Lewis:2013hha}%
  \BibitemOpen
  \bibfield  {author} {\bibinfo {author} {\bibfnamefont {A.}~\bibnamefont
  {Lewis}},\ }\href {\doibase 10.1103/PhysRevD.87.103529} {\bibfield  {journal}
  {\bibinfo  {journal} {Phys. Rev.}\ }\textbf {\bibinfo {volume} {D87}},\
  \bibinfo {pages} {103529} (\bibinfo {year} {2013})},\ \Eprint
  {http://arxiv.org/abs/1304.4473} {arXiv:1304.4473 [astro-ph.CO]} \BibitemShut
  {NoStop}%
\end{thebibliography}%

\end{document}